\documentclass[useAMS,usenatbib]{mn2e}

\input{psfig.sty}
\usepackage{graphicx}
\usepackage{epsfig}
\usepackage{amssymb}
\usepackage{amsmath}
\usepackage{amsfonts}
\usepackage{txfonts}
\usepackage{multirow}
\usepackage{afterpage}
\usepackage{color}

\title[CFHTLenS and RCSLenS $\times$ Planck Lensing]{CFHTLenS and RCSLenS Cross-Correlation with Planck Lensing Detected in Fourier and Configuration Space}
\author[Harnois-D\'{e}raps,  Tr\"oster, Hojjati, van Waerbeke, Asgari  et al.]{Joachim Harnois-D\'{e}raps$^{1,2}$\thanks{E-mail: jharno@roe.ac.uk}, Tilman Tr\"oster$^{1}$, Alireza Hojjati$^1$, Ludovic van Waerbeke$^{1}$, \newauthor 
 Marika Asgari$^2$, Ami Choi$^2$, Thomas Erben$^3$, Catherine Heymans$^2$, Hendrik Hildebrandt$^3$,   \newauthor
 Thomas D. Kitching$^4$, Lance Miller$^5$, Reiko Nakajima$^3$, Massimo Viola$^{6}$,    St\'ephane Arnouts$^7$,   \newauthor  
    Jean Coupon$^8$ \& Thibaud Moutard$^7$\\
$^{1}$Department of Physics and Astronomy, University of British Columbia, 6224 Agricultural Road, Vancouver, B.C., V6T 1Z1, Canada\\
$^{2}$Scottish Universities Physics Alliance, Institute for Astronomy, University of Edinburgh, Blackford Hill, Scotland, UK\\
$^{3}$Argelander-Institut f\"ur Astronomie, Auf dem H\"ugel 71, 53121 Bonn, Germany\\
$^{4}$Mullard Space Science Laboratory, University College London, Holmbury St Mary, Dorking, Surrey RH5 6NT, UK\\
$^{5}$Astrophysics, Department of Physics, University of Oxford, Keble Road, Oxford, OX1 3RH, UK\\
$^{6}$Leiden Observatory, Leiden University, PO Box 9513, NL-2300 RA Leiden, Netherlands\\
$^{7}$Aix Marseille UniversitŽ, CNRS, LAM - Laboratoire d'Astrophysique de Marseille, 38 rue F. Joliot-Curie, F-13388, Marseille, France\\
$^{8}$Astronomical Observatory of the University of Geneva, ch. d Ecogia16, 1290 Versoix, Switzerland.\\
}

\begin{document}

\date{\today}

\pagerange{\pageref{firstpage}--\pageref{lastpage}} \pubyear{2013}

\maketitle

\label{firstpage}

\begin{abstract}
We measure the cross-correlation signature between the Planck CMB lensing map and the  weak lensing observations  
from both the Red-sequence Cluster Lensing Survey  (RCSLenS)  and the Canada-France-Hawai Telescope Lensing Survey (CFHTLenS).  
In addition to a Fourier analysis, we include the first configuration-space detection, based on the  estimators
$\langle \kappa_{\rm CMB} \kappa_{\rm gal} \rangle$ and  $\langle \kappa_{\rm CMB} \gamma_{t} \rangle$.
Combining 747.2 deg$^2$ from both surveys, we find  a detection significance that exceeds {\color{black}$ 4.2\sigma$} in both Fourier- and configuration-space analyses. 
Scaling the predictions  by a free parameter $A$, we obtain  $A^{\rm Planck}_{\rm CFHT}= 0.68\pm 0.31 $ and {\color{black} $A^{\rm Planck}_{\rm RCS}= 1.31\pm 0.33$.}
In preparation for the next generation of measurements similar to these, we quantify the impact of different analysis choices on these results.
 First, since none of these estimators probes the exact same dynamical range, we improve our detection by combining them.   
 Second, we carry out a detailed investigation on the effect of apodization, zero-padding and mask multiplication,
validated on a suite of high-resolution simulations, and find that the latter produces the largest systematic bias in the cosmological interpretation. 
Finally, we show that residual contamination from intrinsic alignment  and the effect of photometric redshift error
are both largely degenerate with the characteristic signal from massive neutrinos, 
however the signature of baryon feedback might be easier to distinguish. {The three lensing datasets are publicly available.} 
\end{abstract} 



\section{Introduction}
\label{sec:intro}


{\color{black} In the context of precision cosmology, cross-correlation analyses between independent probes can provide additional information about our Universe and 
unique insights on systematic effects at play in the data sets. Indeed, residual systematics from the data processing and instrumentation are highly suppressed in this type of analysis,
allowing for potentially unbiased cosmological measurements
For these reasons, the interest in cross-correlations  is growing rapidly, and the types of probes are gaining in diversity.
Recent analyses combined the cosmic microwave background (CMB) lensing data with  quasars distribution \citep{2012PhRvD..86h3006S}, 
  galaxy positions \citep{2015MNRAS.451..849A, Bianchini2015, 2015arXiv150203405O, 2015arXiv150201591P,2015arXiv150705551G,2016arXiv160207384B},  
  the cosmic infrared background \citep{2013ApJ...771L..16H, 2015ApJ...808....7V},
 the $\gamma$-ray sky \citep{2015ApJ...802L...1F} 
 and with low-redshift lensing maps \citep{2015PhRvD..91f2001H, 2015PhRvD..92f3517L, 2015A&A...584A..53K, Kirk15};
 others combined lensing maps with thermal Sunyaev-Zel'dovich (SZ) data \citep{2014PhRvD..89b3508V, 2014JCAP...02..030H}
 and with large scale structure data \citep{2015arXiv150705977D, 2015arXiv150703086B, Bundendiek2015}.

Gravitational lensing by large scale structure is unique among these different cosmological probes in that it is insensitive to galaxy bias and minimally affected by many complicated astrophysical phenomena,
two of the main sources of nuisance in other probes.
It is caused by the deflection of photon trajectories by gravitational potentials encountered along the line of sight.
By measuring the distortions in the  images of lensed objects such as the CMB or distant galaxies, the lensing mass can be reconstructed
\citep[for a review, see][]{2001PhR...340..291B}. 
A cross-correlation between the CMB lensing signal and galaxy distribution was recently used to measure redshift evolution in the galaxy bias 
 \citep{2015arXiv150203405O, 2015A&A...584A..53K, Bianchini2015, 2015MNRAS.451..849A} and in the growth factor \citep{2015arXiv150705551G},
 and to improve the calibration of the cosmic shear data \citep{2016arXiv160207384B}.
In comparison, the lensing-lensing cross-correlations sample only the underlying matter density instead of the galaxies.
The statistical properties of the measurement can therefore be linked to the cosmological parameters in a more direct manner.
This carries strong potential in terms of constraining the mean matter density ($\Omega_{\rm M}$), the level of fluctuations  contained therein ($\sigma_8$),
and investigating other phenomena that can affect the signal, such as the contamination by intrinsic galaxy alignment,  the neutrino mass, and baryonic feedback mechanisms \citep{2014MNRAS.443L.119H, 2015MNRAS.449.2205K}.

The cross-correlation measurement between two independent weak lensing datasets  is still in its infancy,
with only three such analyses so far that cover $\sim$150 $\mbox{deg}^2$ each. 
The first, by \citet[][H15 hereafter]{2015PhRvD..91f2001H},  combines a  convergence map reconstructed from the CMB data collected by the Atacama Cosmology Telescope \citep[][ACT]{2014JCAP...04..014D}
with the weak lensing galaxy catalogue from the Canada-France-Hawaii Telescope Stripe 82  Survey \citep[][CS82]{2014RMxAC..44..202M}. 
The significance was high enough to claim the first lensing-lensing detection ($\sigma=4.2$)
and place a 12\% constraint on $\sigma_8$ at redshift $z\sim0.9$.
The second, by  \citet[][LH15]{2015PhRvD..92f3517L}, combined the CMB lensing map reconstructed from the Planck 2013 and 2015 data \citep{2014A&A...571A..17P,2015arXiv150201591P}, 
with the galaxy lensing catalogue from the CFHT Lensing Survey\footnote{CFHTLenS: {\tt www.cfhtlens.org}}  (CFHTLenS hereafter).
Although the total area was similar to that from H15, their measurement was less significant ($\sigma=2.0$) due to the higher noise level in the Planck lensing map compared to that from ACT.
 The third, by \citet[][K15]{Kirk15}, measured the cross-correlation between the Science Verification Data from the Dark Energy Survey\footnote{DES: {\tt www.darkenergysurvey.org}} (DES-SV) and both the Planck lensing map and the South Pole Telescope lensing map \citep[][SPT]{2012ApJ...756..142V}. Within an overlapping region of 139 deg$^2$, they found hints of the cross-correlation signal with significances of 2.2 and 2.9$\sigma$ respectively.





In this work, we extend these previous measurements by including the recently extracted   weak lensing data  
from RCSLenS\footnote{RCSLenS: {\tt www.rcslens.org}}, the Red-sequence Cluster Lensing Survey 
\citep{Gilbank2011, Hildebrandt2015}. Although this new dataset is shallower than CFHTLenS, CS82 and DES-SV, it covers an area $\sim$5 times larger than either of these.
This larger footprint greatly improves the cross-correlation measurement whose statistical error is dominated by the noise in the CMB lensing maps, which scales as the inverse 
of the area. 
The RCSLenS data has thus far been used in \citet{2015arXiv150703086B} to test the laws of gravity, 
in \citet{Bundendiek2015} to develop optimal cosmological estimators for combined shear and clustering measurements, 
in \citet{2015arXiv151203627K} to place constraints on the matter content and on the dark energy equation of state from the shear-ratio method, 
and in \citet{2015arXiv151203626C} to assess the accuracy of photometric redshift measurement from 
cross-correlation with overlapping  spectroscopic data.

The three previous CMB-lensing vs. galaxy-lensing measurements were performed in Fourier space. 
In this paper, we additionally include the first analysis in configuration-space, based on two complementary   estimators of the cross-correlation signal. 
We show that results from these are in  agreement with the Fourier analysis, 
which demonstrates the robustness of our measurements. 
Moreover, because of the finite size of the lensing surveys, the different estimators do not probe the exact same scales, which means that 
we can benefit from a joint analysis that combines them.  In preparation for the next generation of cross-correlation measurements similar to this one, 
it is therefore essential to start investigating which combination provides the best results. It is also crucial to understand how different choices made in the analysis 
and in the data treatment affect the final measurement. With regards to this, the previous works use different approaches, 
and some details justifying their choices are not clear. 
To fill this gap, we fully explore five distinct methods, or `pipelines', and propagate their impact on the cosmological interpretation.

Before discussing these rather technical details, we first summarize the theoretical background relevant for our measurement in the following section. 
We then present the three datasets and their derived products in Section \ref{sec:data}. 
Section \ref{sec:measurement} contains the core of this work where we describe our five different Fourier measurement strategies, 
our forward modelling approach,  our two configuration-space estimators, plus a series of systematic uncertainty verifications and null tests. 
Results are  detailed and interpreted in Section \ref{sec:cosmology}, and compared against the three previous measurements (H15, LH15 and K15), 
after which we present our conclusions. The appendices contain supplementary material that validates the analysis pipeline.

Two fiducial cosmologies are adopted throughout this paper, mainly for direct visual comparison in some of the figures.
The first  corresponds to the WMAP9+SN+BAO cosmology \citep{2013ApJS..208...19H},
in which the matter density, the dark energy density, the  amplitude of matter fluctuations, the Hubble parameter 
and the tilt of the matter power spectrum are described by  $(\Omega_{\rm M}, \Omega_{\Lambda}, \sigma_8, h, n_s) = (0.2905, 0.7095, 0.831, 0.6898, 0.969)$.
The second corresponds to the Planck 2013 data release \citep{PlanckCollaboration2013}, with
$(\Omega_{\rm M}, \Omega_{\Lambda}, \sigma_8, h, n_s) = (0.308, 0.692, 0.826, 0.678, 0.961)$.
In both cases, we assume a flat cosmology.
}

\section{Background}
\label{sec:th}

This section briefly reviews the theoretical framework from which we extract predictions for our cross-correlation signal.
Gravitational lensing occurs whenever photons travelling from distance sources are deflected from their original trajectories by a foreground mass distribution.
The convergence $\kappa$ at position ${\boldsymbol \theta}$ on the sky can be related to the {\color{black} matter density contrast }$\delta_{\rm m}$ 
distributed  along the line of sight at comoving distance $\chi$ from the observer by:
\begin{eqnarray}
\kappa({\boldsymbol \theta}) = \int_0^{\chi_H} {\rm d}\chi W(\chi) \delta_{\rm m}(\chi,\chi{\boldsymbol \theta}) ,
\label{eq:kappa}
\end{eqnarray}
with 
\begin{eqnarray}
W(\chi) = \frac{3 \Omega_{\rm m} H_{0}^{2}}{2 c^2} \chi g(\chi) (1 + z) 
\label{eq:W-def}
\end{eqnarray}
and
\begin{eqnarray}
g(\chi) = \int_{\chi}^{\chi_H} {\rm d} \chi' n(\chi') \frac{\chi' - \chi}{\chi'}.
\label{eq:g-chi-def}
\end{eqnarray}
In the above expressions, $c$ is the speed of light in vacuum, $H_{0}$ the Hubble parameter, $\chi_H$ is the comoving distance to the horizon and $z$ the redshift.
{\color{black} The term $n(\chi)$ is related to the redshift distribution of the observed  galaxy sources, $n(z)$, by 
$n(\chi) = n(z) \mbox{d}z/\mbox{d}\chi$, and is mainly determined by the depth of the survey and the galaxy selection function. }
We describe in Section \ref{subsec:CFHTLenS_data} how this quantity is extracted from our data.    
Then, using Limber's approximation \citep{Limber1954b},  the angular power spectrum of the $\kappa$ field, $C^{\kappa}_{\ell} $, can be estimated from:
 \begin{eqnarray}
C^{\kappa}_{\ell} = \int_0^{\infty} d\chi W^2(\chi) P(\ell/\chi;z),
\label{eq:limber}
\end{eqnarray}
where $P(k,z)$ is the matter power spectrum and $\ell$ is the angular multipole. 
In the case where two different estimates of the convergence field, $\kappa_i$ and $\kappa_j$, 
are combined in a  cross-correlation analysis, the corresponding theoretical prediction is      
{\color{black} 
the angular cross-spectrum between $\kappa_i$ and $\kappa_j$, 
 which can be estimated as:
 \begin{eqnarray}
C^{\kappa_{i}\kappa_{j}}_{\ell} = \int_0^{\infty} d\chi W^{{i}}(\chi)W^{{j}}(\chi) P(\ell/\chi;z).
\label{eq:limber_cross}
\end{eqnarray}
In the above expression, $W^{i}$ and $W^{j}$ indicate the  lensing kernels defined in equation \ref{eq:W-def} with their respective $g^i(\chi)$ and $g^j(\chi)$. 
These two are generally different because of  their distinct source redshift distributions. 
In this paper we combine maps of convergence $\kappa$ -- or simply $\kappa$ maps -- from two weak lensing galaxy surveys, CFHTLenS and RCSLenS, with $\kappa$ maps obtained from the Planck 2015 data release. 
We refer to these maps as $\kappa_{\rm gal}$ and $\kappa_{\rm CMB}$ respectively, 
 and the two kernels that enter equation \ref{eq:limber_cross} are labelled $W^{\rm gal}$ and $W^{\rm CMB}$.
Once the $n(z)$ is known, $W^{\rm gal}$ can be directly evaluated from equations \ref{eq:W-def} and \ref{eq:g-chi-def}.
In evaluating  $W^{\rm CMB}$, we approximate the source distribution of the CMB photons as a single  redshift  plane at $z_*=1080$, which turns equation \ref{eq:g-chi-def} into:
\begin{eqnarray} 
g^{\rm CMB}(\chi) = 1 - \frac{\chi}{\chi_*}.
\label{eq:g_CMB}
\end{eqnarray} 
}

The angular cross-spectrum is related to the two configuration-space two-point correlation functions  \citep{1991ApJ...380....1M}: 
\begin{eqnarray}
	\xi^{\kappa_{\rm CMB}\kappa_{\rm gal}}( \vartheta) = \frac{1}{2\pi} \int_0^\infty \mbox{d} \ell\ \ell C^{\kappa_{\rm CMB}\kappa_{\rm gal}}_\ell J_{0}(\ell \vartheta) 
	\label{eq:xi1-prediction}
	\end{eqnarray}
	\begin{eqnarray}
	\xi^{\kappa_{\rm CMB}\gamma_t}( \vartheta) = \frac{1}{2\pi} \int_0^\infty \mbox{d} \ell\ \ell C^{\kappa_{\rm CMB}\kappa_{\rm gal}}_\ell J_{2}(\ell  \vartheta)
	\label{eq:xi2-prediction}
\end{eqnarray}
which we also measure from the data.
Here, ${J_0}$ and ${J_2}$ are Bessel functions of order 0 and 2, and the quantity $\vartheta$ represents the angular separation on the sky, not to be confused with the pixel coordinate $\boldsymbol \theta$. 
In the last expression, $\gamma_t$ is the {\color{black} tangential shear, which refers to the component of the shear that aligns tangentially around $\boldsymbol \theta$, averaged in rings of radius $\vartheta$.
Details about measurements of $\gamma_t$ are provided in Section \ref{subsubsec:real_space}.}

\section{The Data Sets}
\label{sec:data}

This section reviews the data sets and procedures with which the  $\kappa_{\rm gal}$ and $\kappa_{\rm CMB}$ maps are constructed.

\subsection{$\kappa_{\rm gal}$ maps}
\label{subsec:CFHTLenS_data}

The CFHTLenS and RCSLenS data sets were both imaged by the MegaCAM camera mounted on the Canada-France-Hawaii Telescope,  located on the Mauna Kea volcano in Hawaii.
The images are then processed by a  reduction algorithm \citep{2013MNRAS.433.2545E}, photometric redshift estimator \citep{2012MNRAS.421.2355H, 2000ApJ...536..571B} 
and shape finder \citep{2007MNRAS.382..315M, 2008MNRAS.390..149K, 2013MNRAS.429.2858M}.
Thorough description and systematic verification of these two lensing data sets are presented in \citet{Heymans2012c}  and \citet{Hildebrandt2015},
and we highlight here their main properties. 

\subsubsection{CFHTLenS shear data}

The CFHTLenS is split into four regions, referred to as the W1-4 fields, that cover a total of 154 deg$^2$ on the sky.
The galaxies are selected from a photometric redshift cut, demanding that the redshift of maximum likelihood, $z_B$,  falls in the range
$0.4 < z_B < 1.1$. Galaxies are also required to have a shape signal weight (called the  {\it lens}fit weight) greater than zero, and to
pass the star-galaxy separation test (i.e. {\tt FIT\_CLASS} = 1).
This results in a sample of 4,760,606 selected galaxies, with ellipticity dispersion $\sigma_{\epsilon}= 0.278$ and effective number density of $\bar{n} = 9.06$ gal/arcmin$^{2}$ 
\citep[assuming the definition of][]{Heymans2012c}.
We estimate the redshift distribution of this sample by stacking the full photometric redshift PDF from all selected galaxies weighted by the {\it lens}fit weight, as in \citet{2013MNRAS.431.1547B}.
This results in a $n_{\rm CFHT}(z)$ distribution that is well fitted by:
\begin{eqnarray}
    n_{\rm CFHT}(z) = a_1 {\rm exp} \bigg[ - \frac{(z - 0.7)^2}{b_1^2}\bigg] + c_1 \times {\rm exp} \bigg[ - \frac{(z - 1.2)^2}{d_1^2}\bigg]
    \label{eq:nz_cfht}
\end{eqnarray}
with $(a_1, b_1, c_1, d_1) = (1.50, 0.32, 0.20, 0.46)$, as shown in   \citet[][hereafter  vW13]{VanWaerbeke2013}.
We show this function in Fig. \ref{fig:nz} as the blue solid line.

\subsubsection{RCSLenS shear data}

\begin{figure}
\begin{center}
\includegraphics[width=3.3in]{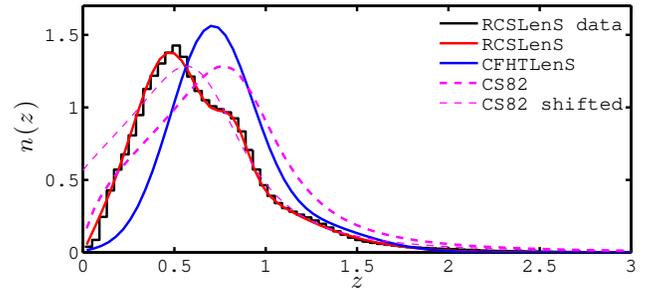}
\caption{Redshift distribution of the selected RCSLenS sources (black histogram), 
compared to the fit function (red solid line, equation \ref{eq:nz_rcs}). 
Also shown are the $n(z)$ of the CFHTLenS (equation  \ref{eq:nz_cfht},  blue solid line) and that of the CS82 survey (thick dashed line, magenta).
The thin dashed line has been shifted by $\Delta z = 0.2$.}
\label{fig:nz}
\end{center}
\end{figure}

The RCSLenS data consists of  14 disconnected regions whose
combined total area reaches 785 deg$^2$. 
Since photometric
information is incomplete for several fields, the redshift distribution
is estimated using the CFHTLenS sample augmented
with  near-IR and GALEX near-UV data. This new photometric sample, called
CFHTLenS-VIPERS, is calibrated against ~60,000 spectroscopic redshifts, 
as described in \citet{2015MNRAS.449.1352C}.
We first apply a magnitude cut $18.0 \le \mbox{mag}_r \le 24$ on the RCSLenS data,
which produces a sample of 10,586,079 galaxies.
 The construction of the $n_{\rm RCS}(z)$ that best describes this sample
 is then obtained by stacking the photometric redshift PDF of the CFHTLenS-VIPERS galaxies that pass the same RCSLenS selection criteria.
In addition to this magnitude cut, two modifications must be included in order to account for differences between CFHTLenS and RCSLenS:
\begin{enumerate}
\item {The CFHTLenS limiting magnitude is approximately 1.5 mag deeper than
RCSLenS. Objects in CFHTLenS-VIPERS are randomly discarded in order for the
$r$-band magnitude counts to match RCSLenS counts.}
\item{ The PDF stacking has to account for the {\it lens}fit weight, since the shear
data are weighted this way. This is accomplished by measuring the
relation between $r$-band magnitude and {\it lens}fit weight in RCSLenS, and applying these weights to the CFHTLenS-VIPERS galaxies based on this. 
We establish this relationship by binning the selected RCSLenS galaxy sample in
narrow magnitude bins and calculating the average {\it lens}fit weight in each bin.}
\end{enumerate}
The  $n_{\rm RCS}(z)$ distribution is then constructed using the CFHTLenS-VIPERS photometric
redshift PDFs of galaxies in the $18.0 \le \mbox{mag}_r \le 24$ magnitude range, using
the magnitude-weighting relation found in (ii)- and the adjusted counts from (i).
Finally, we require again {\tt FIT\_CLASS}=1. 
The source distribution is shown in Fig. \ref{fig:nz} as the black histogram,
and can be fitted by:
\begin{eqnarray}
    n_{\rm RCS}(z) =a_2 \ z \ \mbox{exp}\bigg[\frac{-(z-b_2)^2}{c_2^2}\bigg]+d_2 \ z \ \mbox{exp}\bigg[\frac{-(z-e_2)^2}{f_2^2}\bigg]+  \nonumber \\ 
     g_2 \ z \ \mbox{exp}\bigg[\frac{-(z-h_2)^2}{i_2^2}\bigg]
    \label{eq:nz_rcs}
\end{eqnarray}
where $(a_2, b_2, c_2, d_2, e_2, f_2, g_2, h_2, i_2)$ = 
(3.126,      
-0.419,  
0.979,   
1.678,     
 0.404,  
0.250,   
 0.400,    
 0.813,    
 0.121).
This fit-function is shown in the figure as the red solid curve.
The double bump was not present in the original CFHTLenS selection, and is the result of the improved photometric 
measurement in \citet{2015MNRAS.449.1352C}, compared to vW13.
Note that this functional form differs from that used for CFHTLenS in equation \ref{eq:nz_cfht}, 
but that we show in Section \ref{subsec:nz} that this has  consequence on the results, aside from providing a better fit.
Once the {\it lens}fit weight are assigned, we compute the number density and shape noise of our RCSLenS sample, 
and  obtain $\bar{n} = 4.84$ gal/arcmin$^{2}$ and $\sigma_{\epsilon} = 0.272$, respectively.

\subsubsection{Map reconstruction}

\begin{table}
\caption{Summary of the different properties of the 18 $\kappa$ maps used in this paper.
The different areas listed here are the total unmasked area per field, which add up to 146.5  and 600.7 deg$^2$ for CFHTLenS and RCSLenS footprints respectively.  
 All maps are smoothed with Gaussian filters with dispersion $\sigma$ =  4.24 arcmin. 
When applied, mask apodization reduces the total areas listed here to 130 deg$^2$ for CFHTLenS  and 394.7 deg$^2$ for RCSLenS. 
See Fig. \ref{fig:mask_apodize} for an example. 
}
\begin{center}
\begin{tabular}{lcccc}
\hline
Survey & Fields  & Area (deg$^2$ ) & pixel size (arcmin)& \\ 
\hline
\multirow{4}{*}{CFHTLenS}
 &  W1 & 63.6 & 1.0692 &\\
&  W2 & 20.0 &0.6018  &\\
&  W3 & 41.5 & 0.8610 &\\
&  W4 & 21.2 &0.7188  &\\
\hline
\multirow{14}{*}{RCSLenS}
&  CDE0047 & 55.2& 1.0& \\
&  CDE0133 & 27.8& 1.0& \\
&  CDE0310 & 68.7& 1.0& \\
&  CDE0357 & 27.7& 1.0& \\
&  CDE1040 & 27.6& 1.0& \\
&  CDE1111 & 67.7& 1.0& \\
&  CDE1303 & 13.4& 1.0& \\
&  CDE1514 & 66.0&1.0& \\
&  CDE1613 & 24.9& 1.0& \\
&  CDE1645 & 24.0& 1.0& \\
&  CDE2143 & 71.1& 1.0& \\
&  CDE2329 & 38.9& 1.0& \\
&  CDE2338 & 64.9& 1.0 & \\
&  CSP0320 & 22.8& 1.0& \\
\hline
\end{tabular}
\end{center}
\label{table:FieldProperties}
\end{table}%

The $\kappa_{\rm gal}$ maps reconstructed from the CFHTLenS data are  described in details in  vW13.
These are based on the KS algorithm developed by \citet{1993ApJ...404..441K},
and include careful treatment of the mask and noise properties of the shear catalogue.
Note that the four CFHTLenS mass maps have different areas and resolutions, which are detailed in Table \ref{table:FieldProperties}.
The maps are smoothed with a Gaussian filter of width  $\sigma = 4.24$ arcmin in order to suppress the effect of shot noise.

The same tools are used in the map making of the RCSLenS data, which are described in \citet{Hildebrandt2015}. 
The RCSLenS $\kappa$ maps are also smoothed with the same Gaussian filter used for CFHTLenS maps.  
In contrast with the CFHTLenS maps, the pixel size is uniform for all fields,  exactly 1 arcmin on the side.
The area of each field is calculated from the number and size of  unmasked pixels, which gives a total unmasked area of 146.5 deg$^2$ for CFHTLenS and 600.7 deg$^2$ for RCSLenS. 
Mask apodization (see Section \ref{subsubsec:apo})  reduces these areas to  130 deg$^2$ for CFHTLenS  and 394.7 deg$^2$ for RCSLenS.
This quantity is calculated differently for $\gamma_t$, and involves instead the `mosaic' mask provided at the catalogue level.
The areas are instead 125.4 and 462.6 for the two surveys.

Noise properties of these 18 $\kappa_{\rm gal}$ maps are studied by generating 100 noise realizations per field in which the orientations of the galaxies have been randomized.
We also generate a series of B-mode maps by rotating the ellipticity of each galaxy by 45 degrees, via the transformation $(e_1, e_2) \rightarrow (-e_2, e_1)$ before proceeding with the
KS reconstruction algorithm (see vW13 for details).
The cross-correlation signal between these maps and the CMB lensing data is expected to be consistent with zero, providing an indicator for systematic biases in the 
maps or in the pipeline.
%
To be clear, all these map products serve for the Fourier space and configuration space $\kappa_{\rm CMB} \times \kappa_{\rm gal}$  measurements, 
but not for the  $\kappa_{\rm CMB} \times \gamma_{t}$ measurement, which is performed at the level of galaxy catalogue. 
Flat sky approximation is assumed in the reconstruction of the $\kappa_{\rm gal}$ maps, in the numerical simulations  and in both
cross-correlation estimators involving $\kappa_{\rm gal}$. In contrast, the $\gamma_t$ estimator is constructed with  curved sky coordinates.
Given the sizes of our fields, measurements made assuming curved or flat sky calculations are almost indistinguishable.

\subsection{$\kappa_{\rm CMB}$ maps}
\label{subsec:CMB_data}

The core of our analysis makes use of the full-sky $\kappa_{\rm CMB}$ map provided in the 2015 Planck public data release\footnote{Planck lensing package: {\tt pla.esac.esa.int/pla/\#cosmology}}  \citep{2015arXiv150201591P}.
This map is  reconstructed from a minimum-variance combination of the multi-frequency temperature and polarization maps recorded by the Planck microwave telescope, 
based on the quadratic lensing estimator of \citet{2002ApJ...574..566H}. The combination is performed on the component-separated SMICA maps \citep{2015arXiv150205956P},
which mask   the galactic plane  and point sources prior to the lensing reconstruction.
The public map is provided in multipole space as spherical harmonic coefficients in the range $10 \le \ell \le 2048$,  stored in a $N_{\rm side} = 2048$ {\small HEALPIX} file. 
The map is noise-dominated, which explains why higher $\ell$-modes -- or smaller scales -- are not included. 

We cut out the regions of the sky that overlap with the four CFHTLenS fields and the 14 RCSLenS fields, including an additional large band around each of these
such that the final area of the cutout patch is four times larger.  {\color{black} Indeed, we show in Section \ref{subsubsec:real_space_data_prep}
that this larger cutout can improve the signal-to-noise ratio in the cross-correlation measurement by 10-18\%}.
We also extract maps over these 18 regions from the 100 full-sky $\kappa_{\rm CMB}$ simulations and from the Planck lensing analysis mask map, which are provided in the data release.
%

Note that the removal of point sources from the CMB lensing map has an important effect  on the cross-correlation signal, since
these are often associated with massive clusters that contain a fair fraction of the galaxy lensing signal.
To assess the importance of  this effect, we make use of numerical simulations. We describe our strategy  and report our results in Section \ref{subsec:sims}.


\section{The measurements}
\label{sec:measurement}

This section describes the cross-correlation measurements between CMB lensing and galaxy lensing using three distinct estimators. 
We first describe the $\kappa_{\rm CMB} \times \kappa_{\rm gal}$ analysis in Fourier space, which is similar in essence 
to the measurements from H15, LH15 and K15. Namely, the cross-correlation is obtained from the product of the two Fourier-transposed maps.
{\color{black} We then present the first configuration-space measurements,
which we achieve by measuring two-point correlation functions from  $\{\kappa_{\rm CMB}, \kappa_{\rm gal}\}$ and from  $\{\kappa_{\rm CMB}, \gamma_{t}\}$}. 
Since the techniques involved in Fourier- and configuration-space measurements are quite different, 
we present these measurements in distinct sections.

\subsection{Fourier analysis}
\label{subsec:Fourier_analysis}

{\color{black}
In preparation for the Fourier-space analysis, it is important to properly account for a number of subtle numerical and observational effects,
especially when comparing data and predictions. For instance, data re-binning, map smoothing, masking, zero-padding and apodization all affect the resulting cross-correlation measurement. 
%
%
%
%
There are two possible approaches to include them in the analysis: forward modelling 
the predictions as `pseudo-$C_{\ell}$', 
and backward modelling the measurements to match the theoretical predictions.
In both cases, what requires the most effort is  the mask-induced effect, which involves the construction of a mode-mixing matrix. 
The backward modelling approach furthermore requires its inversion, which appears to introduce a higher level of bias \citep{Asgari2015};
we therefore opted for the forward modelling method.}
%

Among the three previous measurements, the analysis strategies differ considerably. 
H15 opted for backward modelling,  and included data re-binning, map smoothing, mask apodization and a mode-mixing matrix in their pipelines. 
The recent measurement performed by K15 is based on the {\tt PolSpice} software \citep{Szapudi2001}, which is a public code that provides a backward modelling measurement
based on a completely different method from the pipeline used in H15.  
In contrast, LH15 concluded that given the noise levels of their measurement,
only the beam smoothing had to be included in their forward modelling, and mention that mask-induced effects were only important for $\ell>7000$. 
Given the increased area of our measurement,
new requirements on the accuracy of the predictions are more stringent, since the associated uncertainty must always be sub-dominant compared to other sources of error.
We therefore decided to include a full mode-mixing matrix in the forward modelling, which serves at the same time as a preparation for next generation surveys.

\subsubsection{Modelling the pseudo-$C_{\ell}$}
\label{subsec:PCL}


\begin{figure}
\begin{center}
\includegraphics[width=3.0in]{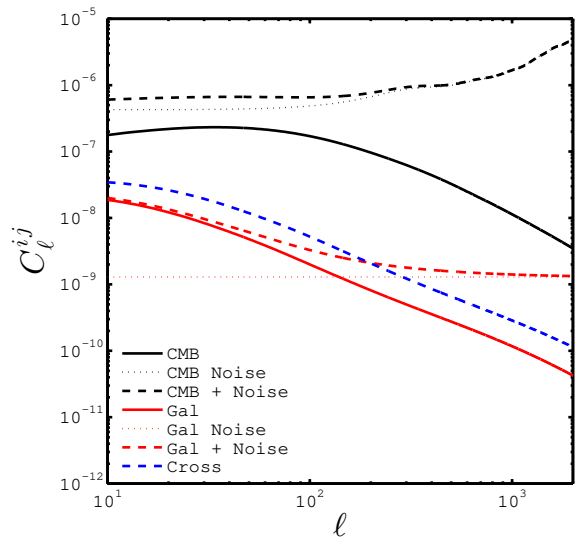}
\caption{Theoretical predictions for the auto-spectra ($ij = \kappa_{\rm CMB}\kappa_{\rm CMB}$ and
$ij = \kappa_{\rm gal}\kappa_{\rm gal}$) and cross-spectrum ($ij = \kappa_{\rm CMB}\kappa_{\rm gal}$). 
These are based on RCSLenS specifications for the shape noise $\sigma_{\epsilon}$, number density $\bar{n}$ 
and redshift distributions $n(z)$. For CFHTLenS, the cross-spectrum is approximately a factor of two higher, 
due to a deeper $n(z)$. 
}
\label{fig:Theory}
\end{center}
\end{figure}

This section describes how we turn the $C_{\ell}$ predictions into pseudo-$C_{\ell}$.
The main components  that need to be modelled are
the  $C_{\ell}$ themselves,  the effect of map/beam smoothing,  the propagated effect of the mask and, finally, the broad binning of the measurement.
The $C_{\ell}$ predictions are obtained by feeding the non-linear matter power spectrum $P(k,z)$ into equations \ref{eq:limber} and \ref{eq:limber_cross}.
To calculate $P(k,z)$, 
 {\color{black} we make use of the {\small CAMB} cosmological code \citep{Lewis2000a}, based on the calibration by \citet{Takahashi2012}.  
  The source redshift distributions $n(z)$ differ between the two galaxy lensing surveys, and are obtained from the fit functions  
 described by equations  \ref{eq:nz_cfht} and \ref{eq:nz_rcs}.
 Predictions for the auto- and cross-power spectra of Planck CMB lensing and RCSLenS  are presented in Fig. \ref{fig:Theory},
 in the WMAP9 cosmology.
The importance of noise in the Planck lensing map is obvious and dominates at all scales, whereas the noise in the galaxy lensing maps 
exceeds the signal for $\ell \gtrsim 150$.
}

The first part of the forward modelling concerns the map smoothing, which is obtained by multiplying the predictions by a Gaussian function whose width matches that of the Gaussian filter used
in constructing the $\kappa_{\rm gal}$ maps (listed in Table \ref{table:FieldProperties}). 
The CMB lensing map provided by the Planck Collaboration includes all $\ell$-modes up to 2048, hence by restricting our measurement to smaller multipoles, 
no modelling is necessary to account for the Planck beam.

The second part concerns the impact of observational masks. Mask-induced effects  in Fourier-based analyses are discussed in detail in many papers 
 \citep[e.g.][]{1994ApJ...426...23F, 2002ApJ...567....2H, Harnois-Deraps2012e, Asgari2015},
and can be split into three components, at least for two-point function measurements. 
First, they introduce an overall downward shift of power in the Fourier transform,  solely due to higher number of zero elements compared to an unmasked map.
This can be easily corrected for with a rescaling {\color{black} of the form $C_{\ell} \rightarrow C_{\ell}/\sum(M)$, where $M$ refers to the mask map, and the sum in the denominator
runs over all pixels.}
Second, masking introduces sharp features in the observation field which, unless smoothed out by apodization, propagate in the power spectrum of the masked map
and greatly enhance the measured values at high-$\ell$.
The third  effect of masking is to cause a coupling between the otherwise independent $\ell$-modes of the underlying $C_{\ell}^{ij}$.
This comes from the fact that the cross-correlation measurement can  expressed as a convolution between the  cross-spectrum of the masks, $C_{\ell}^{M_1 M_2}$, and the $C_{\ell}^{ij}$. 
Mathematically, the convolution can be described by a mode-mixing matrix, $\rm M_{\ell \ell'}^i$, whose calculation is described in Appendix \ref{sec:mode_mixing}.
The mode-mixing matrix was ignored in LH15, hence for comparison purposes, we  include it as an optional segment of our analysis. 

The third and final part of the forward modelling approach consists of rebinning the modelled pseudo-$C_{\ell}$
the same way as the data. Although the original predictions are generally smooth, the mode-mixing matrix can
introduce noisy features, which are made less significant with the rebinning process. 

%
%

Since each of the 18 fields has a distinct mask, and since  the RCSLenS and CFHTLenS  have different source galaxy distributions and smoothing kernels,
 the 18 pseudo-$C_{\ell}$ measurements are not expected to converge to a unique predicted value.
Instead, each field has its own, distinct, forward-modelled  prediction, hence to combine them we must adopt the following  strategy:
For each field individually,
\begin{enumerate}
\item we measure the pseudo-$C_{\ell}^{\kappa_{\rm CMB}\kappa_{\rm gal}}$; 
\item depending on whether the field belongs to RCSLenS or CFHTLenS, we select the $C_{\ell}^{\kappa_{\rm CMB}\kappa_{\rm gal}}$ prediction with the appropriate $n(z)$;
\item we apply the map smoothing kernel and, optionally, the mode-mixing matrix that correspond to this field;
\item we rebin the  measurements and predictions into coarse $\ell$-bands;
\item we construct the covariance matrix  (see Section \ref{subsec:error} for details).
\end{enumerate}
 We finally combine the 18 data vectors (along with their 18 covariance matrices) into one large data vector (and one large covariance matrix)  from which we can constrain  cosmology. 
 In this process, we assume that the 18 fields are generally separated well enough on the sky so that the covariance between them can be ignored. 
 There is actually some overlap between the larger cutouts of two RCSLenS and CFHTLenS fields,
 but this can only have a negligible effect on the covariance.
 The final covariance matrix therefore consists of 18 diagonal blocks, with zeros everywhere else.   
Section \ref{subsubsec:fourier} contains details about these five steps, however we must first describe step zero, 
which concerns the  map preparation.


{\color{black} In the three previous $\kappa_{\rm CMB}\times \kappa_{\rm gal}$ analyses,  the lensing maps  were prepared and analyzed differently.
Indeed, a number  of choices are available with regards to mask apodization (and how this is done),  zero-padding of the maps, 
multiplication of the $\kappa_{\rm CMB}$ and $\kappa_{\rm gal}$ masks, selecting the best sizes for the $\kappa_{\rm gal}$  smoothing kernels, etc. 
All of these impact the measurement, only partial justification is provided in H15, LH15 and K15 to support their choices.
In this section, we seek to fill this gap and present how we optimize our analysis by comparing different 
 pipelines, summarized in Table \ref{table:pipeline}. 
 We investigate their effects all the way down to the signal-to-noise ratio, which we present in Section \ref{sec:cosmology}.}

\subsubsection{Apodization}
\label{subsubsec:apo}

{\color{black} Mask apodization refers to the explicit smoothing of the mask, and is meant to soften the sharp transitions that occur between masked and unmasked regions.
In Fourier analyses, this procedure lowers  the impact of the mask on the measurement, especially for high-$\ell$ modes.}
It also has important effects on the mode-mixing matrix as it significantly suppresses the far off-diagonal elements,
thereby making the estimator  more stable \citep{Asgari2015}. 
At the same time, apodization reduces the area over which the measurement is carried out, thus increasing the statistical noise.
We explore this tradeoff in accuracy/precision by including it as an optional module.

\begin{figure}
\begin{center}
\includegraphics[width=3.5in]{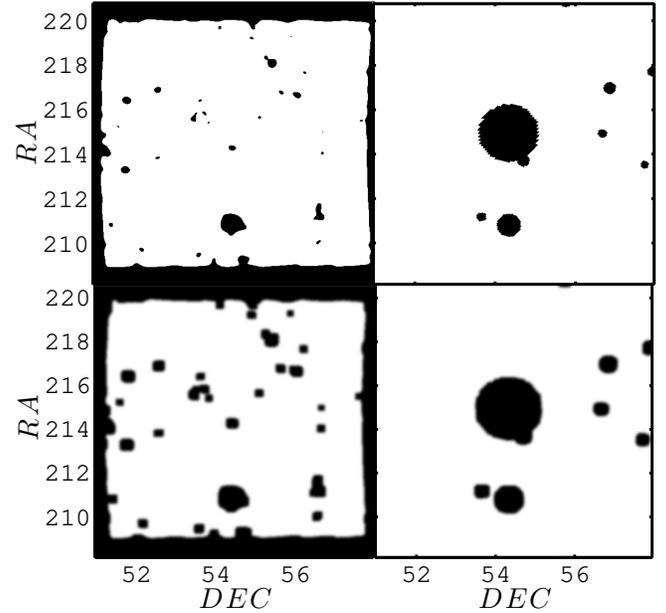}
\caption{Mask of the CFHTLenS field W3, original (top left) and apodized with a Gaussian beam over 5 pixels (bottom left). 
In this case, the area drops by 10\%, but the effect is more dramatic in other regions. The right panels show the Planck masks over the same region, with and without apodization.}
\label{fig:mask_apodize}
\end{center}
\end{figure}

 When apodization is turned on, we identify each masked pixel, set to zero all neighbouring cells separated by 5 pixels or less,
 and convolve the resulting mask by a Gaussian filter with a width of about 2 cells,
such that the transition between zeros and ones occurs over these 5 cells. 
The pixels that were originally masked acquire a slight offset after this and we force them back to zero.  
The apodized masks are finally re-applied on the original $\kappa_{\rm gal}$ and $\kappa_{\rm CMB}$ maps by a simple multiplication.
A visual example of this procedure is shown in Fig. \ref{fig:mask_apodize},
 in the case of the CFHTLenS W3 field, which clearly illustrates the loss of area.

\subsubsection{Zero-padding of $\kappa_{\rm gal}$}
\label{subsubsec:zero-pad}

 A common practice in Fourier analyses is to zero-pad the maps, which consists of surrounding the data with a band of pixels set to zero,
typically doubling the size of the data in each dimension.   
This technique avoids aliasing when manipulating non-periodic maps, 
plus it allows for an interpolation between the Fourier modes that can sometimes increase the smoothness of the measurement. 

In the case of cross-correlations, the two maps are not necessarily equal in  size, 
and one can perform the Fourier analysis by adding layers of zeros to the smaller map until it matches the size of the larger map.
In our case, the Planck $\kappa_{\rm CMB}$ map covers the full sky, 
whereas the 18 $\kappa_{\rm gal}$ maps are scattered all around, with sizes varying between 14 and 78 deg$^2$. 
If we were to follow the three previous analysis strategies, we would carve out these exact 18 footprints from the Planck lensing map. 
In fact, we have the freedom to choose $\kappa_{\rm CMB}$ maps of any size,
and there is a potential gain in signal-to-noise that can be achieved by cutting out {\it larger} maps.
Indeed, a non-zero correlation length implies that regions of $\kappa_{\rm CMB}$ outside the survey footprint are correlated with regions of $\kappa_{\rm gal}$ inside of it. 
This is best explained with a simple toy example.


We are interested in cross-correlating two maps -- in our case $\kappa_{\rm CMB}$ and $\kappa_{\rm gal}$ --
which we label  $M_1(\boldsymbol \theta)$ and $M_2(\boldsymbol \theta)$ to keep the discussion  general.
For simplicity, we also neglect small scales cutouts from the observed maps in this example, and only consider the large scale footprints.
%
Let us suppose that $M_1$  extends over a large region, $A$, while $M_2$ is  measured only over a smaller region, $B$, with $B \subseteq A$.
{\it If} we decided to measure the cross-correlation {\it only} over the part of the sky that perfectly overlaps (region $B$),
we would reject any area from $M_1$ that is outside $B$, and our pseudo-$C_{\ell}$ estimator would be written as:
\begin{eqnarray}
\langle M_{1, \ell} M_{2,\ell}^* \rangle_{B} = {\rm Re}\langle \!\! \int \!\!\!\! \int_B M_{1}(\boldsymbol \theta) M_{2}(\boldsymbol \theta') e^{i \boldsymbol \ell (\boldsymbol \theta - \boldsymbol \theta')} \mbox{d}\boldsymbol \theta \mbox{d} \boldsymbol \theta'  \rangle
\label{eq:BB}
\end{eqnarray}
The square brackets here refer to the polar integration, and the integration is carried solely over region $B$. 
We refer to this as the `adjusted' maps estimator.
Alternatively, we could decide to measure the cross-correlation over {\it all} the available data, i.e. 
\begin{eqnarray}
\langle M_{1, \ell} M_{2,\ell}^* \rangle_{A,B} = {\rm Re}\langle \!\! \int_A \!\! \int_B M_{1}(\boldsymbol \theta) M_{2}(\boldsymbol \theta') e^{i \boldsymbol \ell (\boldsymbol \theta - \boldsymbol \theta')} \mbox{d}\boldsymbol \theta \mbox{d}\boldsymbol \theta'  \rangle
\label{eq:AB}
\end{eqnarray} 
where now the two integrals run over different regions $A$ and $B$.
As mentioned before, the numerical implementation of this measurement involves zero-padding $M_{2}$ such that it matches $M_{1}$ in size.
Then both integrals can be  carried over region $A$, and we refer to this as the `large' maps estimator.
 
In order to find out  whether there is a difference between these two measurements,
we subtract equation \ref{eq:BB} from equation \ref{eq:AB}:
\begin{eqnarray}
 \langle M_{1, \ell} M_{2,\ell}^* \rangle_{A,B}  - \langle M_{1, \ell} M_{2,\ell}^* \rangle_{B}  = {\rm Re}\langle \!\! \int_{A-B} \!\!\!\! M_{1}(\boldsymbol \theta) \!\! \int_{B} \!\!  M_{2}(\boldsymbol \theta') e^{i \boldsymbol \ell (\boldsymbol \theta - \boldsymbol \theta')} \mbox{d}\boldsymbol \theta \mbox{d} \boldsymbol \theta'  \rangle
 \label{eq:A_minus_B}
\end{eqnarray}
This is effectively computing the correlation between $M_2$ {\it inside} region $B$ with $M_1$ {\it outside} region $B$. 
When the two maps are uncorrelated, this obviously vanishes. 
However, any correlation that exists between the two maps would make this subtraction non-zero,
provided that the correlation length exceeds one pixel.
{\color{black} When we replace $M_1$ and $M_2$ by  $\kappa_{\rm CMB}$ and $\kappa_{\rm gal}$
and realize that the correlation length predicted by equation \ref{eq:xi1-prediction} extends at least to one degree, 
it becomes clear that the `large' maps estimator should have a better signal-to-noise, although the level of improvement is not obvious.

Technically, only the $\kappa_{\rm gal}$ map is zero-padded in this procedure --  on the contrary, the area selected from the $\kappa_{\rm CMB}$ map is larger, 
thereby adding a large amount of {\it non-zero} elements. Failing to find a better name to describe this procedure, we still refer to it  as `zero-padding',
with the understanding that it is not an accurate description of what is going.
This zero-padding strategy  has not been used in the previous CMB lensing-galaxy lensing analyses, 
and was not even an option for the ACT/CS82 and SPT/DES-SV data since their footprints exactly match.
To quantify the actual gain that it provides, we perform our analysis on both adjusted $\kappa_{\rm CMB}$ maps (exact $\kappa_{\rm gal}$ footprints) 
and larger maps  (4 times the area of the $\kappa_{\rm gal}$ footprints, zero-padding the $\kappa_{\rm gal}$ maps), and propagate the impact on the cosmological constraints.


\subsubsection{Mask Combination}
\label{subsubsec:bm}

\begin{table}
\caption{Different forward modelling pipelines P examined in this paper to carry out the pseudo-$C_{\ell}$ measurement. 
LH15 used pipeline P5,
while H15 and K15 used two different backward modelling versions of pipeline P5.
As described in Section \ref{subsubsec:zero-pad}, what we mean by `zero-padding' differs from the traditional usage, as it refers
to the analysis of  $\kappa_{\rm CMB}$ cutout maps  that are 4 times larger than the $\kappa_{\rm gal}$ footprints.
Only the latter is zero-padded, whereas we keep the full $\kappa_{\rm CMB}$  values  in the padded region. 
`Joint mask' means {\color{black} multiplying} the masks from both maps, and re-applying the result on both maps.
}
\begin{center}
\begin{tabular}{cccccc}
\hline
Pipeline &apodization & zero-pad & joint mask \\
\hline
P1                            &   no               &  no           &  no            \\
P2                            &   yes             &  no           &  no             \\
P3                            &   no               &  yes         &  no             \\
P4                            &   yes             &  yes         &  no             \\
P5                            &   yes             &  no         &  yes           \\
\hline
\end{tabular}
\end{center}
\label{table:pipeline}
\end{table}%

The last step in the map preparation  concerns the method by which the masks (from  the $\kappa_{\rm CMB}$ and $\kappa_{\rm gal}$ maps)
are incorporated in the measurement.
Despite the potential gain in signal-to-noise ratio that can be accessed by following the method described in Section \ref{subsubsec:zero-pad}, 
previous studies by H15,  LH15 and K15 have opted for  a strategy based on equation \ref{eq:BB} according to which the analyses require `adjusted' maps. 
Moreover, they constructed a `joint mask' from the product of both masks, and applied it back on all their maps. 
As briefly mentioned at the end of  Section \ref{subsec:CMB_data}, combining masks in this way has severe consequences on the 
cross-correlation signal, since the point sources that are masked in $\kappa_{\rm CMB}$ correspond to 
regions of high signal in $\kappa_{\rm gal}$. There is therefore a strong correlation between the $\kappa_{\rm CMB}$ mask
and the $\kappa_{\rm gal}$ maps, which biases the measurement unless properly accounted for.
We explore this approach as well, mainly for comparison purposes, but 
%
%
%
%
%
we otherwise treat both masks separately. 

To organize our numerous measurement strategies, we refer to a particular choice of apodization, zero-padding and joint-masking as a `pipeline'.
We consider five different pseudo-$C_{\ell}$ pipelines, which are labelled P1-P5 and detailed  in Table \ref{table:pipeline}.
From our numerical simulations presented in Section \ref{subsec:Fourier_validate}, we find that only P5 significantly suffers from  mask-induced effects, 
hence our results do not include the mode-mixing calculations unless explicitly specified.


\subsubsection{Measuring the pseudo-$C_{\ell}$}
\label{subsubsec:fourier}

\begin{figure*}
\begin{center}
\begin{tabular}{cc}
\includegraphics[width=3.0in]{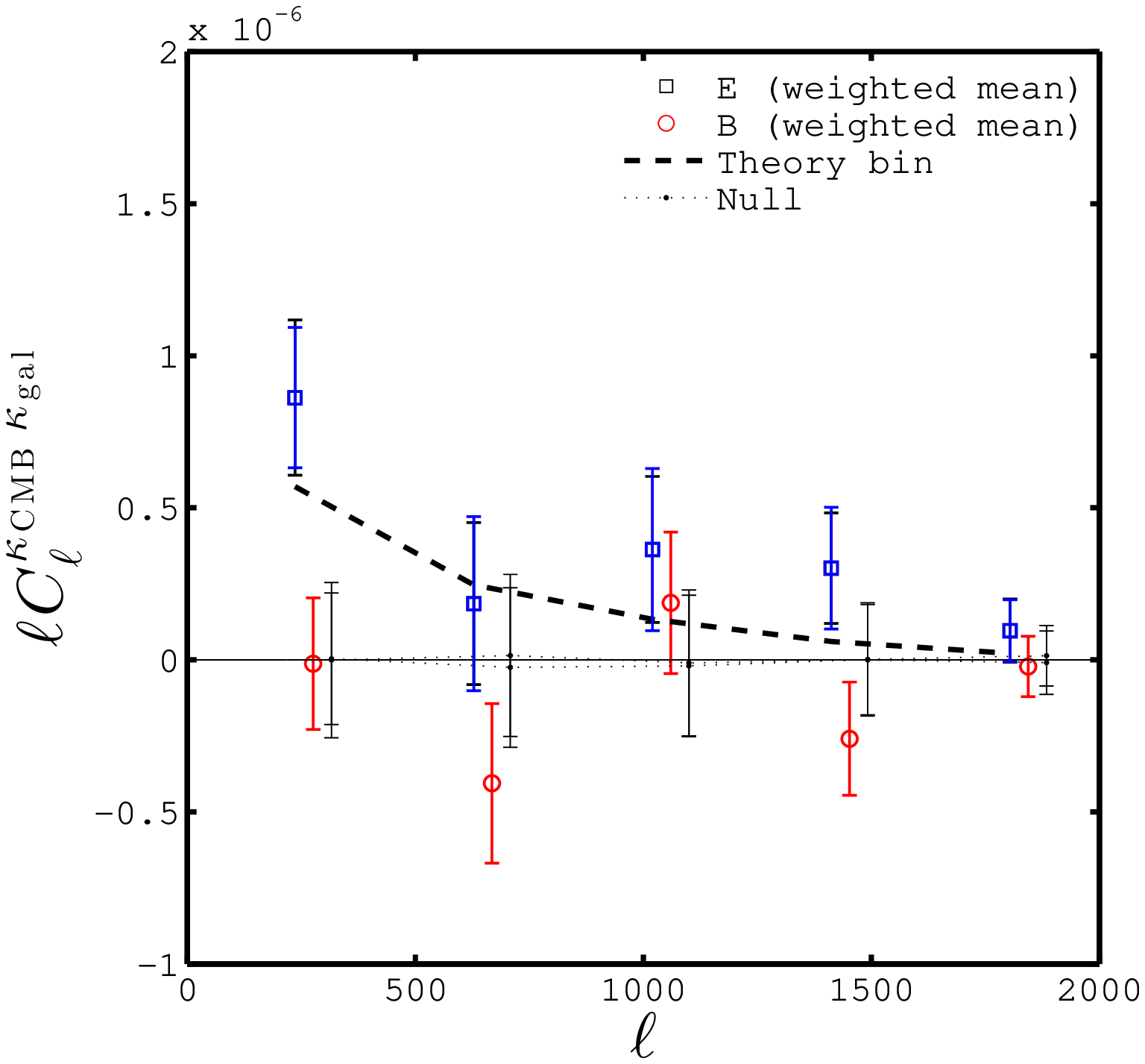}                         & \includegraphics[width=3.0in]{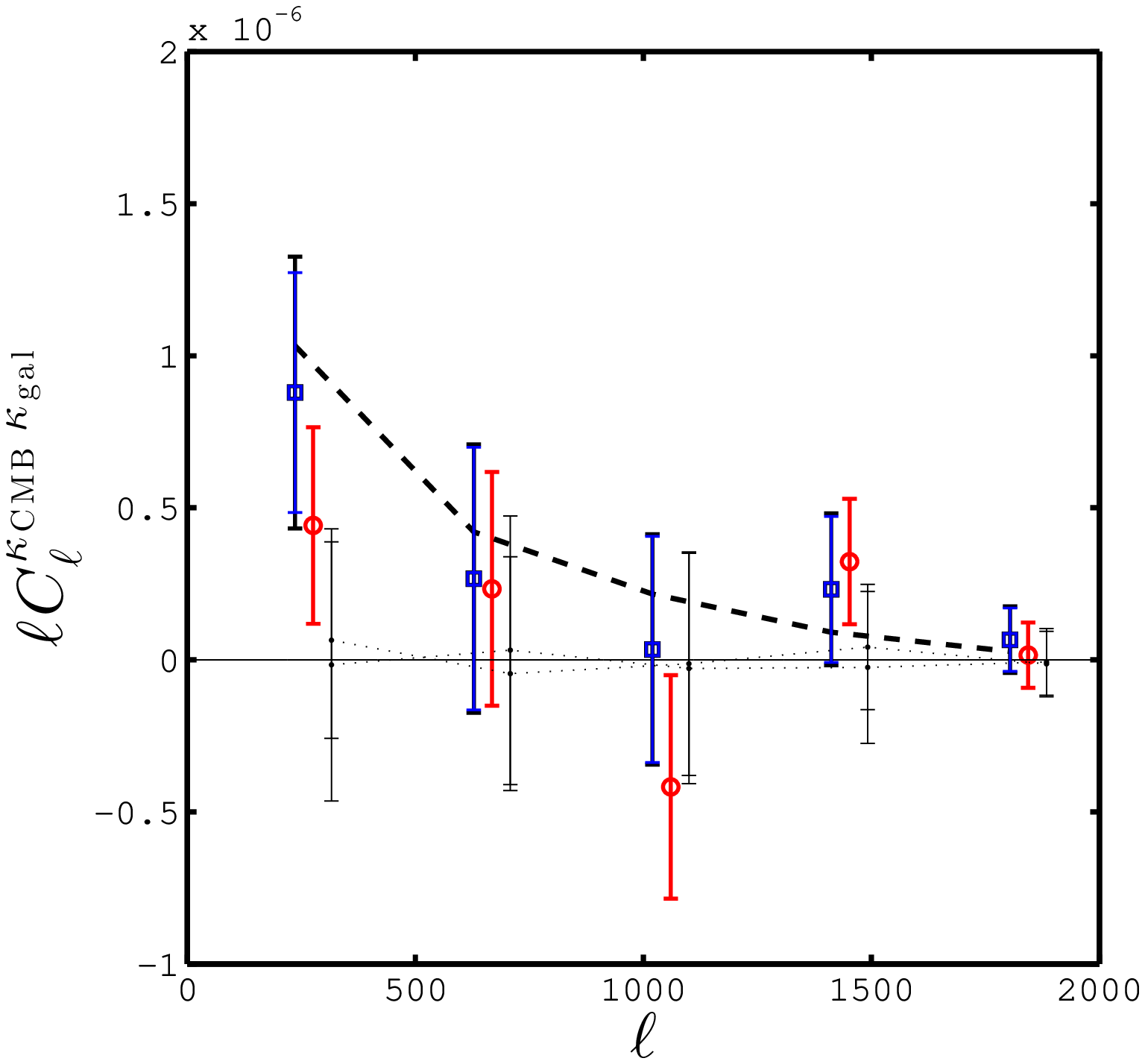} \\
\end{tabular}
\caption{Area-weighted average of the cross-spectrum between the $\kappa_{\rm CMB}$ maps from the Planck 2015 data release and the $\kappa_{\rm gal}$ maps 
from the RCSLenS (left) and CFHTLenS (right). 
These pseudo-$C_{\ell}$ measurements are carried out with pipeline P1, which has the highest signal-to-noise ratio; we compare the other pipelines summarized in Table \ref{table:pipeline} in Fig. \ref{fig:cmb_x_kappa_RCS_all_P}. 
In both panels, the black error bars are obtained from the cross-correlation between the 100 CMB lensing maps and the E-mode $\kappa_{\rm gal}$ maps, 
while the blue errors show the analytical Gaussian predictions (equation \ref{eq:GaussVar}).
The correlation between the points is negligible.
The three null tests shown here are CMBLenS simulations x RCSLenS data (E- and B-modes), and CMBLenS data x RCSLenS random catalogues, as described in Section \ref{subsec:null}. }
\label{fig:cmb_x_kappa_RCS}
\end{center}
\end{figure*}

For each pipeline listed in Table \ref{table:pipeline}, we measure the pseudo-$C_{\ell}$ in the following way.
We first compute the amount of unmasked area common to both maps for each field $i$, $\sum(M^i)$, 
in preparation for the $C_{\ell} \rightarrow C_{\ell}/\sum(M)$ rescaling mentioned in 
Section \ref{subsec:PCL}.
{\color{black} We then Fourier transform the galaxy lensing and CMB lensing  maps, 
complex-multiply the results and reject the imaginary part. 
Denoting observed quantities with a `hat', this operation estimates $\hat{C}^{\kappa_{\rm CMB}\kappa_{\rm gal}, i}_{\boldsymbol \ell }$.
}
A finely-binned cross-spectrum is then found by  angle-averaging this quantity in annuli of width $\Delta \ell^i = 2\pi/\theta_{\rm max}^i$, 
where $\theta_{\rm max}^i$ is the angular size of the largest dimension for the field $i$. 
Noting that the maps are not strictly square, the density of the Fourier modes is slightly anisotropic, which our calculation takes into account. 
So far this measurement  knows nothing  about masks: it treats zero-valued pixels as any other pixels,
and the resulting $\hat{C}^{\kappa_{\rm CMB}\kappa_{\rm gal}, i}_{\boldsymbol \ell }$ are highly suppressed for heavily masked fields.
We therefore rescale the results as  $\hat{C}^{\kappa_{\rm CMB}\kappa_{\rm gal}, i}_{\ell} \rightarrow \hat{C}^{\kappa_{\rm CMB}\kappa_{\rm gal}, i}_{\ell}/\sum(M^i)$ to correct for this. 

Following LH15, we next calculate the average\footnote{This is a `normal' average, not a noise-weighted or area-weighted mean.} of the measurements inside five coarse bins that are  linearly spaced in the range  $\ell=\{30, 2000\}$,  with width $\Delta\ell_c=394$ and centred on $\ell_c$ = (227, 621, 1015, 1409, 1803). 
We repeat this  pseudo-$C_{\ell}$ measurement on the 18 fields and with methods P1-P5 for the $\kappa_{\rm CMB}\times \kappa_{\rm gal}$ data, 
as well as  for a number of supplementary maps that serve for null tests and error estimates that we detail in  Section \ref{subsec:null}.

We present our cross-correlation measurements  in Fig. \ref{fig:cmb_x_kappa_RCS}  for RCSLenS (left) and CFHTLenS (right).
We only show here the results from pipeline P1, but include  P1-P5 in Appendix \ref{sec:pipelines}.
In both panels, the thick black squares represent the $\hat{C}_{\ell}^{\kappa_{\rm CMB}\kappa_{\rm gal}}$ measurement, 
while the thick red circles and the black points are null tests described in Section \ref{subsec:null}. 
The thick dashed lines represent the  pseudo-${C}_{\ell}^{\kappa_{\rm CMB} \kappa_{\rm gal}}$ predictions in the WMAP9 cosmology.  
The gain in precision is clearly seen in  RCSLenS measurements compared to CFHTLenS and is purely due to the larger area coverage.
 The predicted  $\ell$-dependence of the signal becomes apparent, although its shape is poorly resolved.
{\color{black} However, one of the main advantages of the Fourier analysis over the two configuration-space analyses is that the data points shown here are very weakly correlated.}
The error bars are estimated by combining 100 Planck lensing simulations with the measured $\kappa_{\rm gal}$, as detailed in Section \ref{subsec:error}.


 \subsection{Configuration-space analysis}
\label{subsubsec:real_space}

The main advantage of the configuration-space measurement is that the effect of masking is much milder than in Fourier space. 
It affects the reconstruction of the $\kappa_{\rm gal}$ map, since it is the shear map that are masked.	
For the $\gamma_t$ estimator, it simply reduces the observation area.
In addition, there is no need for mask apodization nor mode-mixing matrices, which makes the analysis much simpler. 
The measurements from all  14 RCSLenS fields are expected to converge to the same value, and the same can be said for the four CFHTLenS fields.
Only the $n(z)$ and the map smoothing differ between both surveys, which still needs to be modelled.
The combination of the different fields become a simple weighted mean over the individual measurements.
In fact, forward and backward modelling become more directly connected; we choose the former such as to be consistent with the Fourier analysis.
However, these configuration-space analyses require a slightly different preparation of the data.
 
  \subsubsection{Data preparation}
\label{subsubsec:real_space_data_prep}

 {\color{black} Although this is not a requirement, we choose to down-weight the noise of the $\kappa_{\rm CMB}$ map by applying a Wiener filter:  
\begin{eqnarray}
	\kappa^{\rm CMB,  WF}_{\ell m} = \frac{C^{\kappa_{\rm CMB, th}}_\ell}{C^{\kappa_{\rm CMB, th}}_\ell + N_{\rm CMB}}\ \kappa_{\ell m} ^{\rm CMB}
\end{eqnarray}
where $C^{\kappa_{\rm CMB, th}}_\ell$ is the best-fit auto-power spectrum of the lensing map, and $N_{\rm CMB}$ is the total (instrumental + statistical) noise spectrum\footnote{
This filtering is similar to that of \citet{2015arXiv150201591P} and ignores off-diagonal contributions in the signal (from $C_{\ell}^{TE}$)
and in the noise (from residual mask-induced mode-mixing). }, 
both  provided in the public Planck data release. This filtering is accomplished on the full-sky map (i.e. before carving out the field footprints)
and therefore this part of the analysis -- and only this part --  uses spherical harmonic transforms as opposed to Cartesian Fourier transforms}. 
We have checked different filtering procedures, as well as no filtering at all, and the significance of the measured cross-correlation  signal is not strongly affected.

As mentioned above, apodization is no longer necessary, however the zero-padding technique presented in Section \ref{subsubsec:zero-pad}
can provide advantages here as well.  To illustrate this, let us re-examine the toy example where maps $M_1$ and $M_2$ are  measured  over regions $A$ and $B$, respectively.
From a configuration-space perspective, 
{\color{black}  
the difference between the estimators  for the  `large' and the `adjusted' maps (equation  \ref{eq:A_minus_B}) becomes: 
\begin{eqnarray}
 \langle M_{1}(\theta+\Delta \theta) M_{2}(\theta) \rangle_{A,B}  - \langle M_{1}(\theta+ \Delta \theta) M_{2}(\theta) \rangle_{B} 
 \end{eqnarray}
where $\theta \in B$ in both terms,  $(\theta + \Delta \theta) \in A$ in the first term, but  $(\theta + \Delta \theta) \in B$
in the second term. After the subtraction, what is left is a contribution from $\theta \in B$ while $(\theta + \Delta \theta) \notin B$,
i.e. the correlation between $M_{1}$ and the part of $M_{2}$ that is outside region $B$.
Again, with  $M_1 \rightarrow\kappa_{\rm CMB}$, $M_2 \rightarrow\kappa_{\rm gal}$ or $\gamma_t$, 
and recalling that the correlation length of these two estimators stretches beyond a degree (from  equations \ref{eq:xi1-prediction} and \ref{eq:xi2-prediction}, shown in Fig. \ref{fig:cmb_x_kappa_RCS_real}),
it is possible to improve the analysis by analyzing the `large' $\kappa_{\rm CMB}$ maps as opposed to the `adjusted' maps.
We carried out the measurements on both types of maps and found that the signal-to-noise ratio (SNR, see Section \ref{subsec:results}) is improved by 10-18\% in the large maps, 
compared to the adjusted maps. We therefore opted for the former in our analysis.

After examination of the predictions for our two configuration-space cross-correlation estimators,
 it appears that the cross-correlation signal extends to 3 degrees, but not much beyond.
We therefore include a  region that is 3 degrees thick on each side of the CFHTLenS fields, and extend this to 4 degrees for the RCSLenS. 
Because the largest measured angular bin from RCSLenS (CFHTLenS) is 3 (2) degrees, no information is lost by not using a larger padding.}
We otherwise proceed exactly as for the map preparation in the Fourier analysis: we surround the $\kappa_{\rm gal}$ maps
with layers of  zero elements until their size matches that of the $\kappa_{\rm CMB}$ maps.
This is even more direct for the $\gamma_t$ estimator, which is computed directly from the galaxy catalogue: we simply feed the `adjusted'  or the  `large' $\kappa_{\rm CMB}$ maps in the estimator.
More details about this are provided in the next section.

  \subsubsection{The estimators}
\label{subsubsec:real_space_est}

\begin{figure*}
\begin{center}
\begin{tabular}{cc}
\includegraphics[width=3.0in]{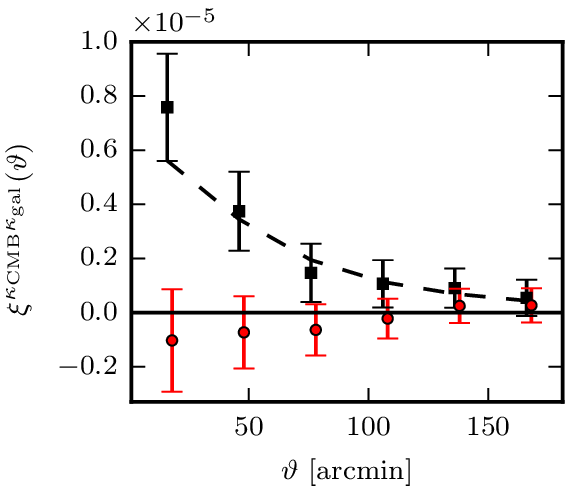}      & \includegraphics[width=3.0in]{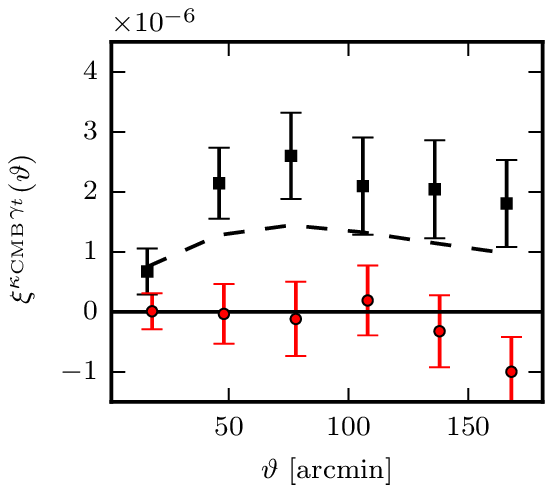} \\
\end{tabular}
\caption{
Configuration-space cross-correlation between the Planck 2015 and RCSLenS lensing data, plotted against WMAP9 predictions described by equations
\ref{eq:xi1-prediction} and \ref{eq:xi2-prediction}.
The error bars are computed from 100 CMB lensing simulations, which confirm that the points are highly correlated.
The Planck CMB maps have been Wiener filtered to reduce the noise levels.
Black squares show Planck cross E-modes, red circles  show Planck cross B-modes, as in Fig. \ref{fig:cmb_x_kappa_RCS}.}
\label{fig:cmb_x_kappa_RCS_real}
\end{center}
\end{figure*}

 The configuration-space $\kappa_{\rm CMB} \times \kappa_{\rm gal}$ estimator   of the two-point correlation function is computed as follows:
\begin{eqnarray}
	\label{equ:kappa-estimator}
	\xi^{\kappa_{\rm CMB}\kappa_{\rm gal}}( \vartheta) = \frac{\sum_{ij} \kappa_{\rm CMB}^i \kappa_{\rm gal}^j \Delta_{ij}(\vartheta)}{\sum_{ij}  \Delta_{ij} (\vartheta)}. 
\end{eqnarray}
The sum runs over all pixels $i$ at positions $\boldsymbol \theta_i$ on the  $\kappa_{\rm CMB}$ map and pixels $j$ on the corresponding $\kappa_{\rm gal}$ map. 
The term $\Delta^{ij}$ controls the binning, which is organized in 6 broad bins of  width $\Delta$ = 30 arcmin each (we use 4 bins for CFHTLenS given the higher noise levels) with:
\begin{eqnarray}
\Delta^{ij}(\vartheta)= \begin{cases} 
                                        1 {\rm , if } \left|{\boldsymbol \theta}_i - {\boldsymbol \theta}_j \right | < \vartheta \pm \frac{\Delta}{2} \\
                                        0 {\rm , otherwise} 
                                      \end{cases}
\label{eq:bin_operator}
\end{eqnarray}
{{\color{black} The sum is performed with a { k-d tree} algorithm to speed up the computation by only checking pairs that are not separated more than the largest bin.
}}

The second configuration-space  estimator that we use is this paper is a generalization of the galaxy-galaxy lensing estimator,
which is typically extracted by stacking the tangential shear signal from background galaxies around a discrete set of foreground objects, usually foreground galaxies.
For the $\kappa_{\rm CMB} \times \gamma_{t}$ estimator, we perform a similar stacking, but this time we stack on all pixel positions instead of foreground galaxies, and weight the sum by the pixel values. 
The  $\kappa_{\rm CMB} \times \gamma_{t}$  estimator is given by:
\begin{eqnarray}
	\label{equ:shear-estimator}
	\xi^{\kappa_{\rm CMB}\gamma_t}(\vartheta) = \frac{\sum_{ij} \kappa_{\rm CMB}^i e_t^{ij} w^j \Delta_{ij}(\vartheta)}{\sum_{ij}  w^j \Delta_{ij}(\vartheta)}\frac{1}{1+K(\vartheta)} \,
\end{eqnarray}
where $\Delta_{ij}(\vartheta)$ is the binning operator described by equation \ref{eq:bin_operator}.
In this case, the sum runs over all pixel $i$ in the $\kappa_{\rm CMB}$ map,  and all galaxies $j$ in the CFHTLenS or RCSLenS survey. 
Here $e_t^{ij}$ is the tangential component of the ellipticity of a galaxy $j$ (at position ${\boldsymbol \theta}_j$) with respect to the pixel $i$ in $\kappa_{\rm CMB}$ (at position ${\boldsymbol \theta}_i$). 
The B-mode signal is extracted by replacing $e_t^{ij}$ by the cross-shear $e_x^{ij}$ in the above expression.
For technical details about the definition of the ellipticity components $e_t$ and $e_x$, see for example  \citet{2015MNRAS.452.3529V}.
The quality of the shape measurement is determined by the {\it lens}fit weight \citep{2013MNRAS.429.2858M} and is denoted by $w^j$.
The trailing  factor $(1+K(\vartheta))^{-1}$ accounts for the multiplicative bias correction: 
 \begin{eqnarray}
	\frac{1}{1+K(\vartheta)} = \frac{\sum_{ij}  w^j \Delta_{ij}(\vartheta)}{\sum_{ij}  w^j (1+m^j)\Delta_{ij}(\vartheta)} \ .
\end{eqnarray}
{\color{black} It is straightforward to extend this estimator to other types of cross-correlation measurements,
simply  by replacing $\kappa_{\rm CMB}$ by another map \citep[for an application on  tSZ $\times \gamma_{t}$, see][]{Hojjati2016}.}


 We present in Fig. \ref{fig:cmb_x_kappa_RCS_real}  the results from these two measurements on the RCSLenS fields,
 against theoretical predictions from the WMAP9 cosmology.  
 The data and predictions are in good agreement for both the $\xi^{\kappa_{\rm CMB} \kappa_{gal}}$ and $\xi^{\kappa_{\rm CMB} \gamma_t}$ measurements.
  The later prefers slightly  higher values than WMAP9,  but this preference is weak, especially when  considering the fact 
 that points here are highly correlated.

\subsection{Covariance Estimation}
\label{subsec:error}

We describe in this section our strategy to construct an accurate covariance matrix, which, for our Fourier analysis, largely follows the methods of H15,  LH15 and K15.   

\subsubsection{Fourier-space covariance}
\label{subsubsec:error_fourrier}

The uncertainty about the  measurement $\hat{C}^{\kappa_{\rm CMB}\kappa_{\rm gal}}_{\ell}$
is evaluated from the expression:
\begin{eqnarray}
\widehat{\rm Cov}^{\kappa_{\rm CMB} \kappa_{\rm gal}}_{\ell \ell'} = \langle \Delta \hat{C}^{\kappa_{\rm CMB} \kappa_{\rm gal}}_{\ell}  \Delta \hat{C}^{\kappa_{\rm CMB} \kappa_{\rm gal}}_{\ell'} \rangle,
\end{eqnarray}
where again the `hat' symbols refer to `observed' quantities.
 Knowing that the noise $N_{\rm CMB}$ is much larger than the signal $\kappa_{\rm CMB}$ \citep[see][and Fig. \ref{fig:Theory}]{2015arXiv150201591P}, we can write:
\begin{eqnarray}
\widehat{\rm Cov}^{\kappa_{\rm CMB} \kappa_{\rm gal}}_{\ell \ell'} \simeq \langle \Delta \hat{C}^{N_{\rm CMB} \kappa_{\rm gal}}_{\ell}  \Delta \hat{C}^{N_{\rm CMB} \kappa_{\rm gal}}_{\ell'} \rangle.
\label{eq:Cov_approx}
\end{eqnarray}
%
From this, it follows that the complete covariance can be captured by cross-correlating the observed $\kappa_{\rm gal}$ maps with CMB lensing noise maps,
or alternatively with CMB lensing simulations, from which only the noise component will be measured. 
For this purpose, we use the 100 simulations  provided by the Planck release.
 The covariance  extracted from these 100 cross-correlation measurements serves  as our estimation of the error about 
$\hat{C}^{\kappa_{\rm CMB}\kappa_{\rm gal}}_{\ell}$. 

Note that the map masking and smoothing is already included this estimate (and in equation \ref{eq:Cov_approx}), 
therefore the error extracted this way is exactly what is needed for the forward modelling approach.
In a backward modelling approach however, one would need to deconvolve the mode-mixing matrix from this estimate by solving%
\footnote{To the best of our knowledge, this step was not included in H15.  
It probably did not impact heavily the results nor the conclusions, but might have altered by a small amount the final covariance matrix that enters the $\chi^2$ calculations, and therefore the significance  as well.}:
\begin{eqnarray}
\widehat{\rm Cov}^{\kappa_{\rm CMB} \kappa_{\rm gal}} = {\rm M}{\rm Cov}^{\kappa_{\rm CMB} \kappa_{\rm gal}} {\rm M}^{T}.
\label{eq:CovBackward}
\end{eqnarray}
Since the scales involved are mostly in the linear regime, it is possible to verify our error estimate against the  Gaussian prediction:
\begin{eqnarray}
\widehat{\rm Cov}^{\kappa_{\rm CMB} \kappa_{\rm gal}}_{\ell \ell', G} = \frac{\delta_{\ell \ell'}}{f_{\rm sky}(2\ell + 1)\Delta \ell} \bigg( \hat{C}^{\kappa_{\rm CMB}}_{\ell} \hat{C}^{\kappa_{\rm gal}}_{\ell} + \bigg[\hat{C}^{\kappa_{\rm CMB}\kappa_{\rm gal}}_{\ell}  \bigg]^2\bigg). 
\label{eq:GaussVar}
\end{eqnarray}
In the above expression, $\Delta \ell$ is the bin width, $\delta_{\ell \ell'}$ is the Kronecker delta function, and $f_{\rm sky}$ is the sky fraction commonly covered by both fields, including masking and apodization (when applicable). 
By inserting the observed auto-spectra $\hat{C}^{\kappa_{\rm CMB} }_{\ell}$ and $\hat{C}^{\kappa_{\rm gal}}_{\ell}$ as opposed to the theoretical values, these Gaussian  predictions naturally take into account the statistical  noise, the effect of masks and smoothing.
The cross-spectrum  term $\hat{C}^{\kappa_{\rm CMB}\kappa_{\rm gal}}_{\ell} $ can be replaced by the theoretical value $C^{\kappa_{\rm CMB}\kappa_{\rm gal}}_{\ell}$
since the effects of masking are much lower and the two noise contributions are uncorrelated. 
The different terms entering equation \ref{eq:GaussVar} are presented in Fig. \ref{fig:TheoryErr}.
We see therein that in the absence of noise, the cross-term is an order 10-50\% correction to the leading term, whereas it is 
currently a negligible $<10\%$ correction on the error. 
In this Gaussian  calculation, we define the observed convergence map power spectrum as: 
\begin{eqnarray}
   \hat{C}^{\kappa_{\rm gal}}_{\ell} = C^{\kappa_{\rm gal, th}}_{\ell} + \frac{\sigma_{\epsilon}^2}{\bar{n}},
\end{eqnarray}
where the galaxy intrinsic shape dispersion is given by  $\sigma_{\epsilon}$ = 0.278 and 0.272 {\color{black}} for the CFHTLenS and RCSLenS respectively,
 and the mean number density by $\bar{n}$ =  8.30 and 4.25 gal/arcmin$^2$ (see Section \ref{subsec:CFHTLenS_data}).
 Since the covariance scales {\it linearly} with the auto-spectra and with the inverse of the sky coverage, 
 the error bars on the RCSLenS measurements are $\sim \sqrt{2}$ smaller than those from CFHTLenS (the expected signal is about twice smaller due to the shallower $n(z)$, 
  but benefits from four times the area).

\begin{figure}
\begin{center}
\includegraphics[width=2.5in]{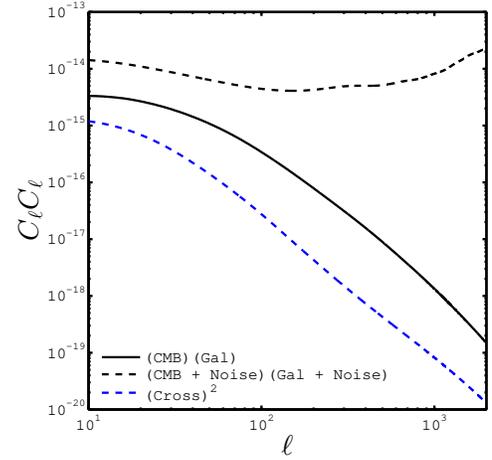}
\caption{Theoretical predictions for the two terms  that enter the Gaussian error predictions of equation \ref{eq:GaussVar}, shown in the WMAP9 cosmology for the RCSLenS survey. 
The  black solid (dashed) line corresponds to the first  term   -- $\hat{C}^{\kappa_{\rm CMB}}_{\ell} \hat{C}^{\kappa_{\rm gal}}_{\ell}$ -- without (with) the noise contribution.
The blue line represents the second term -- $\left[\hat{C}^{\kappa_{\rm CMB}\kappa_{\rm gal}}_{\ell}  \right]^2 $ -- which is always subdominant. 
The noise-free error for CFHTLenS is about twice that of  RCSLenS given its higher signal, but that difference is reduced when noise is included in the predictions. }
\label{fig:TheoryErr}
\end{center}
\end{figure}

The Gaussian predictions are shown as the blue error bars in Fig. \ref{fig:cmb_x_kappa_RCS}, whereas the error extracted from the 100 CMB lensing simulations is shown in black.
Both error estimates are in close agreement in all pipelines, often better than 10\%. 
 Following H15, LH15 and K15, we use the latter in our error analysis, which are computed for each field separately. 
 These contain non-zero off-diagonal elements that play a minimal role, as seen in Fig. \ref{fig:CrossCoeff} for RCSLenS field CSP0320.
 As expected from the Gaussian predictions and shown here, the correlations between different Fourier modes are always negligible.
  The off-diagonal elements are nevertheless important carriers of noise properties. As shown in \citet{Taylor2013}, the fractional noise in a covariance
 matrix of $N_d^2$ elements, estimated from $N_s$ simulations,  can be estimated as $\sqrt{2/(N_s-N_d-2)}$. 
 In a one-parameter measurement  -- in our case the amplitude of the cross-correlation signal --
 this error propagates through to an {\it extra} covariance term on the parameter, which inflates the covariance by $(1+1)/(N_s-N_d)[1 + 2/(N_s - N_d)] = 0.022$  \citep{2014MNRAS.442.2728T}.
Because this  2\% effect is small compared to the other error terms, we do not include it here. 
 The noise in the covariance matrix has a second effect: it produces an error {\it on} the error, which can often be catastrophic.
 Given our numbers of simulations and data points, the fractional error on our error reaches  $\epsilon = \sqrt{2/N_s + 2(N_d/N_s^2)} =14.5\%$. 
 Since it is non-negligible, we decided to be conservative and modify our final uncertainty to include the upper $1\sigma$ limit
(details can be found in Section \ref{subsec:results}).
  Note that the three previous paper omitted to include this extra error, hence care should be taken when comparing results.

\subsubsection{Configuration-space covariance}
\label{subsubsec:error_real}

We estimate the covariance about our configuration-space estimators in a similar way, 
i.e. from 100  $\kappa_{\rm CMB}\times \kappa_{\rm gal}$ measurements carried out with the Planck lensing simulations.
The diagonal elements are used to estimate the error bars in Fig.  \ref{fig:cmb_x_kappa_RCS},
while the cross-correlation coefficients are shown in Fig. \ref{fig:r_real-space} for the joint $\{\xi^{\kappa_{\rm CMB}\kappa_{\rm gal}}, \xi^{\kappa_{\rm CMB}\gamma_{t}} \}$ data vector. 
There is a high level of correlation between pairs of data points within the same estimator (lower left and upper right blocks),
while pairs that come from both estimators  (off-diagonal blocks) are subject to a strong correlation or anti-correlation, depending on the angular bins they belong to.
 {\color{black} This is expected as the different points measure mostly the same matter structures,
 and the two configuration-space estimators probe slightly different scales. 
 }

\begin{figure}
\begin{center}
\includegraphics[width=2.4in]{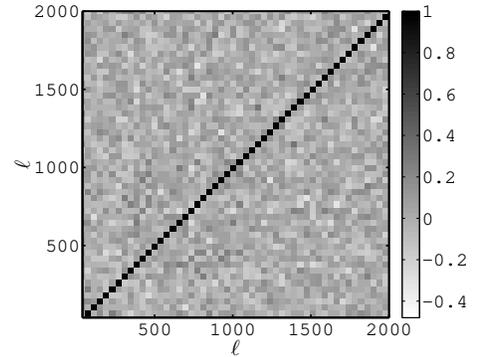}
\caption{Cross-correlation coefficient matrix measured from the RCSLenS field CSP0320, defined as $r_{ij} = \mbox{Cov}_{ij}/\sqrt{\mbox{Cov}_{ii} \mbox{Cov}_{jj}}$.
Within noise, this is consistent with the Gaussian diagonal predictions.
Measurements of $r_{ij}$ made on other fields are very similar to this one.}
\label{fig:CrossCoeff}
\end{center}
\end{figure}

\begin{figure}
\begin{center}
\includegraphics[width=2.5in]{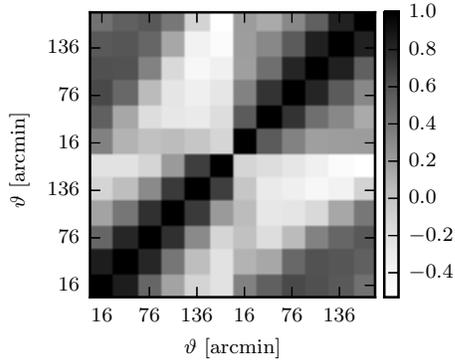}
\caption{Cross-correlation coefficient matrix about the configuration-space estimators, measured from all RCSLenS fields. 
The first six bins  along each axis represent the correlation  about $\xi^{\kappa_{\rm CMB}\kappa_{\rm gal}}$,
while the last six show $\xi^{\kappa_{\rm CMB}\gamma_t}$.}
\label{fig:r_real-space}
\end{center}
\end{figure}


\section{Validation }
\label{sec:validation} 
 
 This section describes a series of tests performed first on the analysis pipelines, then on the data products,
 that are meant to validate our results.

\subsection{Numerical simulations}
\label{subsec:sims}

\begin{figure*}
\begin{center}
\includegraphics[width=2.0in]{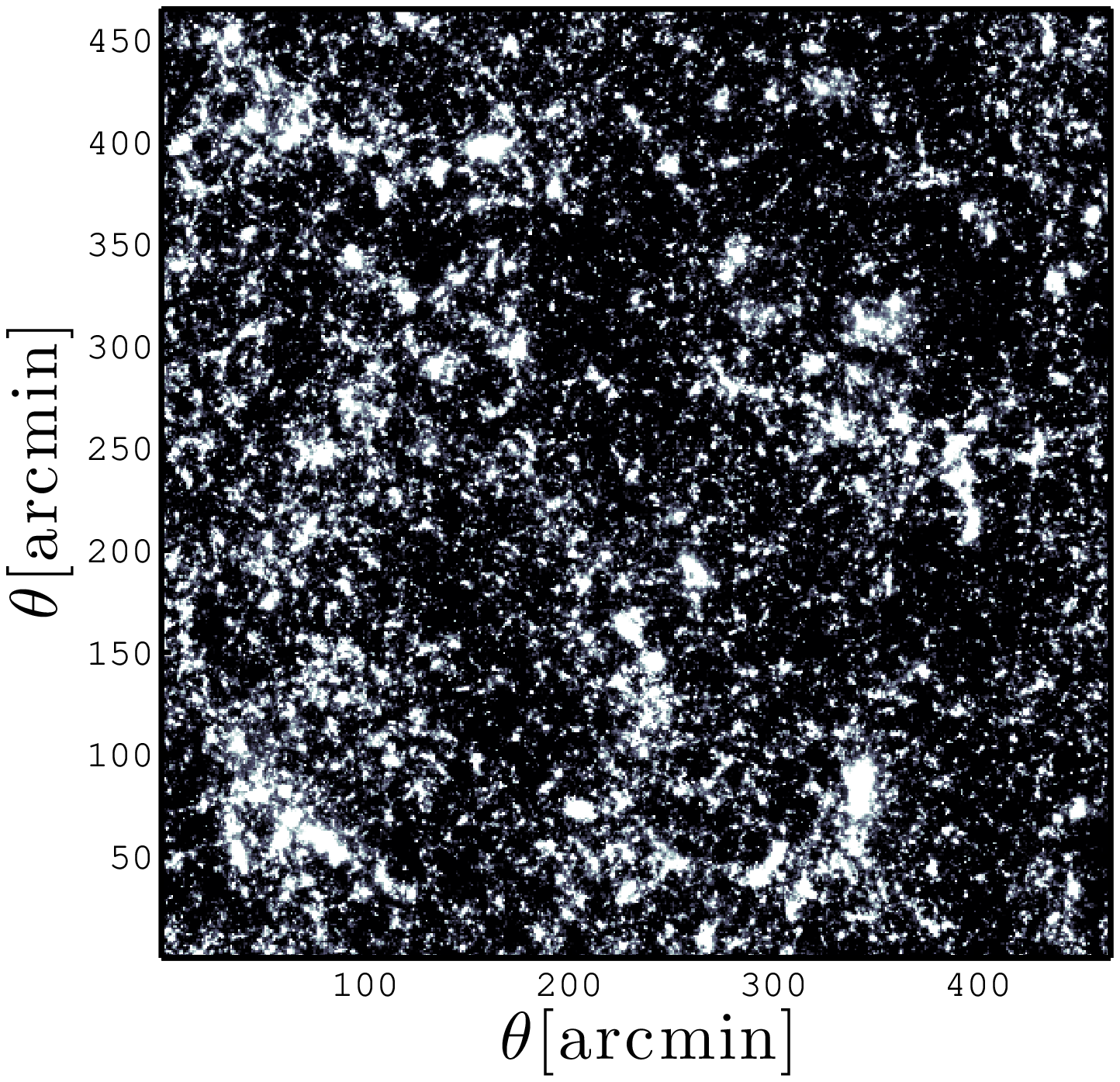}
\includegraphics[width=2.0in]{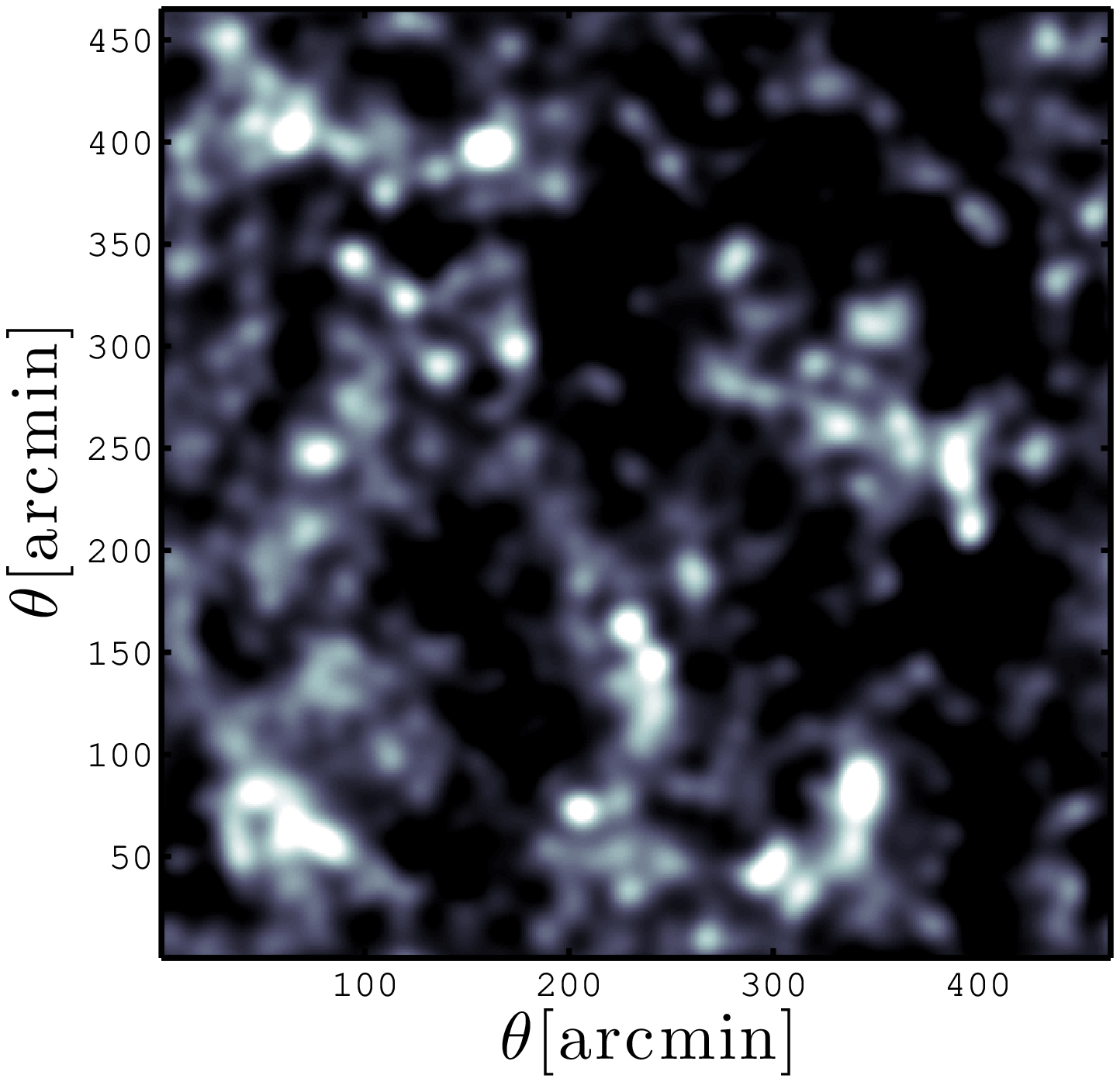}
\includegraphics[width=2.0in]{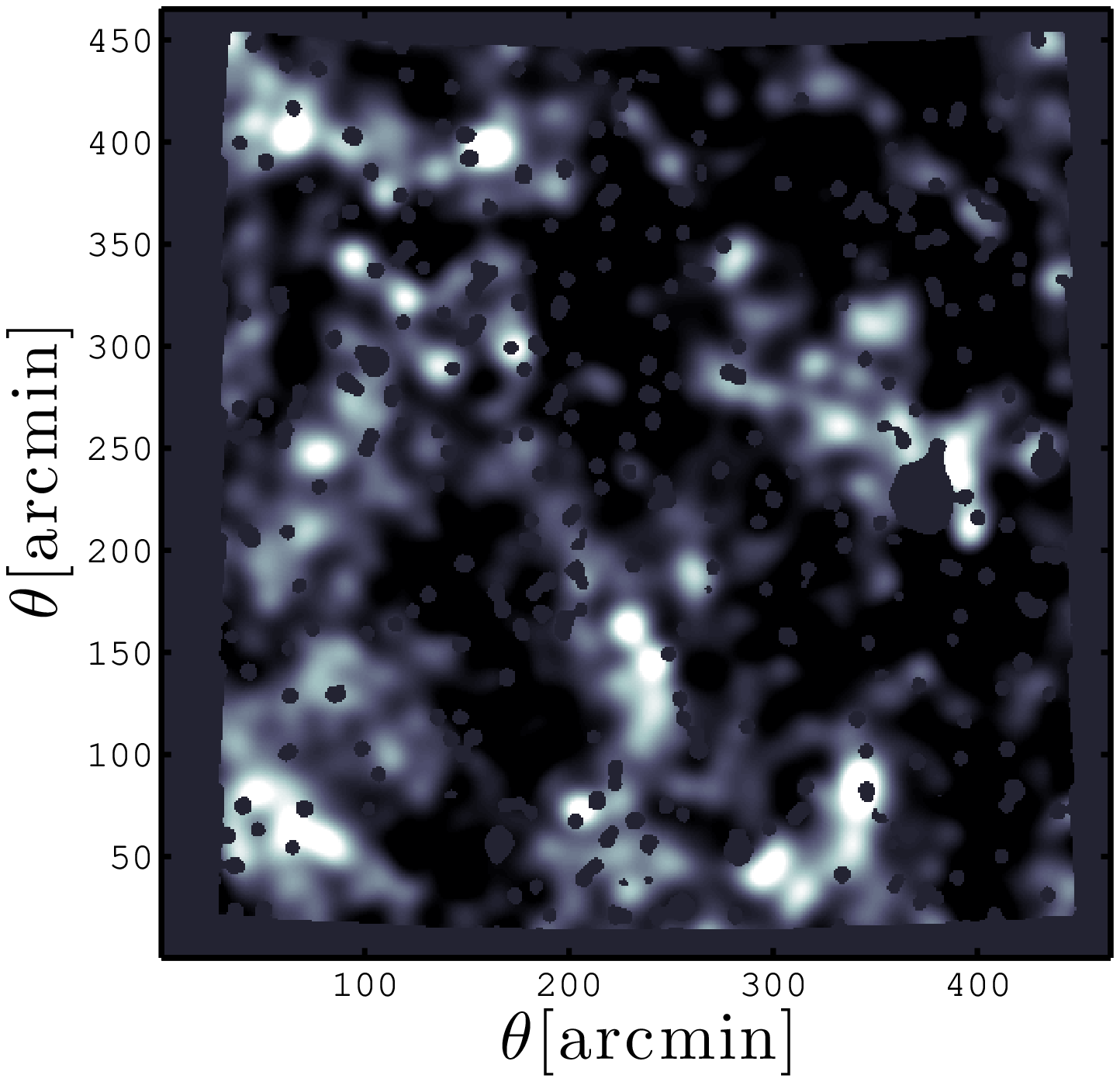}
\includegraphics[width=2.0in]{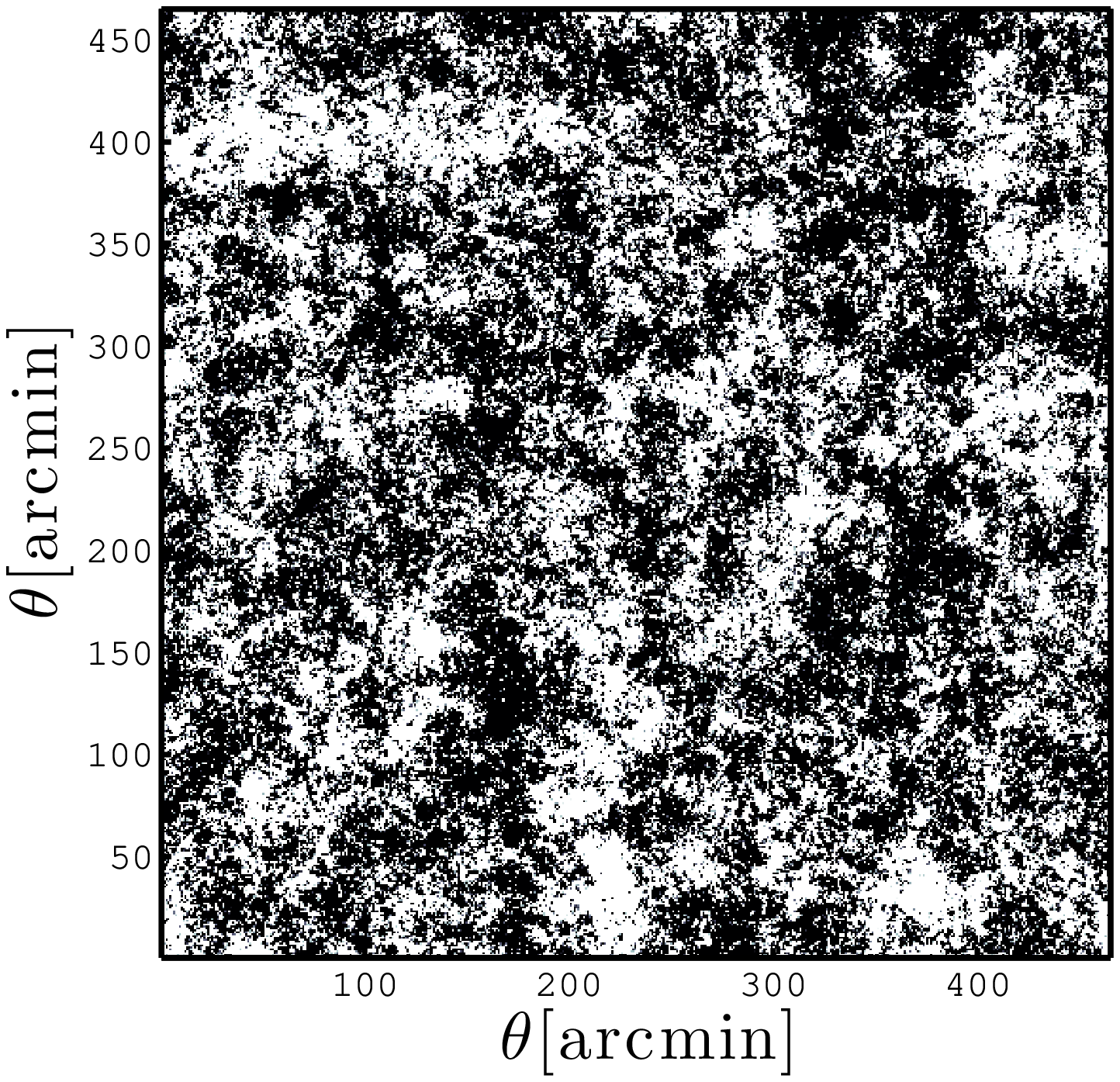}
\includegraphics[width=2.0in]{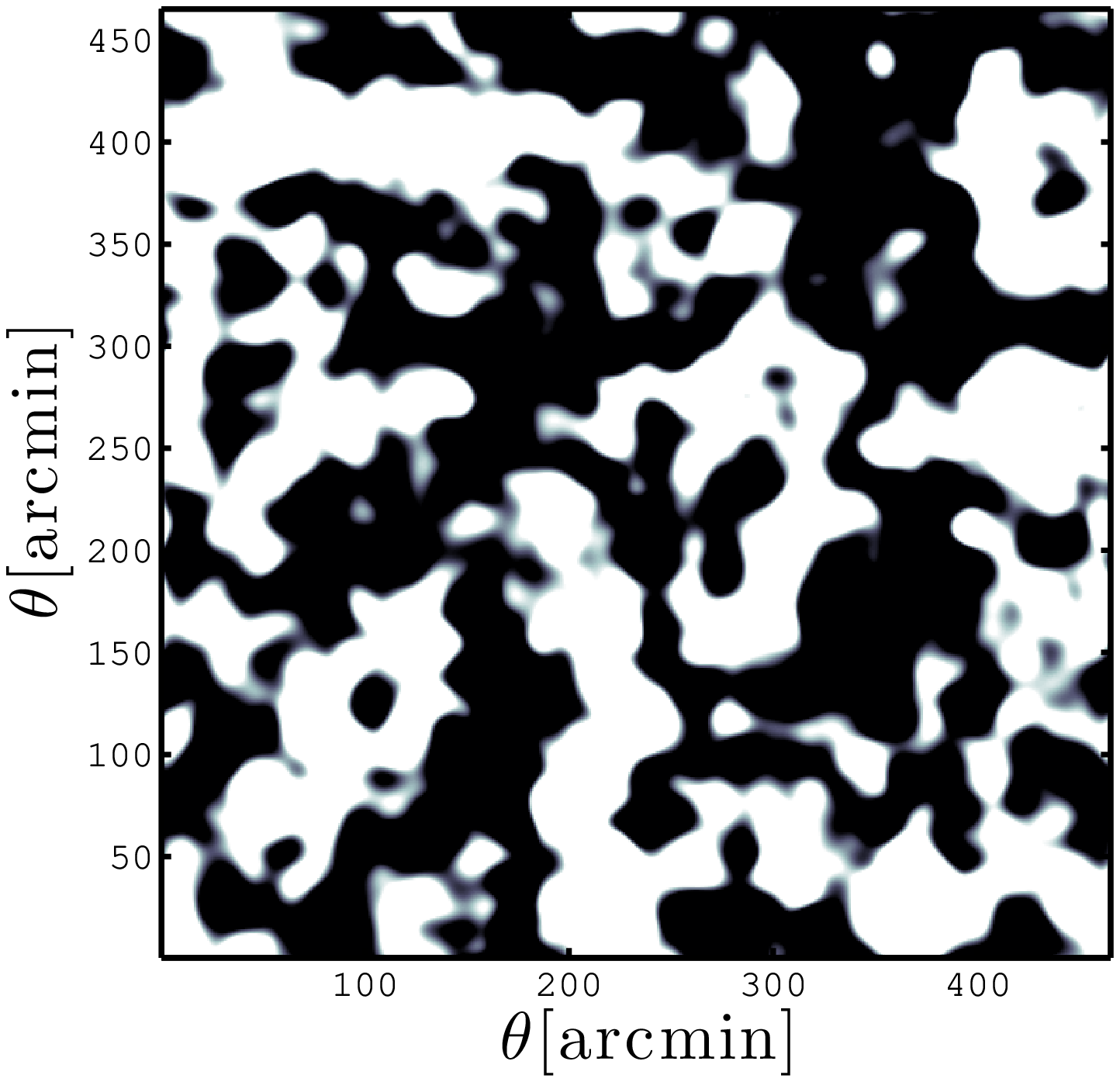}
\includegraphics[width=2.0in]{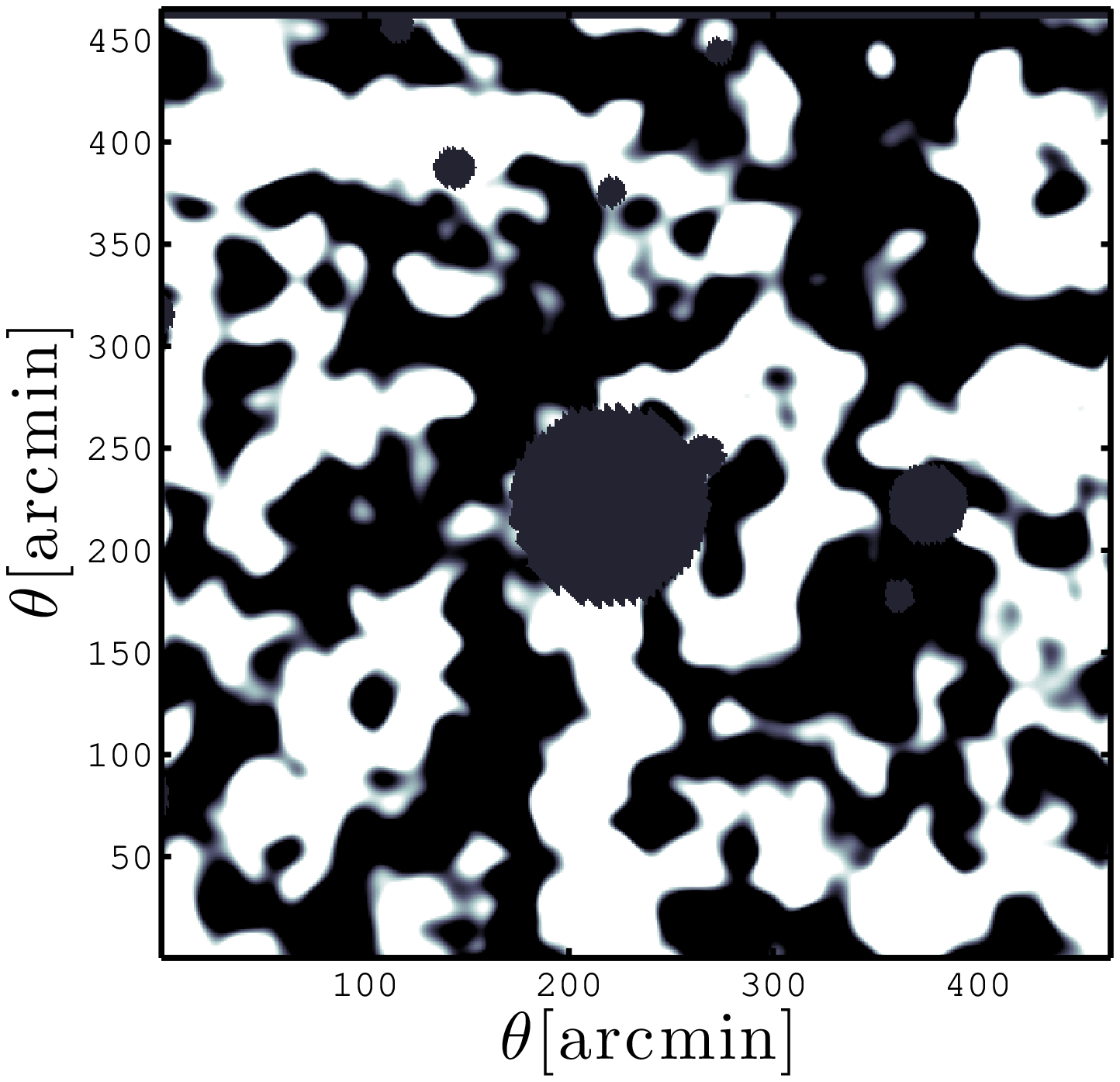}
\caption{(top left:) One of the 50  simulated  $\kappa_{\rm gal}$ maps from the SLICS series, shown with full $6000^2$ pixel resolution and  for $z_s = 0.58$. 
                In our validation pipeline, the map is first  convolved with a 6 arcmin Gaussian beam to mimic the resolution of the data maps (top middle panel), 
                then we apply the W3 mask (top right). 
                The three bottom panels show the same procedure, this time for the $\kappa_{\rm CMB}$ map as $z=1080$, using the mask from the Planck  lensing analysis that covers the W3 footprint.
                }
\label{fig:mask_sim}
\end{center}
\end{figure*}

We verify our Fourier pipelines
against numerical simulations, for which the cosmology is known, and the observational mask and smoothing can be switched on and off.
This way, it is possible to validate each segment of our forward modelling procedure. The map reconstruction algorithm has been thoroughly tested in vW13, 
and it is actually the cross-correlation measurement itself that is investigated. 

For this purpose, we use 100 lines-of-sights (LOS) from the SLICS-LE  suite described in \citet{Harnois-Deraps2015a}.
These are dark matter only simulations that were produced from the public {\small CUBEP$^3$M} $N$-body code \citep{Harnois-Deraps2013b},
based on the WMAP9 cosmology.
In the original SLICS-LE suite, each light cone is ray-traced through a series of 18 mass planes in the redshift range ${\{0,3 \}}$, covering 60 deg$^2$ with a resolution of $6000^2$ pixels.
We extend the light cones to the surface of last scattering at $z_*$ by filling the high-redshift region  with 10 mass planes separated by 517 $h^{-1}$Mpc.
In order to cover a large opening angle at the high redshift end, we opted for mass planes that are 1034$h^{-1}$Mpc on the side (exactly twice the separation distance),
such that the full map can be calculated by collapsing the density field from exactly one half the simulation volume. 
Aside from this change in box size -- the SLICS are 505$h^{-1}$Mpc on a side -- 
the cosmology and the number of simulation elements are unchanged. 

Since this high-redshift extension of the SLICS contributes only minimally in the final cross-correlation measurement, 
there is no need to accurately resolve the non-linear  structure therein. 
Therefore we use a  numerical  solver solely based on linear perturbation theory\footnote{This solver is implemented numerically 
in the initial condition generator of the $N$-body code described in \citet{Harnois-Deraps2013b}.} 
to create mass maps at the 10 prescribed redshifts: $z_{\rm high}$ = 
  3.498,      
  4.544,    
  5.993,    
  8.086,    
   11.27,     
  16.49,    
   25.90,     
  45.63,    
  98.89  and    
   346.6.
As these maps are noisy due to the discreteness of the particles, we convolve  these 10 maps with a beam very close to the actual  smoothing scale of the $\kappa_{\rm gal}$ maps.
Note that the central  purpose of these simulations is to validate the signal pipelines, not to estimate the uncertainty.
As such, there is no need to create more than one of these $z>3$ extensions.  
Instead, we graft  this single high-redshift section to 100 independent LOS, and ray-trace the full light cones up to $z_*$ in the Born and flat sky approximations. 

To reproduce the measurement under investigation, we include one simplification, which is to assume 
that all galaxy sources are located on a single redshift plane at $z_s = 0.58$. This choice of $n(z)$ is simpler than adopting a broader distribution,
yet sufficient to validate our pipelines.  
We then convolve both the simulated  $\kappa_{\rm CMB}$ and $\kappa_{\rm gal}$  maps with a 6 arcmin beam
and recover a smoothing scale comparable to that present in the data\footnote{This convolution
is performed in addition to the convolutions done on each of the $z>3$ mass planes mentioned in the main text. 
The latter becomes obsolete in the measurement, but is still useful for map visualization purposes.}.
After this smoothing, there is no need to keep the full resolution of the simulation maps, hence
we lower the pixel count to approximately match that of the observations. 
Finally, we apply the actual observation masks. 
For testing purposes, we use the W3 footprint, which falls nicely within one 60 deg$^2$ simulated LOS, 
but we verified that our results hold for other fields.
This procedure is shown  in the top three panels of Fig. \ref{fig:mask_sim} for $\kappa_{\rm gal}$,
and in the three bottom panels  for $\kappa_{\rm CMB}$.

\subsection{Verification of Fourier pipelines}
\label{subsec:Fourier_validate}

We next proceed with the cross-spectrum analysis of these mock observations, presented in Figs. \ref{fig:sim_validate} and \ref{fig:power_mask_sim}.
We first verify the agreement between the cross-correlation predictions and the unmasked, un-smoothed simulated maps,
such as those presented in the left-most panels of Figure \ref{fig:mask_sim}.
This agreement is well established in  Fig. \ref{fig:sim_validate}  where the theory falls well within the $1\sigma$ region of the simulations at all relevant scales. 
Smoothing is then applied on the $\kappa$ maps and the resulting cross-spectrum is shown as black open circles in the three panels of  Fig.  \ref{fig:power_mask_sim}  (labelled $\kappa_{\rm CMB} \times \kappa_{\rm gal}$ in the legend). 
This serves as our reference measurement for the mask pipeline comparisons,
hence we replicate these results in all panels. We do not show the smoothed predictions to avoid overcrowding the figures, but the match is as good as that of  Fig. \ref{fig:sim_validate}.

\begin{figure}
\begin{center}
\includegraphics[width=3.0in]{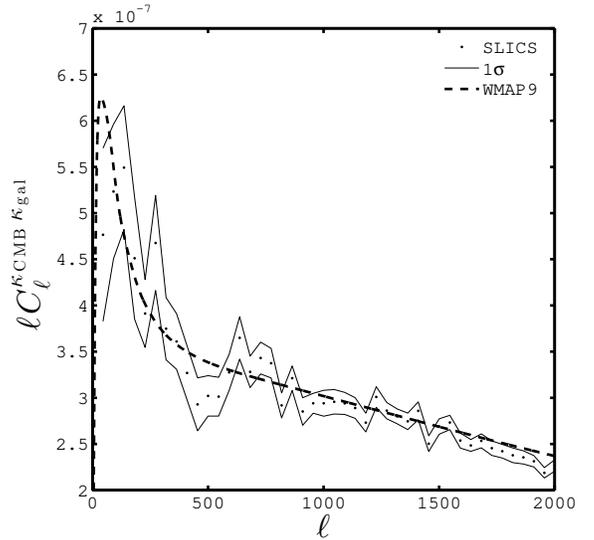}
\caption{Cross-spectrum between  the simulated $\kappa_{\rm gal}$ and $\kappa_{\rm CMB}$ maps averaged over 50 LOS. 
                One such pair is presented in the  top and bottom left panels of Fig. \ref{fig:mask_sim}.
                No mask nor smoothing are applied yet.
                The black symbols show the simulations results, the thin lines  represent the error on the mean ($1\sigma$-scatter/$\sqrt{50}$, 
                which overlaps nicely with the input WMAP9 cosmology (dashed line).}         
\label{fig:sim_validate}
\end{center}
\end{figure}

\begin{figure}
\begin{center}
\includegraphics[width=3.0in]{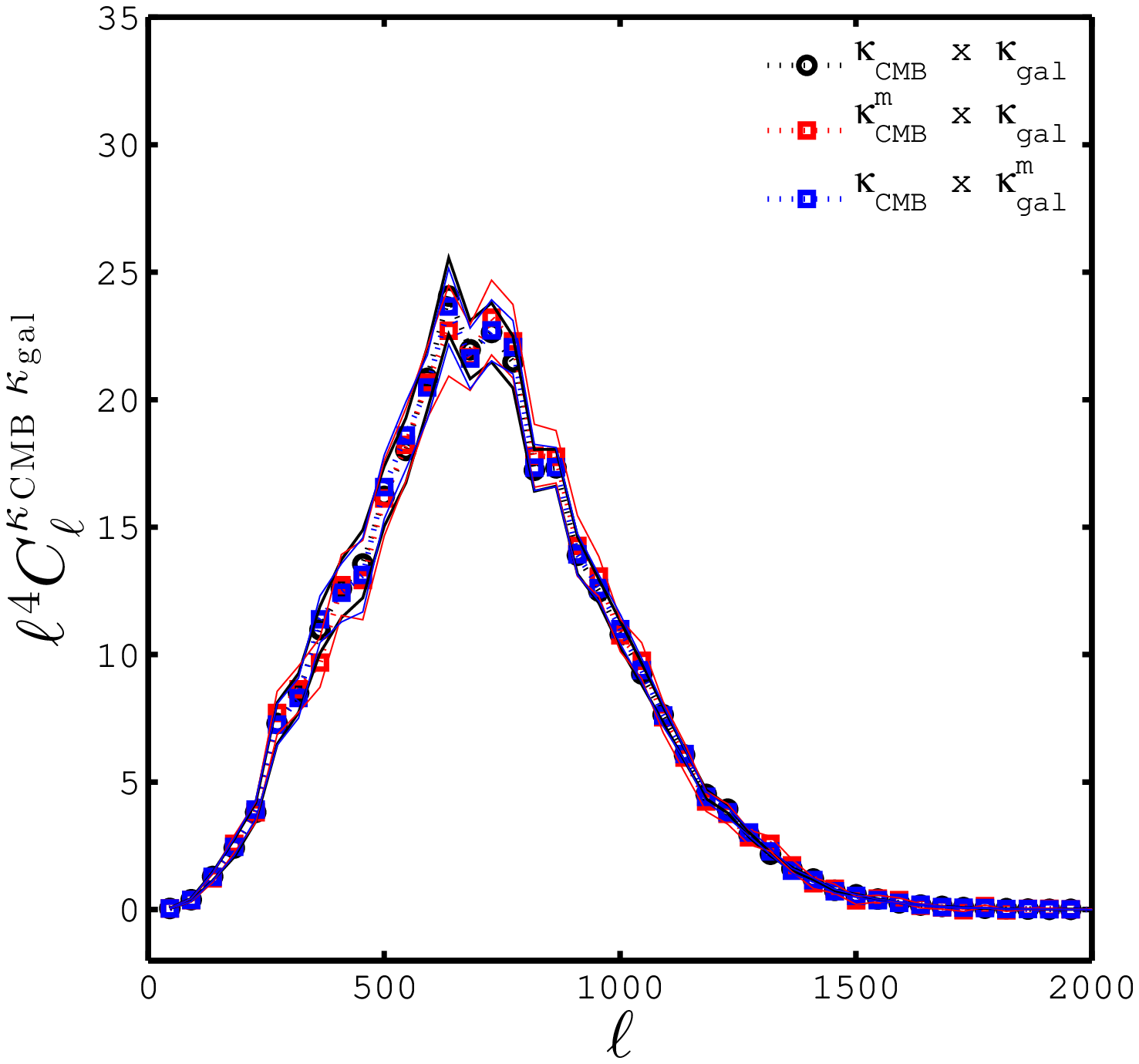}
\includegraphics[width=3.0in]{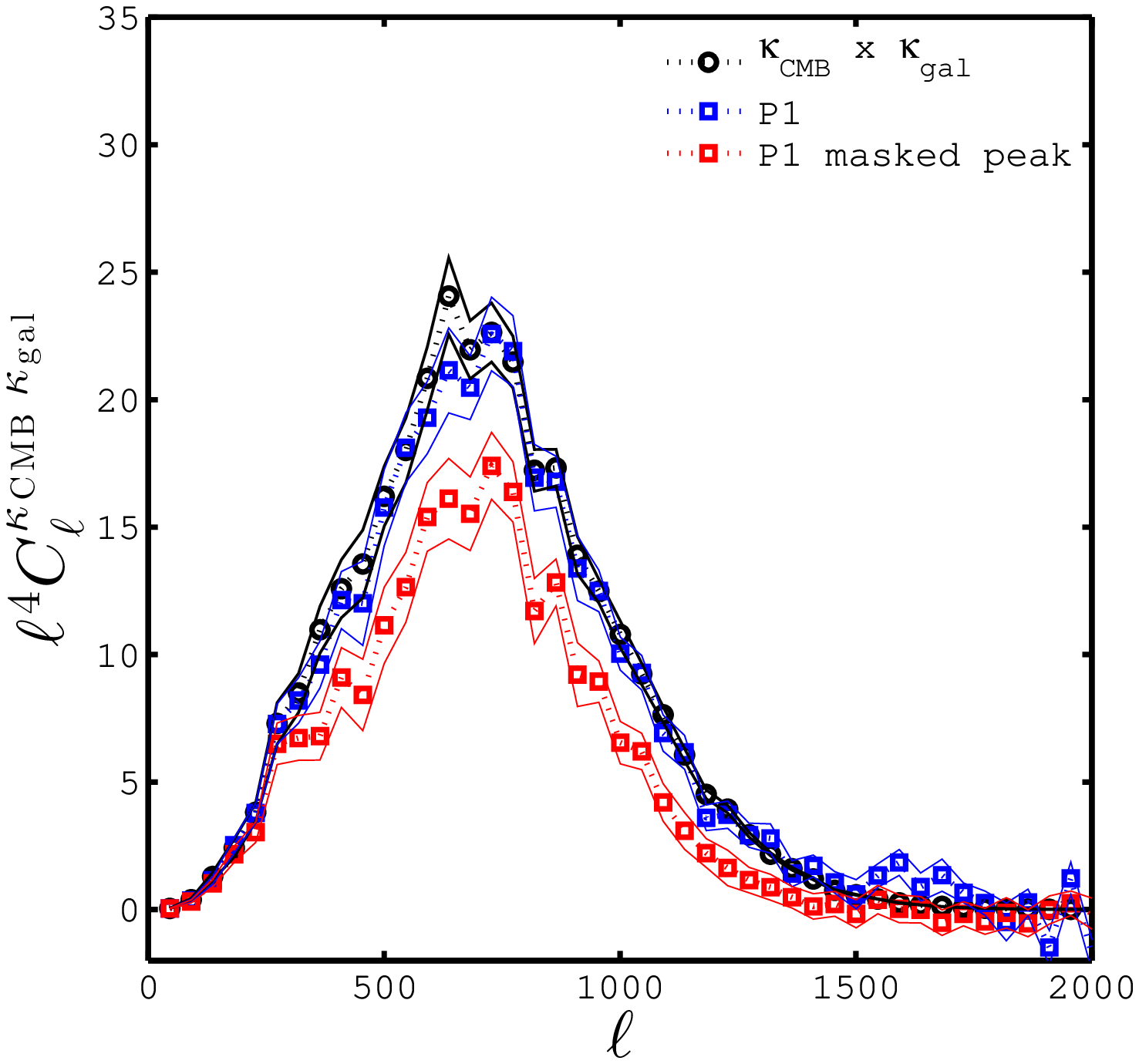}
\includegraphics[width=3.0in]{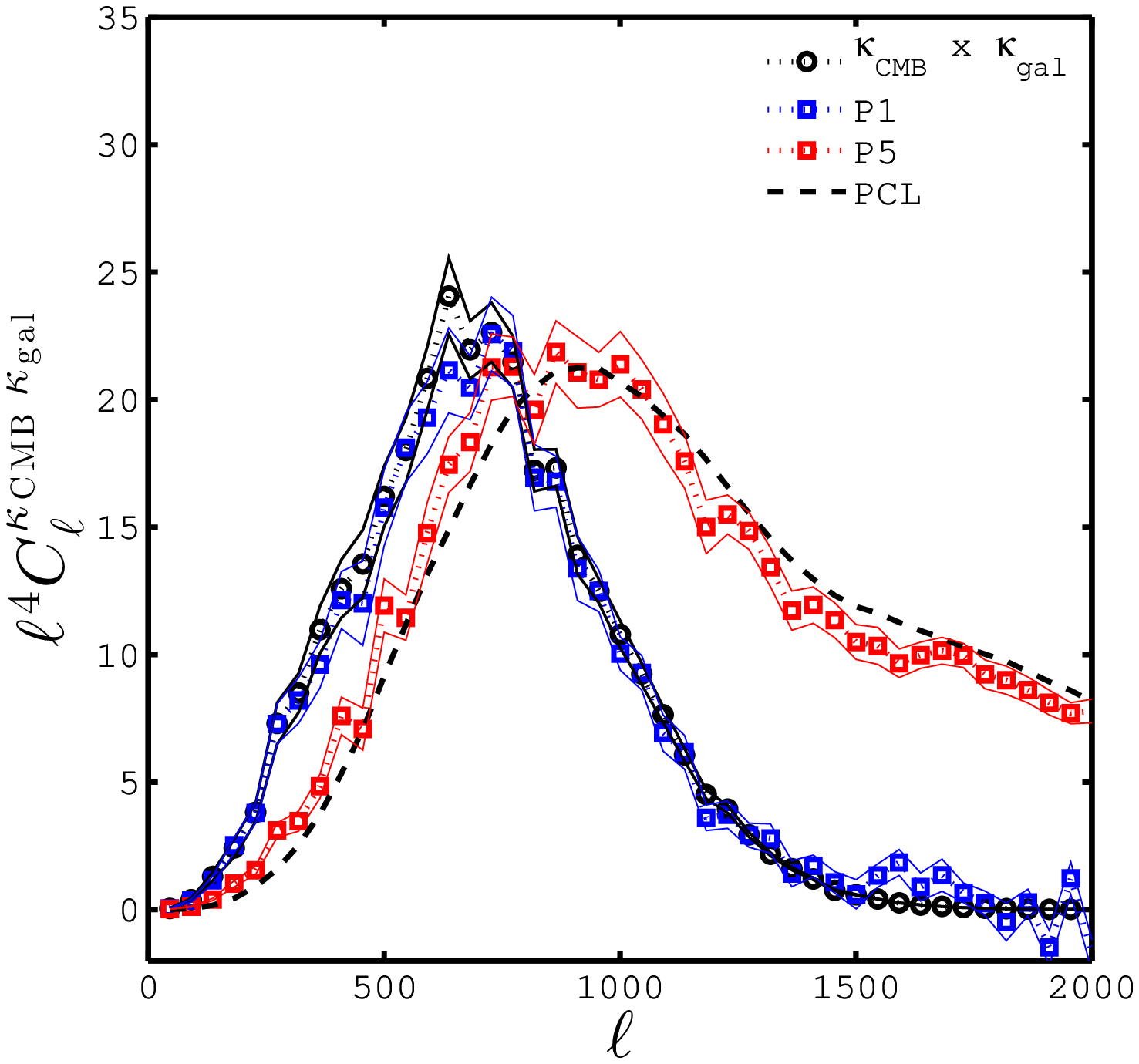}

\caption{Cross-spectrum between  simulated  $\kappa_{\rm gal}$  and $\kappa_{\rm CMB}$  maps,  analyzed with different pipelines. 
                In all panels, the black symbols show the `true' signal, with no mask applied. 
                 The thin lines  represent the error on the mean. 
                In the top panel, the red and blue symbols show the effect of applying the W3 mask
                on {\it either} $\kappa_{\rm gal}$ {\it or} $\kappa_{\rm CMB}$, but not on both. 
                 The middle panel shows the pseudo-$C_{\ell}$ pipeline P1 (blue), 
                 and a version of P1 in which the highest  peaks from $\kappa_{\rm gal}$ were deliberately masked from $\kappa_{\rm CMB}$ (red).
                The bottom panel compares P1,  P5 and the pseudo-$C_{\ell}$ forward model corresponding to  P5, labelled `PCL'.}
\label{fig:power_mask_sim}
\end{center}
\end{figure}

In the top panel of Fig. \ref{fig:power_mask_sim}, we systematically investigate  the effect of masking on these simulations.
First, we compute the cross-spectrum between the unmasked $\kappa_{\rm gal}$ and masked $\kappa_{\rm CMB}$ maps (shown in red, labelled $\kappa_{\rm CMB}^m \times \kappa_{\rm gal}$)
 and between the unmasked $\kappa_{\rm CMB}$ and masked $\kappa_{\rm gal}$ maps (shown in blue, labelled $\kappa_{\rm CMB} \times \kappa_{\rm gal}^m$). 
 The results are rescaled by $1/\sum(M)$,  according to the discussion in Section \ref{subsubsec:fourier}.
The agreement between  the mean value from these two measurements and the unmasked case (black) 
demonstrates that these two estimators are unbiased.

We next consider the case where both maps are masked, each  with their own respective masks,
thereby reproducing the pseudo-$C_{\ell}$ analysis with pipeline P1.  
The results are shown with the blue symbols in the central panel, which also trace faithfully the predictions up to $\ell\sim 2000$.
We reproduce the same level of agreement with pipelines P2-4.
There are slight undershoots and overshoots visible at $\ell\sim 600$ and $1500$ respectively, which comes from chance alignment of similar structures in the two masks or from data-mask interaction.
Ignoring the mode-mixing matrix (described in  Appendix \ref{sec:mode_mixing}) is therefore equivalent to modelling the blue symbols with a theoretical curve passing through the black region,
which is accurate in this case.

One  interesting conclusion we can draw from this calculation is that {\it if} one of the maps has no mask {\it and if}  the other mask does not correlate with the maps, then the pseudo-$C_{\ell}^{\kappa_{\rm CMB}\kappa{\rm gal}}$ measured this way is unbiased, as shown from either the blue or red curves from the top panel, and the mode-mixing matrix reduces to the identity matrix. 
{\color{black} This conclusion does not strictly hold for the measurement from our data sets. 
Whereas the Planck masks do not correlate with the $\kappa_{\rm gal}$ simulations, 
they correlate with the $\kappa_{\rm gal}$ data. This coupling is caused by the exclusion of point sources from the Planck lensing maps, 
which often correspond to high-density regions in the foreground lensing maps.
Instead, what is needed is a mock Planck mask that includes cutouts from the densest regions of the foreground maps. 
We investigate this effect in the simulations, where we construct a CMB lensing mask exclusively from  regions with $\kappa_{\rm gal} > 1.8\times10^{-4}$.
On each LOS, this selects only a handful of peaks with a distribution that qualitatively matches that of the actual Planck masks. 
Results are presented as the red symbols in the middle panel of Fig. \ref{fig:power_mask_sim}.
The $\kappa_{\rm CMB} \times \kappa_{\rm gal}^m$ measurement (not shown) is left unchanged, but now 
 $\kappa_{\rm CMB}^m \times \kappa_{\rm gal}^m$  (red, labelled `P1 masked peak') are systematically lower by a factor of 2 for $\ell<1000$,
even after at the area rescaling.
For higher multipoles, this bias seems to increase at first, but then gets noisier at the same time, 
which prevents us from making accurate statements. 
Unfortunately, this bias is very sensitive to the exact value of the threshold we used in the mask construction.
We set it  to $1.8\times10^{-4}$, but this numerical value was based solely on visual arguments. 
A more precise analysis will need to be performed with hydrodynamical simulations, where the point source 
cluster selection can be accurately reproduced. As such, our results are only indicative that this masking induced effect does 
suppress the cross-correlation signal at all scales, possibly by as much as  a factor of 2. 
If our interpretation is correct, this could largely explain the low amplitude found by LH15. 
}

Previously, both LH15 and K15 opted for the joint-masks approach, our P5,  where the two masks are multiplied, apodized, then  
re-applied on both maps prior to the cross-correlation.
This P5 analysis method is shown with the red symbols in the bottom panel of Fig. \ref{fig:power_mask_sim},
compared to P1 in blue.
 The introduction of power by masking is more important than in the other panels, and in the case of the W3 mask, this affects all modes with $\ell>800$.
 There is also a significant underestimation of power happening even at lower $\ell$-modes. 
 Again, ignoring the mode-mixing matrix is equivalent to modelling the measurements (blue for P1, red for P5) with the black symbols.
 This can be safely ignored for P1, but would lead to significant biases in the case of P5.  
In addition, forward modelling pipeline  P5 also comes with an extra challenge: it is now the combined-mask auto-spectrum that enters the mode-mixing matrix (equation \ref{eq:mixmatrix}), 
which is of higher amplitude compared  to the cross-spectrum of two different masks.
As such, debiasing the estimator with the mode-mixing matrix becomes both crucial and difficult. 
Unless this is done properly however, we see from  this comparison that  P5 is the least accurate among our pipelines.  

{\color{black} We present the full forward model predictions in the bottom panel (dashed line labelled `PCL'), 
which should be compared to the `P5' simulation measurement. 
In principle, the PCL line is the quantity that would go in the modelling component of the $\chi^2$ analysis.
The overall shape of the P5 symbols is well recovered by forward model.
Unfortunately there are discrepancies between the two curves, including a systematic under estimation for $\ell<500$,
which  prevent us from extracting accurate cosmological information with that method at the moment.
This is likely due to residual numerical effects in our implementation of the mode-mixing matrix
that we could not track down.
The overall results from our paper are not affected by this  since it concerns only an extension to the P5 pipeline,
 hence we leave a  careful re-examination of this numerical calculation for future work. }

We recognize that it has clear advantages compared to the other methods  when it comes to the inversion of the mode-mixing matrix, as done in H15: it is positive everywhere, 
which makes the deconvolution less noisy.
%
Finally note that in  Fig. \ref{fig:power_mask_sim}, the error bars presented are about the mean, computed from 50 SLICS simulations,
which are equivalent to 3000 deg$^2$. The 1$\sigma$ scatter in all these measurements increases rapidly for $\ell > 1000$,
which calls for large survey area.

\subsection{Verification of  configuration-space pipelines}
\label{subsec:real_validate}

We test our two configuration-space estimators  on 50 of SLICS simulations and present the results in Fig. \ref{fig:SLICS_real}. 
Similarly to the Fourier space analysis of the mock data, the $\kappa_{\rm CMB}$ maps  are smoothed, but this time with a 20 arcmin Gaussian beam that reproduces 
the effect of the Wiener filter on the data; 
the $\kappa_{\rm gal}$ are smoothed with a 6 arcmin beam. First,  the amplitude of the $\xi^{\kappa_{\rm CMB}\kappa_{\rm gal}}$ estimator is lower than the predictions, 
but only by a small amount. 
Second, there seems to be a small mismatch between the theory and the measured $\xi^{\kappa_{\rm CMB}\gamma_t}$
for $\vartheta \sim 100'$, but this has a weak significance. More importantly, these discrepancies are  smaller than the current statistical accuracy of our measurement, 
hence they should not affect our results. Aside from this, there is a good agreement between the simulations 
and the WMAP9 predictions.

\begin{figure}
\begin{center}
\includegraphics[width=2.7in]{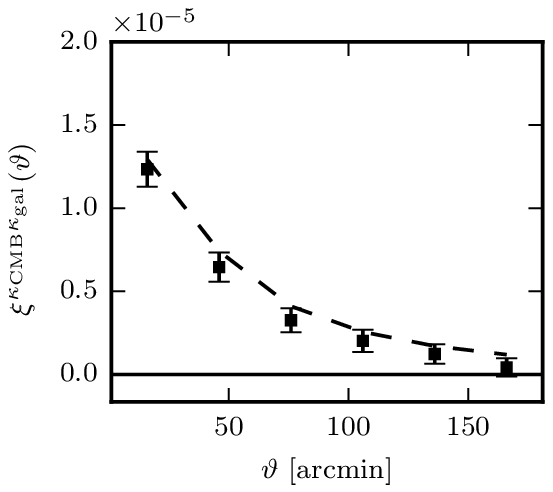}
\includegraphics[width=2.7in]{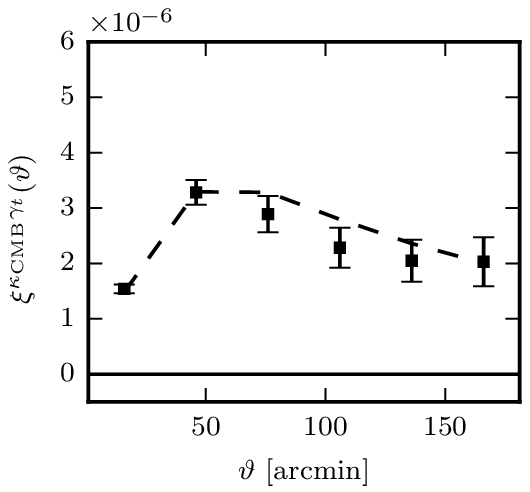}
\caption{Configuration-space estimators verified on the SLICS $N$-body simulations, compared to the theoretical WMAP9 predictions. 
The error bars are about the mean, i.e. divided by $\sqrt{50}$.}
\label{fig:SLICS_real}
\end{center}
\end{figure}


\subsection{Null tests for systematics residuals}
\label{subsec:null}

We perform a series of null tests that assert the quality of our maps and serve to flag potential systematic effects,
even though these are highly suppressed in cross-correlation measurements. 
We repeat most of these tests for each of the pipelines, and on each field.

The first consists of rotating the galaxy shapes by 45 degrees before reconstruction, with the mapping  $(e_1, e_2) \rightarrow (e_2, -e_1)$, thereby turning the (gravity) E-modes 
into (systematics) B-modes.
This procedure was described in vW2013 and any measurements performed on these B-mode maps are expected to be consistent with noise. 
Non-zero signal would point to residual systematics in the data that could leak in the analysis.

We begin with a verification that the E- and B-mode mass maps are mostly uncorrelated, by comparing their cross-spectra to their respective E- and B-
auto-power spectra. We present this measure in Fig. \ref{fig:ExB}, and see that the amplitude is generally lower than $5.0\times10^{-9}$, 
whereas both the E and B auto-spectra 
are more than two orders of magnitude higher, at all multipoles. 
We therefore conclude that the level of correlation between E and B is consistent with zero.
This is not the same as claiming that the $\kappa_{\rm gal}$ maps are free of B-modes, since that contamination would not be detected in that test.
It rather shows that {\color{black} if any residual contamination from the B-modes leaked into the E-mode maps, they do not correlate with the true E-modes, 
hence should disappear in cross-correlations measurements}. For measurements and further discussions of B-modes in CFHTLenS 
and RCSLenS, see \citet{Heymans2012c} and \citet{Hildebrandt2015}.

\begin{figure}
\begin{center}
\includegraphics[width=2.3in]{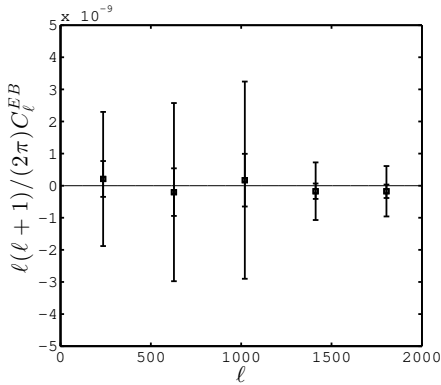}
\caption{
Cross-spectrum between the E- and B-mode $\kappa_{\rm gal}$ maps measured from the 14 RCSLenS fields  with the pseudo-$C_{\ell}$ pipeline P1.
The outer error bars represent the $1\sigma$ scatter between the fields, the inner ones are about the mean. 
}
\label{fig:ExB}
\end{center}
\end{figure}

To verify this, we compute the cross-correlation between the Planck lensing maps and the B-mode maps,
shown as thick red circles in Fig. \ref{fig:cmb_x_kappa_RCS}.
The level of B-modes in the cross-correlation is clearly lower than the E-modes for all multipoles.
Although this figure shows the results from the P1 pipeline, we verified that this was also the case for P2-P5.
Additionally, this B-mode cross-correlation measurement is shown to be consistent with zero in Section \ref{subsec:results} from a $\chi^2$ calculation, 
which is indicative of a truly successful test. This confirms that if there are true B-modes in the E-mode maps, 
they do not significantly survive the cross-correlation.  
We perform this null test on the two configuration-space estimators, shown with red points in Fig. \ref{fig:cmb_x_kappa_RCS_real},
which are also consistent with the null hypothesis. 

The next null test consists of rotating the galaxies randomly before performing the map reconstruction and the cross-correlation,
which we repeat 100 times. 
By construction, these randomized $\kappa_{\rm gal}$ realizations are  consistent with the noise levels of the $\kappa_{\rm gal}$ maps.
Therefore we expect that their cross-correlation with  $\kappa_{\rm CMB}$ should be consistent with zero as well.
We perform this test on  pipeline 1, 2 and 5, and show the results as  black points in Fig. \ref{fig:cmb_x_kappa_RCS}.
These are all in excellent agreement with the null expectations. 
There is no need to carry this test for P3-4, since zero-padding can not  alter the noise levels.

Finally, we perform a cross-correlation of 100 CMB lensing simulations, provided by the Planck 2015 data release, with both the $\kappa_{\rm gal}$ E-modes and B-modes maps, 
also shown as thin black points in Fig. \ref{fig:cmb_x_kappa_RCS}.
We see that the mean value is well centred on zero in all pipelines.

From these different tests, we conclude that the level of residual systematics that affect our signal  must be much smaller
than the statistical error uncertainty. The CFHTLenS and RCSLenS data sets are in excellent condition for
cross-correlation science, and no further action needs to taken to recalibrate the measurement.


\section{Cosmological Inference}
\label{sec:cosmology}

We present in this section our cosmological interpretation of this cross-correlation analysis.
We  describe the significance of our measurement and compare our results with H15, LH15 and K15.
We also discuss the potential contamination by intrinsic alignment, the effect of uncertainty in $n(z)$ and on the $m$-correction, 
and finally explore the constraining power on various cosmological parameters.

  \subsection{Significance}
\label{subsec:results}

\begin{table}
\caption{Summary of $\chi^2$ and SNR values obtained with the five different pipelines P. 
The CFHTLenS dataset has  19 degrees of freedom (5 bins x 4 fields - 1 free parameter), whereas the RCSLenS dataset has 69 (5 bins x 14 fields - 1 free parameter). 
This is directly reflected in the significantly different $\chi_{\rm min}^2$ and $\chi^2_{\rm null}$ values between both surveys. 
The quantities $A^{\rm WMAP9}$  and  $A^{\rm Planck}$ are the best-fit amplitudes that scale the theoretical signal, which differ for each pipeline and survey.
 The values listed here include the covariance debiasing factor $\alpha$ and the extra error $\epsilon$ due to the noise in the covariance (see main text for more details). 
}
\begin{center}
\begin{tabular}{llccccc}
\hline
    &  P  & $\chi^2_{\rm min}$ & $\chi^2_{\rm null}$ & SNR & $A^{\rm WMAP9}$  & $A^{\rm Planck}$  \\
\hline
\multirow{5}{*}{CFHTLenS}
     &  P1 & 13.74 & 18.75 & 2.24 & 0.75 $\pm$ 0.34 & 0.68 $\pm$ 0.31 \\
     &  P2 & 12.41 & 15.80 & 1.84 & 0.64 $\pm$ 0.35 & 0.58 $\pm$ 0.32 \\
     &  P3 & 20.01 & 24.66 & 2.16 & 0.77 $\pm$ 0.36 & 0.70 $\pm$ 0.33 \\
     &  P4 & 18.32 & 21.68 & 1.84 & 0.66 $\pm$ 0.37 & 0.61 $\pm$ 0.33 \\
     &  P5 & 14.69 & 18.64 & 1.99 & 0.68 $\pm$ 0.34 & 0.62 $\pm$ 0.31 \\
\hline
\multirow{5}{*}{RCSLenS}
    &  P1 & 45.27 & 61.50 & 4.03 & 1.44 $\pm$ 0.36 & 1.31 $\pm$ 0.33\\
    &  P2 & 50.63 & 67.79 & 4.14 & 1.47 $\pm$ 0.36& 1.33 $\pm$ 0.33 \\
    &  P3 & 42.62 & 55.55 & 3.60 & 1.36 $\pm$ 0.40 & 1.24 $\pm$ 0.34\\
    &  P4 & 47.43 & 62.02 & 3.82 & 1.54 $\pm$ 0.41 & 1.39 $\pm$ 0.37\\
     &  P5 & 59.17 & 66.84 & 2.77 & 1.04 $\pm$ 0.39 & 0.95 $\pm$ 0.34\\
\hline
\multirow{5}{*}{Combined}
     &  P1 & 57.71 & 76.00 & 4.28 &  1.07 $\pm$ 0.25 & 0.98 $\pm$ 0.22 \\
     &  P2 & 62.23 & 79.14 & 4.12 &  1.04 $\pm$ 0.26 & 0.95 $\pm$ 0.22 \\
     &  P3 & 60.46  & 75.94 & 3.94 & 1.05 $\pm$ 0.26 & 0.95 $\pm$ 0.25 \\
     &  P4 & 64.61 & 79.26 & 3.83&  1.05 $\pm$ 0.28 & 0.96 $\pm$ 0.25\\
     &  P5 & 70.34 & 80.93 & 3.26& 0.84 $\pm$ 0.26 & 0.77 $\pm$ 0.24\\ 
\hline
\end{tabular}
\end{center}
\label{table:stat}
\end{table}%

\begin{table}
\caption{Summary of $\chi^2$ and SNR values obtained with the two configuration-space statistics. 
The CFHTLenS and RCSLenS dataset are measured directly on  4 and 6 bins respectively,
which explains the different $\chi^2$ compared to the Fourier analysis. 
With one free parameter ($A$), the number of degrees of freedom is $\nu=3$ and $\nu=5$
for the two surveys, when the estimators are analyzed separately.
For the joint $\kappa_{\rm CMB}\kappa_{\rm gal}+ \kappa_{\rm CMB}\gamma_t$  analysis,  
we have $\nu=7$ and $\nu=11$ respectively.
The last row shows the joint estimator analysis for the  combined surveys.
The error on $A$ includes the covariance debiasing factor $\alpha$ and the extra error $\epsilon$ due to the noise in the covariance.}
\begin{center}
\begin{tabular}{llccccc}
\hline
                               &     & $\chi^2_{\rm min}$ & $\chi^2_{\rm null}$ & SNR & $A^{\rm WMAP9}$  & $A^{\rm Planck}$  \\
\hline
\multirow{3}{*}{CFHTLenS}
    & $\kappa_{\rm gal}$   & 1.13   &  5.81& 2.16  &  0.74 $\pm$ 0.36 & 0.67 $\pm$ 0.33 \\
    &  $\gamma_t$             &  0.49  &  5.97 &  2.34 &  0.88 $\pm$ 0.41 & 0.81 $\pm$ 0.37 \\
    &  both                            & 1.56  &  7.93  &  2.53 &  0.72 $\pm$ 0.30 & 0.65 $\pm$ 0.28\\
\hline
\multirow{3}{*}{RCSLenS}
    & $\kappa_{\rm gal}$  & 2.23  &  19.06 & 4.10 &  1.30 $\pm$ 0.34 &  1.18 $\pm$ 0.31 \\
    &  $\gamma_t$             &  4.40 &  17.70 & 3.65 &  1.43 $\pm$ 0.42 & 1.30 $\pm$ 0.39 \\
    &  both                           &  9.16 &   27.84 & 4.33 &  1.24   $\pm$ 0.31 &  1.12 $\pm$ 0.28 \\
\hline
\multirow{1}{*}{Combined}
    &  both                         &   12.85 & 34.17 & 4.62 &  1.04 $\pm$ 0.24 & 0.94 $\pm$ 0.21 \\
\hline
\end{tabular}
\end{center}
\label{table:stat_real}
\end{table}%

The significance of our Fourier-space measurement is quantified by a signal-to-noise ratio (SNR) metric that examines deviations from a pure noise realization,
closely following the methods described in LH15,  H15 and K15.
Under the assumption that the fields are independent, we can ignore their field-to-field covariance and merge the 18 data and
prediction vectors; the 18 covariance matrices are then organized into a large block-diagonal matrix.
We compute the $\chi^2$ statistic as:
\begin{eqnarray}
\chi^2 = {\rm x}\mbox{Cov}^{-1} {\rm x}^{\rm T}
\label{eq:chi}
\end{eqnarray}
 with
 \begin{eqnarray}
 {\rm x} = \hat{C}^{\kappa_{\rm CMB}\kappa_{\rm gal}} - A \widetilde{C}^{\kappa_{\rm CMB}\kappa_{\rm gal}}. 
 \label{eq:chi2}
\end{eqnarray}
In equation \ref{eq:chi2}, $\hat{C}^{\kappa_{\rm CMB}\kappa_{\rm gal}}$ is the observed pseudo-$C_{\ell}$,
$\widetilde{C}^{\kappa_{\rm CMB}\kappa_{\rm gal}}$ is the modelled pseudo-$C_{\ell}$, 
and a matrix summation is implied. 
We include a free parameter $A$ that modifies the overall amplitude of the predictions,
and can eventually be mapped to combinations of $\sigma_8$, $\Omega_{\rm M}$ and other sources of systematic effects.  
Recall that the predictions are different for each field due to differences in smoothing scales, masks and $n(z)$ distributions. 
The configuration-space measurements are processed similarly, with the exception that the fields are already merged in each survey, leading to $\chi^2$ 
with fewer degrees of freedom.

The inverse covariance matrix that enters equation \ref{eq:chi} is debiased with the correction factor $\alpha = (n-p-2)/(n-1) = 0.94$ described in \citet{Hartlap2007a}, 
where $p$ is the number of data bins and $n$ is the number of simulations used in the covariance estimation. 
With $p=5$ and $n=100$, the inverse covariance (or equivalently the $\chi^2$) must be lowered by $\alpha^{-1} - 1 =  6\%$. 
This debiasing scheme is not exact, as noted by \citet{2016MNRAS.456L.132S}; when sampled from simulations, the inverse of a covariance matrix should be debiased
by marginalizing over an inverse-Wishart distribution of the true covariance matrix.
Discrepancies between this and the \citet{Hartlap2007a} method increase rapidly with lower values of $\alpha$. 
With $\alpha  = 0.94$, this would lead to a correction on the $\chi^2$ calculated from the inverse covariance in the range $[0-20]\%$,
with larger corrections being applied on larger $\chi^2$, 
hence would lower the significance of our  measurement by a few percent. 
We note that H15 and LH15 did not debiased their inverted covariance, while K15 used the  Gaussian estimator (equation \ref{eq:GaussVar})  
verified against $N$-body simulations that were debiased with the \citet{Hartlap2007a} scheme.
We must therefore  keep in mind that ignoring the \citet{2016MNRAS.456L.132S} correction might slightly affect the significance of our conclusions,
as well as that of H15, LH15 and K15.

%
We then define 
SNR $= \sqrt{\chi^2_{\rm null} - \chi^2_{\rm min}}$, where $\chi^2_{\rm null}$ is computed by setting $A = 0$, 
and the best-fit value for $A$ is found at $\chi^2_{\rm min}$.
The error on $A$ is obtained from the one-parameter minimum $\chi^2$ method\footnote{In the minimum $\chi^2$ method with one parameter, 
the $1\sigma$ error on $A$ is found by varying  values of $A$ until $\chi^2_{\rm min} - \chi^2_{A} = 1$ \citep[see, e.g.][]{2003psa..book.....W}.}.
As discussed in Section \ref{subsubsec:error_fourrier}, the error on $A$ must be modified by a factor $(1 + \epsilon/2)$ as a result of the propagated uncertainty coming from the noise
in the covariance matrix \citep{2014MNRAS.442.2728T}.
 We therefore have $\sigma_A \rightarrow \sigma_A(1 + \epsilon/2) = \sigma_A(1 + 0.07)$, which, for instance, maps $\sigma_A=0.3$ to $\sigma_A=0.32$.   
The $\chi^2$ and SNR calculations are performed on the coarsely binned  measurement, where the inversion of the covariance matrix benefits from data compression.
With 5 $\ell$-bins per fields, we expect $\chi^2_{\rm min} \sim 19$ for CFHTLenS, $69$ for RCSLenS, and $89$ for the combined set.
The results from the Fourier-space analyses are shown in Table \ref{table:stat} for all pipelines,  assuming both WMAP9 and Planck cosmologies.

The measured values of $\chi_{\rm min}^2$ seem surprising low at first, as one typically expects $\chi^2/\nu\sim 1$, where $\nu$ is the numbers of degrees of freedom. 
Once we factor in the expected uncertainty  on the measured $\chi^2$ however, this apparent inconsistency is relaxed.
Indeed, the error on the $\chi^2$ measurement  can be calculated as $\sigma_{\chi^2} = \sqrt{2\nu}$,  
which yields $\sigma_{\chi^2} = 11.7$ and $6.1$ for the Fourier-space analyses of RCSLenS and CFHTLenS
respectively.
Within $2.2\sigma$, all measured  $\chi_{\rm min}^2$ from RCSLenS are consistent with $\chi^2/\nu\sim 1$,
and CFHTLenS are consistent within $1.1\sigma$.
This suggests that the error bars on our RCSLenS measurements are slightly overestimated, 
which implies that our results  are conservative.
%
%
%
%
The SNR from RCSLenS roughly doubles compared to that of CFHTLenS, which is solely due to the difference in survey area. 
Indeed, since the covariance scales as the inverse of the area (neglecting changes in $n(z)$), we can estimate the
gain in SNR between the two surveys from $\sqrt{A_{\rm RCS}/A_{\rm CFHT}} = \sqrt{600.7/146.5} = 2.02$,
which is very close to what we  observe.

We find that the CFHTLenS  data  prefer lower  values of $A$, which we loosely interpret as favouring the WMAP9 cosmology, 
whereas the RCSLenS data prefer higher values, seemingly closer to the Planck predictions. 
One must be cautious here since many factors are at play that could explain why this is happening even without requiring a change in cosmology,
as we discuss in the following sections.

One of the most important result from Table  \ref{table:stat} is the observed variations in both SNR and $A$.
 For the CFHTLenS data, P1 and P3 have the highest values in both of these quantities, and
 coincide with pipelines in which mask apodization was {\it not} applied.
 This suggests that this procedure has an important impact and should be carefully examined. 
 For RCSLenS, the SNR recovered from the different pipelines vary from {\color{black} 2.77} (P5) to {\color{black} 4.14} (P2), and the values of $A^{\rm Planck}$ vary from {\color{black} 0.95$\pm$0.34} (P5) to 
 {\color{black} 1.39 $\pm$ 0.37} (P4).
 Pipeline P5 presents the lowest value of $A$, most likely  due to the masking of the point sources in the Planck mask.
  These are regions that carry a significant fraction  of the weak lensing signal.  
  Removing  them from the analysis suppresses the cross-power at all scales, which translates in a lower $A$,
  as noted in LH15 and confirmed by our simulations in Section \ref{subsec:Fourier_validate}. 
    A comparison of the  $C_{\ell}^{\kappa_{\rm CMB}\kappa_{\rm gal}}$ measurements for P1-P5 (leading to the values entered in  Table  \ref{table:stat}) is presented in Appendix \ref{sec:pipelines}.
{\color{black} It is unfortunate that the expected gain in SNR from pipeline P3 and P4 (with the larger, zero-padded maps) is not directly seen in the Fourier analysis,
contrarily to the configuration space. 
This can be attributed to the fact that zero-padding introduces an additional window function, and that a full 
forward modelling of this quantity is required in order to recover the gain expected from equation \ref{eq:A_minus_B}. 
This window function should be incorporated in the mode-mixing matrix formalism, which would then be applied at the end of P3 and P4.
Since these two pipelines show no sign of  bias, we do not want to introduce inaccuracies in the modelling, 
hence we decided not to investigate this any further.}

We next  combine all 18 fields from both surveys to extract a single $A$.
We use $n_{\rm CFHT}(z)$ and $n_{\rm RCS}(z)$ accordingly, and follow the steps listed in Section \ref{subsec:PCL}.
We update the covariance debiasing term $\alpha$ and the extra sampling variance term $\epsilon$
for a measurement made with 10 data points.
The results are shown in the lower section of Table  \ref{table:stat} (labelled `Combined'), where we see 
an overall increase in significance, up to SNR = {\color{black}4.28} for the pseudo-$C_{\ell}$ pipeline P1.
We also recover a best fit value for $A$ that is in excellent alignement with both the WMAP9 and Planck predictions, 
however we are not able to distinguish between the two.

We repeat this measurement with the configuration-space estimators
and present the results  in Table \ref{table:stat_real} for $\xi^{\kappa_{\rm CMB}\kappa_{\rm gal}}$, $\xi^{\kappa_{\rm CMB}\gamma_t}$ 
and for the combination of both data vectors. With 6 bins per estimator in the RCSLenS measurements, the covariance de-biasing factors becomes a 3.5\% correction on the SNR, 
and 6.5\% for the joint $\xi^{\kappa_{\rm CMB}\kappa_{\rm gal}} + \xi^{\kappa_{\rm CMB}\gamma_t}$  analysis.  
We recover the same trends as the Fourier analysis: CFHTLenS prefers the lower values of $A$, RCSLenS prefers higher values,
and the actual values of $A$ agree within $1\sigma$ with the Fourier analyses. We also note that $\gamma_t$ prefers 
higher $A$, as already seen in Fig. \ref{fig:cmb_x_kappa_RCS_real}. {\color{black} This has a very low significance given the size of the error bars, and is likely due to the different physical scales probed by the estimators. 
By construction, for a fixed $\vartheta$ the estimator $\kappa_{\rm CMB}\kappa_{\rm gal}$ is more sensitive to larger angular scales
compared to $\kappa_{\rm CMB}\gamma_t$, largely because of the $J_2$ term that appears in equation \ref{eq:xi2-prediction}. 
Consequently, in order to correctly probe the $\kappa_{\rm CMB}\gamma_t$, the measurement needs to reach larger angles.
We verified in Section \ref{subsec:real_validate} that both $\kappa_{\rm CMB}\kappa_{\rm gal}$ and $\kappa_{\rm CMB}\gamma_t$ give consistent results on simulations,
hence the difference cannot be attributed to a mis-calibration of the estimators.  
Because all points are highly correlated and the preference is only a $1\sigma$ effect, we do not investigate this any further. }
As for the Fourier-space analyses, the $\chi^2$ values are lower than expected.
However, the same consistency applies  to the configuration-space measurements, 
In which the measured values are found to be within $2\sigma$ of the expected values. Here, $\nu=5$ and 3 for the RCSLenS and CFHTLenS surveys,
from which we get $\sigma_{\chi^2} = 3.2$ and $2.4$. For the combined $\xi^{\kappa_{\rm CMB}\kappa_{\rm gal}} + \xi^{\kappa_{\rm CMB}\gamma_t}$  statistics,
$\nu$ = (12-1) and (8-1) respectively.

As mentioned earlier, the SNR measured in the RCSLenS and CFHTLenS with all three estimators are similar and consistent within $\sim1\sigma$. 
Additionally, we find a mild gain in combining both  $\xi^{\kappa_{\rm CMB}\kappa_{\rm gal}}$ and $\xi^{\kappa_{\rm CMB}\gamma_t}$,
 due to the slightly different dynamical regions that are probed. 
We see this reflected in the SNR values, which show increases of about 6\%, 
 achieving the highest significance of our paper this way, with SNR=4.62 for CFHTLenS and RCSLenS combined.
%
Due to the limited number of CMB lensing simulations however, we do not explore the combination with Fourier analysis,
but this should certainly be considered in the analysis of upcoming survey.

It may seem surprising that the error on $A$ seems almost unchanged between CFHTLenS and RCSLenS, 
 since the error about the cross-correlation measurement is more precise by a factor of 2.
 What needs to be taken into account in the survey comparison is that the value of $A$ are quite different,
 hence it is the {\it fractional error} about $A$ that needs to be examined. 
 From both Tables  \ref{table:stat} and  \ref{table:stat_real}, we can see that the quantity $\sigma_A/A$ has dropped by almost a factor of two,
which is how the improvement should be assessed (this scaling is naturally reflected in the SNR metric). 
 
As an additional verification, we asserted from a $\chi^2$ calculation that the B-mode measurements
are all consistent with zero. Given a WMAP9 cosmology, the preferred amplitude is $\left| A\right| <0.15$ in all pipelines, with a detection significance of SNR $\lesssim$ 0.7.
In comparison, the preferred amplitude for E-modes is closer to  unity, with a significance of SNR $\sim$2-4 depending on the survey.
This quantitative test  gives us additional confidence that our measurement is not affected by B-mode contamination.











\subsection{Propagation of uncertainty about $n(z)$}
\label{subsec:nz}



As discussed in H15,  the cross-correlation  signal is significantly affected by the uncertainty on the source distribution.
For instance, if some sources with higher true redshift leak into the low-redshift part of the $n(z)$ due to a wrong photometric estimation, 
it will cause a significant mis-match between the predictions and measurement. 
The converse is equally true, and most of the challenge resides in modelling accurately the high-redshift tail of the distribution.
This tail is the least accurate regime of the photo-$z$ measurements, but at the same time it is the section of the $n(z)$ that 
maximally contributes to the signal -- mostly due to the shape of the geometric kernel $W^{\rm CMB}$.

 \citet{2015arXiv151203626C} found evidence from spec-$z$/photo-$z$ cross-correlations 
that  the stacked PDF calculated from the BPZ software was not quite accurate, such that  both the CFHTLenS and RCSLenS $n(z)$ were biased.
The redshift of the  peak in $n_{\rm RCS}$ is at worst contained within $\Delta z = ^{+0.24}_{-0.1}$ of the value 
found from the weighted stacked PDF.  
This was obtained in the context of tomographic analyses where the full redshift distribution is split  in a number of narrow redshift bins,
while most of this bias cancels in broader bands \citep{Hildebrandt2015}. 
Consequently, we do not apply any peak shift in the data, but instead examine how our uncertainty on the $n(z)$ affect the predictions, as described below.
It was shown in H15 that while holding the cosmological parameters fixed, variations on the peak of the $n(z)$  and on the high-redshift tail
could lead to changes of order $10\% - 20\%$ in the amplitude of the predictions, 
a precision level that likely also applies to CFHTLenS. 
However, the fact that H15 used a COSMOS-matched sample to extend their redshift distribution to higher $z$ probably introduced additional calibration errors.
For our CFHTLenS selection criteria, a  10\% precision is more likely.
The RCSLenS survey is shallower, hence variations in $n(z)$ coupled with a steeper slope in  $W^{\rm CMB}$ could decrease the precision even more.

In comparison to the the analysis of \citet{2015arXiv151203626C}, the redshift of our RCSLenS sample is known to a higher accuracy, thanks to the GALEX UV measurement.
When compared to the VIPERS spectroscopic data, 
\citet{2016arXiv160205917M} found a systematic bias in the mean photo-$z$ estimation of the CFHTLenS+GALEX data. 
Although negligible for $z_{phot}<1$,  the difference $z_{phot} - z_{spec}$ increases linearly and reaches 50\% by $z_{phot} = 1.8$. 
This bias is caused by mis-identified high-redshift outliers that have a matched $z_{spec}$ much closer to 0.5. 
The rate of outlier $\eta_{\rm out}$ is consistent to zero up to $z_{phot} = 1.0$, then is well modelled as 
\begin{eqnarray}
\eta_{\rm out}(z_{phot}) = 0.75(z_{phot} - 1). 
\end{eqnarray}
These VIPERS-CFHTLenS  galaxies are broadly distributed in the $i$-band magnitudes range  22-24. 
RCSLenS $r$-band data has a similar depth, and it reasonable to assume that the same level of outlier rate applies to the RCSLenS galaxies.
We next split $n(z)$ into two terms: 1) the `correct' photo-$z$ term that contributes a fraction $(1-\eta_{\rm out}(z))n(z)$ to the full redshift distribution, and
2) the `outliers' term, that contributes a fraction  $\eta_{\rm out}(z) n(z)$. 
The population of outliers peaks between $z=1.0$ and $2.0$,  and we redistribute it with the transform: $\eta_{\rm out}(z) n(z)\rightarrow \eta_{\rm out}(z/3) n(z/3)$.
This  maps the second term on to a smooth peak centred on $z=0.5$, closely matching the observed $z_{spec}$ distribution of the outliers.
We finally add this modified second term to the first and insert this new $n(z)$ in our predictions. 
This procedure is illustrated in the bottom panel of  Fig. \ref{fig:nz_zmax_cut}:
the two bell-shape curves show the distribution of outliers before (right) and after (left) this mapping, which
transforms the black solid $n(z)$ to the black dashed $n(z)$. 
This modification is then propagated into the prediction in the top and middle panels, from which we can see that this represents 
a $15\%$ systematic uncertainty on the amplitude of the signal, caused by the outliers. We also see that the scale dependence of this systematic bias can be safely neglected.

Another source of uncertainty in the $n(z)$ comes from the choice of functional form, which is mostly weighted by the low- and medium-redshift section of the distribution,
then extrapolated to high redshifts with some arbitrary power law. By construction, the fit will never convincingly match the sparse data, even though the galaxies at 
high redshift are those that contribute the most to the cross-correlation signal. It then becomes debatable whether one should use functions or  the actual $n(z)$ histogram when calculating the 
predictions. Since our RCSLenS $n(z)$ derives from CFHTLenS+VIPERS photometry, it is also probable that some features seen in the high redshift end of the 
distribution do not exist in the data, causing yet another bias in the results. 
In light of this, we examine the consequence of modifying the high-redshift tail on the predictions.
We do so by rejecting the high-redshift part of the $n(z)$ with three cuts, at $z=1.8$, $z=2$ and $z=3$, rescaling at each time to preserve the area under the curve.
When these cuts are applied, we find from Fig. \ref{fig:nz_zmax_cut} that the amplitude of the signal drops by $<3.5\%$ in the worst case, 
with very little angular dependence for $\ell>200$. This means that it can be captured as a small multiplicative factor in front of the theoretical prediction,
and is subdominant compared to the photometric redshift  outliers. 


\begin{figure}
\begin{center}
\includegraphics[width=2.7in]{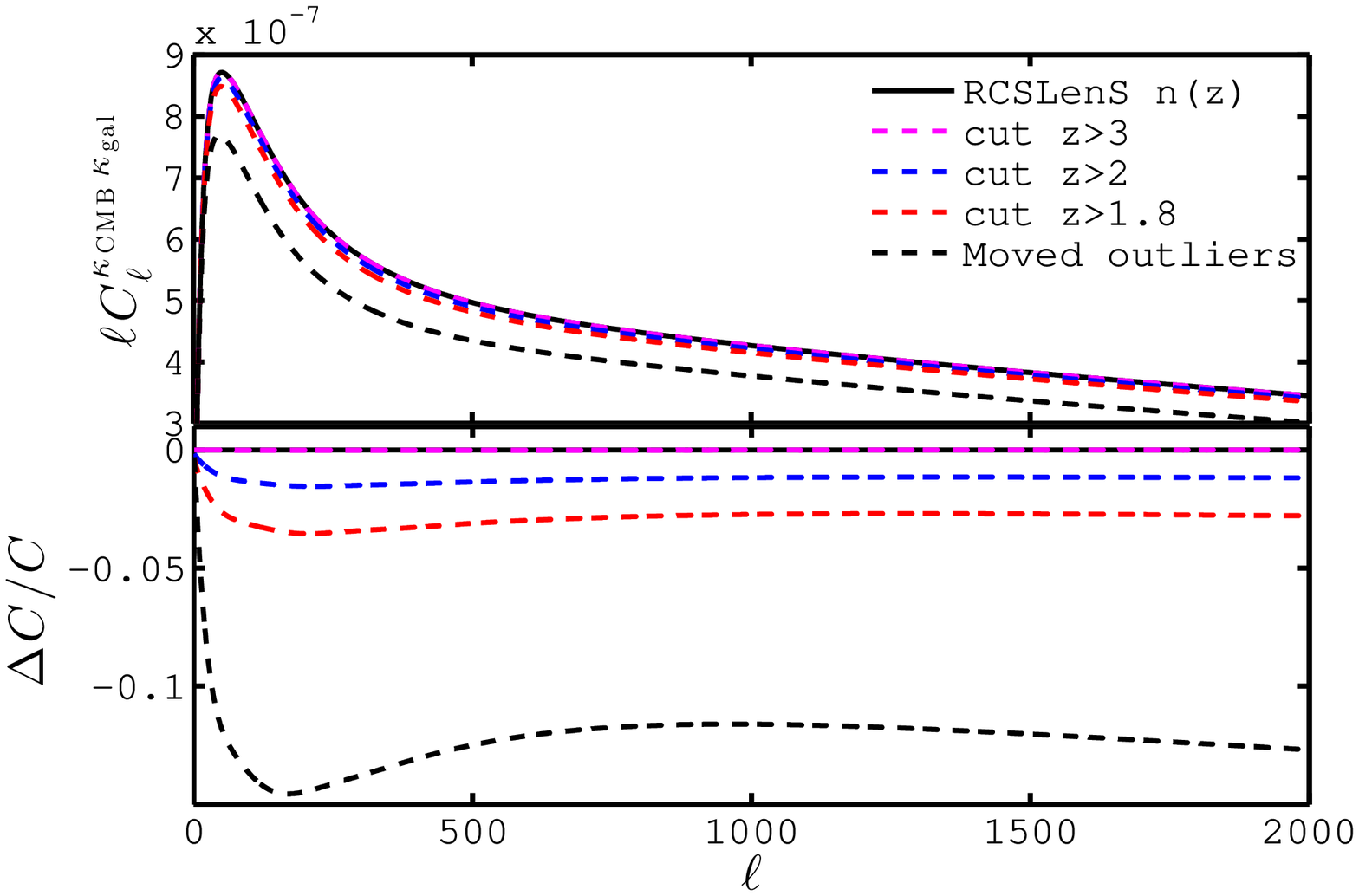}
\includegraphics[width=2.5in]{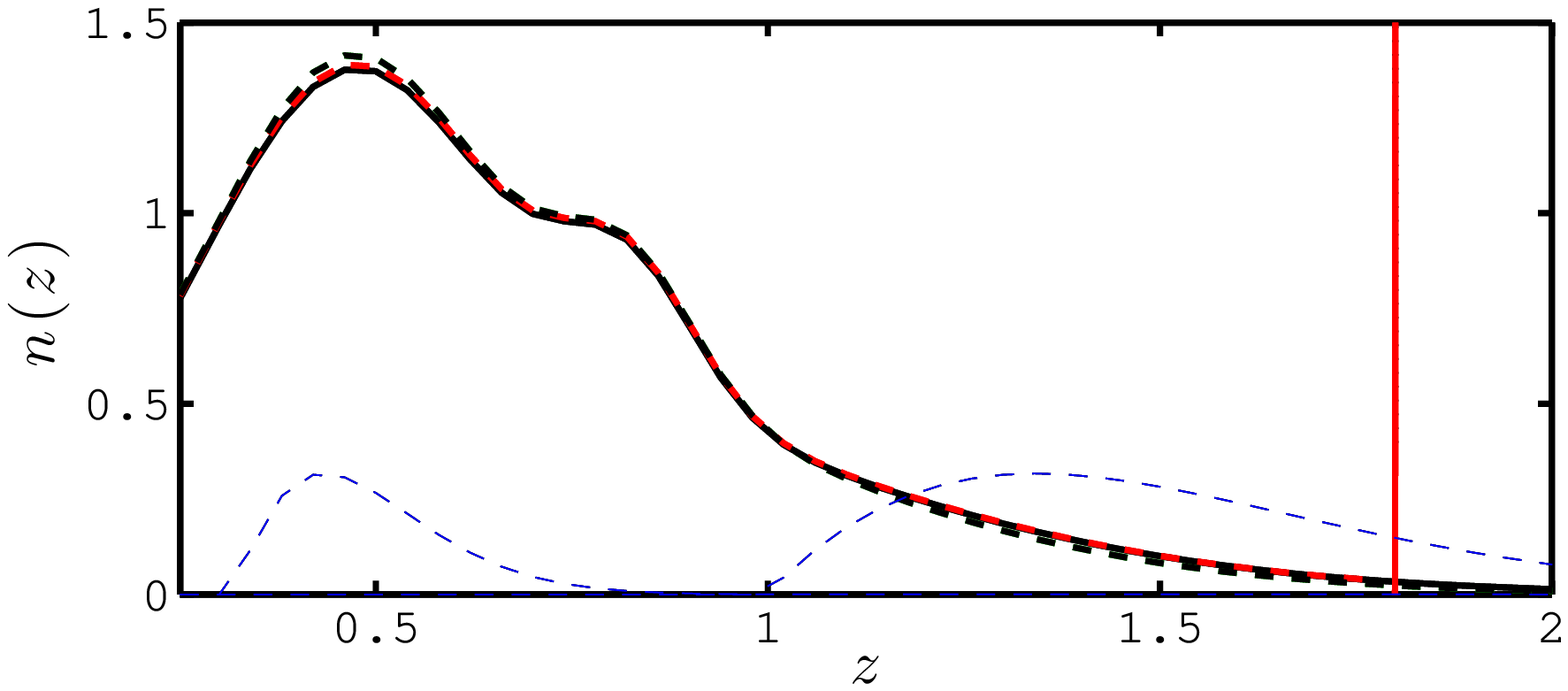}
\caption{(top:) Cross-spectrum predictions for a series of modifications applied to $n(z)$:
the effect of excluding the high-redshift tail with different cuts (magenta, blue and red dashed), and
the reassignment of the catastrophic outliers (black dashed).
(middle:) Fractional difference of the cross-correlation signal with respect to the fiducial RCSLenS $n(z)$, for each curve shown in the top panel.
(bottom:) Source redshift distribution for two of these modifications. The vertical line shows the redshift cut at $z=1.8$ that produces the red dashed curve.
The two dashed blue bell-shape  curves  represent the distribution of outliers before (right) and after (left) the correction that produces the black dashed curve. 
Their amplitudes are multiplied by a factor of 15 in this plot.   }
\label{fig:nz_zmax_cut}
\end{center}
\end{figure}

{\color{black}  It was suggested recently by \citet{2016arXiv160105720L} that an incorrect calibration of the shear signal data could possibly result in  large systematic effects in the cross-correlation measurement.
Indeed, the multiplicative correction $m$-term  can be subject to residual systematics effect that could directly leak into  the signal as  $A \propto (1+m)$.
If a wrong $m$ was leading the systematic budget in our measurement, a corrected $m$  would have to be allowed to take values as high as $m=0.73$ and as low as $m=-0.27$ to ensure $A = 1$
in all pipelines and for both cosmologies.  \citet{2016arXiv160105720L} and \citet{2016arXiv160207384B} remove priors on $m$ and allow even more extreme values.
%
%
However these $m$ values would be in addition to the $m$ calibration that we have already applied to our analysis based on the image simulations presented in  \citet{2013MNRAS.429.2858M}.
Whilst these have their limitations \citep[see][for a discussion]{2015MNRAS.454.3500K}, it is unlikely that the measured $m$ calibration could be so biased.  
In contrast, biases on photometric redshifts described in this section are of greater importance. 
Therefore, we do not allow for residual biases on $m$ not captured by the image simulation in  \citet{2013MNRAS.429.2858M},
as these are expected to be at the level of less than $\sim$5\% \citep{2015MNRAS.454.3500K}, resulting in  $\sim$5\% errors on $A$.
We apply the $m$-correction recommended by the CFHTLenS and RCSLenS
teams, and any residual biases would have to be propagated on $A$. }

To conclude this section, the impact of all $n(z)$-induced systematic effects are included as a 15\% uncertainty on the cosmological information extracted from the amplitude term $A$ defined in equation \ref{eq:chi}; no uncertainty on the $m$-correction is explicitly included.

\subsection{Contamination by intrinsic alignment}
\label{subsec:IA}

{\color{black}
In addition to the $\kappa_{\rm CMB} \kappa_{\rm gal}$ signal, CMB lensing correlates with the intrinsic alignment (IA) of foreground galaxies.
Galaxies located in the same cluster are connected by tidal forces, which also play a role in determining their intrinsic orientation 
\citep[for a recent review, see][]{2015SSRv..193..139K}. 
This   $\kappa_{\rm CMB}$-IA correlation therefore  contributes as an additive  secondary signal
whose shape and amplitude depend on  $n(z)$.
as  shown in \citet{2014MNRAS.443L.119H} and \citet{TroxelIshak14}.
 Unfortunately,  there is a large uncertainty in the modelling aspects of IA, as summarized in \citet{2015SSRv..193...67K}.


 For source redshift distributions similar to CS82, \citet{2014MNRAS.443L.119H} found that this effect   
 could account for 15-20\% of the $\kappa_{\rm CMB} \kappa_{\rm gal}$ measurement, 
 which would be {\it higher} in absence of IA\footnote{\citet{TroxelIshak14} use a different $n(z)$ which makes this comparison less direct.}.    
 Further investigation by \citet{2015MNRAS.453..682C} revealed that the strength of the  contamination also depends on the galaxy type. 
For $n(z)$ similar to that of  CS82, they found that IA from red galaxies alone could contribute to  $\sim$10\% of the signal; 
the blue galaxies are still consistent with no alignment, but given the current uncertainty, it could also double the total $\kappa_{\rm CMB}$-IA.

In both of these studies, the contamination decreases (increases) for deeper (shallower) surveys: \citet{2015MNRAS.453..682C}  estimated that lowering the peak of  $n(z)$ by $\Delta z = 0.1$
increases the contamination  by 15\%. Another 20\% increase can also be encountered if the high-redshift tail is given more weight.
We show the $n(z)$ from CS82, CFHTLenS and RCSLenS in Fig. \ref{fig:nz}, and find that  the peak of the CFHTLenS source distribution is lower than that of CS82
by about $\Delta z = 0.1$, but that its high-redshift tail is also lower. These two differences act in opposite directions, such that  $\kappa_{\rm CMB}$-IA in both surveys
should be similar. In comparison, the RCSLenS is much shallower and its source distribution  peaks at $z=0.5$. 
Surprisingly, we find that it is in fact very similar to the  CS82 $n(z)$ once shifted by $\Delta z = 0.2$ (labelled `CS82 shifted' in  Fig.  \ref{fig:nz}), especially at the high-redshift end.
From the peak-shift dependence described in  \citet{2015MNRAS.453..682C} and assuming the non-linear model of intrinsic alignement \citep{2007NJPh....9..444B}, 
we can estimate that the  $\kappa_{\rm CMB}$-IA contamination from the red galaxies
in RCSLenS should contribute to about 13\% of the signal. 


According to \cite{2014MNRAS.443L.119H} and \citet{2015MNRAS.453..682C},  the overall scale dependence of  
$\kappa_{\rm CMB}\mbox{-IA}/(\kappa_{\rm CMB}\kappa_{\rm gal} + \kappa_{\rm CMB}\mbox{-IA})$ is relatively flat, with at most $\sim50\%$ deviations about its mean in the linear IA model, 
 and possibly as low as $\sim20\%$ in the non-linear IA model. 
Given the current uncertainty in our measurement, we are not sensitive to deviations from flatness, hence we can safely treat IA  as a multiplicative  contribution to our signal.  
In other words, we effectively treat the amplitude parameter $A$ from equation \ref{eq:chi} as  a sum of two terms: $A= A_{\rm th} + A_{\rm IA}$.
From the above discussion, we can estimate the $A_{\rm IA}$ terms to be $0.13$ and $0.09$ for RCSLenS and CFHTLenS respectively.
In the SNR calculation,  we do not correct the data nor the predictions for the IA part, which means that 
either the data are biased low, or the models are biased high. 
If we opted for the former option, we would increase the SNR of the measurement, but
this seems artificial at this point.}

We must also mention that \citet{2015arXiv151002617L} found that within the tidal torque alignement model, the $\kappa_{\rm CMB}$-IA contamination acts in the opposite way,
namely the contamination tends  to increase the signal, which would be {\it lower} without IA. 
This clearly illustrates the need for development in the field of IA before one can draw more precise conclusions from our cosmological measurement.
In the meantime, any interpretation that rely on the merged  $A_{\rm th}$ and $A_{\rm IA}$ quantities should be robust against this uncertainty.



\subsection{Comparison with previous results}

H15 reported the first detection of the galaxy lensing CMB lensing cross-correlation by combining the data from the ACT 
and the CS82 surveys. Due to the high resolution of the ACT lensing maps, it was possible to extract the signal with a significance of $4.2\sigma$,
over a contiguous unmasked area of 121 deg. sq. This area is comparable in size to the sum of the 4 CFHTLenS fields. 
The redshift distribution from CS82 peaks at $z\sim0.7$ with mean at $0.9$, and is obtained from a COSMOS \citep{2007ApJS..172....1S} matched sample, leading
to important uncertainty in the exact $n(z)$ due to sampling variance. Marginalizing over this, they found $A^{\rm Planck} =0.78 \pm 0.18$ 
and  $A^{\rm WMAP9} =0.92 \pm 0.22$, which translates into a $\sim 12\%$ constraint on $\sigma_8(z=0.9)$.
They identified the uncertainty in $n(z)$ and  the unknown level of contamination by intrinsic alignment (see Section \ref{subsec:IA}) as their
dominant limiting factor.

The analysis performed by LH15 examined the cross-correlation between all four CFHTLenS fields and the Planck full sky lensing maps. 
The galaxy selection function differs from the one used in this paper, as they included $z_B>1.3$ galaxies,
with $n(z)$ modelled from COSMOS. This choice was motivated by an increase in SNR, even though it introduced an uncertainty on $n(z)$ in the range 10-20\%. 
They found evidence for the signal with a significance SNR=$1.9\sigma$ and $2.0$, for the 2013 and 2015 Planck releases respectively. 
The difference in significance with H15 can be largely attributed to the higher 
noise level present in the Planck lensing maps, compared to ACT. 
Following an analysis similar to H15, they found $A^{\rm Planck} = 0.44\pm0.22$
and $A^{\rm WMAP9} = 0.52\pm0.26$, from the Planck 2015 data. 
The low values of $A$ were unexpected and attributed to uncertainty in $n(z)$ distribution $(\simeq 10\%)$, masking of point sources in the CMB lensing analysis $(\simeq 15\%)$ 
and intrinsic alignment $(<10\%)$.
Nevertheless, from a simple parameterization of the cosmology dependence of the signal, they were able to measure  $\sigma_8 (\Omega_{\rm M}/0.27)^{0.41} = 0.63^{+0.14}_{-0.19}$. 


From our measurement of the Planck 2015 lensing - CFHTLenS cross-correlation, 
we are able to recover their results with our method P5 - which is the closest match to that of  LH15 -
and obtain $A^{\rm WMAP9} = 0.63\pm0.31$, and $A^{\rm Planck} = 0.58\pm0.28$.
As seen from Table \ref{table:stat}, pipeline P5 favours values of $A$ that are lower by  as much  as 15\%, compared to P1, which has the highest SNR.
This is exactly the level of  bias that was claimed by LH15 to be caused by cluster masking in the Planck map.
Recall that P5 is the only one for which the Planck mask is applied on the $\kappa_{\rm gal}$ maps,
thereby rejecting the important contribution from many clusters in the cross-correlation. 
This bias is even more pronounced in the RCSLenS measurement.  
A more detailed analysis of this bias could be further investigated from hydrodynamical simulations
or within the halo model, but such a study falls outside the scope of this paper. 
{\color{black} Results from our $N$-body simulations detailed in Section \ref{subsec:sims} suggest that masking the densest regions could reduce the signal by as much as a  factor of 2,
which would explain and release the tension with the Planck cosmology observed in LH15. The fact that we do not multiply the masks in P1-4 makes a large difference,
since the clusters that are masked in the $\kappa_{\rm CMB}$ are un-masked in $\kappa_{\rm gal}$ and can therefore contribute to the signal, at least partially.
Indeed, Fig. \ref{fig:cmb_x_kappa_RCS_real} shows that there is a significant signal even beyond a degree, thereby exceeding the angular size of the largest masked point sources
in the Planck CMB lensing maps. }
Aside from this, all our CFHTLenS measurements agree with both Planck and WMAP9 predictions to within $1.5\sigma$.
We obtain similar results from our configuration-space  estimator $\kappa_{\rm CMB} \kappa_{\rm gal}$, with 
$A^{\rm WMAP9} = 0.61\pm0.32$, and $A^{\rm Planck} = 0.55\pm0.29$.

{\color{black}

From the measurements of DES $\times$ Planck and SPT lensing,  K15 reported 
$A^{\rm Planck}_{\rm Planck} = 0.86\pm0.39$ and $A^{\rm Planck}_{\rm SPT} = 0.88\pm0.30$, respectively\footnote{The superscript still refers to the cosmology adopted, while the subscript refers to the CMB lensing data set used in the measurement.}.
As opposed to LH15, they find no evidence of tension with the Planck cosmology.
In their analysis, the bright clusters were also masked from the SPT lensing map, however these regions were filled 
from a Wiener filter interpolation, such that the mask has no small scales structures in it.  
The cross-correlation measurement was carried out with the {\tt PolSpice} software \citep{Szapudi2001}, which deals properly with 
distinct masks without multiplying them. This strategy is therefore similar to our method P1.

 A clear feature  seen in both Tables \ref{table:stat} and \ref{table:stat_real} is that RCSLenS favours higher values of $A^{\rm Planck}$ 
 compared to Planck $\times$ CFHTLenS, SPT$\times$ DES and Planck $\times$ DES, generally higher than unity. 
 This is not too surprising since all these measurements are extracted from different parts of the sky, and fluctuations are expected.  
 All our measurements are consistent with $A=1$ within $1.3\sigma$. 
 

}

\subsection{Sensitivity to baryons and massive neutrinos}
\label{subsec:nu_b}

\begin{figure}
\begin{center}
\includegraphics[width=2.7in]{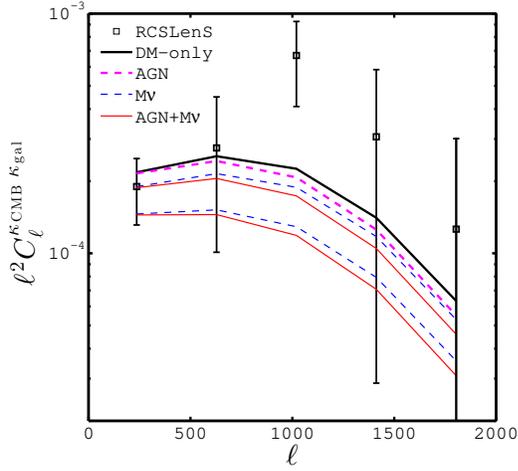}
\caption{Effect of baryon feedback and massive neutrinos on the cross-correlation signal, compared to a $\Lambda$CDM model (WMAP9) with massless neutrinos and no feedback (`DM-ONLY' model).
The dotted magenta line shows the `AGN' model extracted from the OWL simulations \citep{vanDaalen2011},  
while the dashed blue lines  represent, from top to bottom, the effect of  massive neutrinos with $\Sigma m_{\nu}$ = 0.2 and 0.6eV. 
The 0.4eV models fall in the middle hence are not shown.  
Solid red lines show the combined effect. The RCSLenS measurement shown here has been processed with pipeline P2, and models are scaled by the best fit amplitude $A$ = 1.47 (see Table \ref{table:stat}). The error bars shown are estimated from the full covariance matrix.}
\label{fig:b_nu}
\end{center}
\end{figure}

{\color{black}
The CMB lensing $\times$ galaxy lensing measurements  discussed in this paper can serve as an additional probe, providing
an extra constraint on the parameters that describe intrinsic alignment, redshift distributions and cosmology.
While H15 and LH15 have already set constraints on $\Omega_{\rm m}$ and $\sigma_8$,  
 we examine here the possibility to jointly weigh the neutrinos 
and constrain galaxy evolution models through their feedback on the total matter density.  

For our predictions, we following the strategy of \citet{HWVH2015}, which includes the joint neutrino+baryon effects 
with two independent multiplicative terms.
First, baryonic effects on the cross-correlation functions are implemented via their impact on the matter power spectrum, $P(k,z)$, 
as measured\footnote{For the  $P(k,z)$ from the OWLS, see: {\tt http://vd11.strw.leidenuniv.nl}} in \citet{vanDaalen2011}.
These are based on the OverWhelmingly Large (OWL) simulations \citep{Schaye2010}, 
an ensemble of cosmological hydrodynamical simulations that include realistic astrophysical processes and feedback mechanisms coming from supernovae, AGN, stellar winds, etc. \citep{Schaye2010}. Among the available models we selected the one named `AGN'. 
According to  \citet{vanDaalen2011}, this model provides the best fit for a number of observations. 
We compute the relative effects compared to the OWL `DM' model, a dark matter only universe, and define the baryon feedback bias as:
\begin{eqnarray}
b^2_{\rm AGN}(k,z) \equiv \frac{P^{\rm AGN}_{\rm OWL}(k,z)}{P^{\rm DM}_{\rm OWL}(k,z)},
 \label{eq:b_feedback}
\end{eqnarray}

Second, we include  massive neutrinos with total mass $M_{\nu} = \sum m_{\nu}$ = 0.2, 0.4 and 0.6eV in a similar manner, using their implementation in the {\small CAMB} numerical code \citep{Lewis2000a}
to compute the massive neutrino bias relative to the  $\sum m_{\nu}=0.0$eV case:
\begin{eqnarray}
b^2_{M_{\nu}}(k,z) \equiv \frac{P^{{\rm DM}+M_{\nu}}_{\rm CAMB}(k,z)}{P^{\rm DM}_{\rm CAMB}(k,z)},
\label{eq:nu_feedback}
\end{eqnarray}

 We model the impact of massive neutrinos and baryon feedback on our fiducial model from these two biases: 
\begin{eqnarray}
  P^{{\rm AGN}+M_\nu}(k,z) = P^{\rm fid}(k,z) \times b^2_{M_{\nu}}(k,z) \times b^2_{\rm AGN}(k,z). 
  \label{eq:pk_tot}
\end{eqnarray}
This assumes that both effects can be treated separately, which is known to hold at  least up to $k=8 h{\mbox{Mpc}}$ \citep{Bird2012}.
The same conclusion are reached from a halo model approach in \citet{2016arXiv160202154M}.
By feeding the results from equation \ref{eq:pk_tot} into equation \ref{eq:limber_cross}, we propagate these two effects onto  $C_{\ell}^{\kappa_{\rm CMB}\kappa_{\rm gal}}$.

Fig. \ref{fig:b_nu} shows the impact of massive neutrinos and baryonic feedback on the cross-spectrum, relative to a $\Lambda$CDM universe.
We see that the presence of  massive neutrinos (dashed blue) affects all scales, 
causing a suppression that varies from 10\% to 50\% in the mass range  that we probe, and is  maximal at  $\ell\sim2000$. 
This effect is partly degenerate with a decrease in $\Omega_{\rm M}$ and $\sigma_8$, as well as with the contamination
by intrinsic alignment and incorrect estimation of the $n(z)$, although these do not have the exact same scale dependence. 

Similar conclusions can be drawn for the prospect of constraining baryonic feedback models with this measurement.
The distinctive features of the AGN model are mostly  
felt at $\ell > 1000$, which has a characteristic tilt that might be easier to capture than the neutrino mass. 
%
%
In the optimistic scenario where $\Omega_{\rm M}$ and  $\sigma_8$  are fixed by the large scale measurements or other cosmological probes, 
the shape of the cross-correlation with respect to a $\Lambda$CDM universe can serve to constrain baryonic and neutrino physics.
{\color{black} For these two effects to be measured however, the analysis needs to include very small angular scales
that are usually excluded when considering baryon feedback as nuisance.}
}

Also plotted in Fig. \ref{fig:b_nu} is the RCSLenS data using pipeline P2, which shows the highest SNR.
Given the current level of uncertainty, it is not possible to place constraints on any of these effects,
but it becomes clear that they cannot be neglected in upcoming surveys without biasing low  
 the best-fit values of $\sigma_8$ and $\Omega_{\rm M}$. 
In order to reject some of the models shown here, the size of the error bars will need to shrink by at least a factor of 3,
calling for survey combinations whose common area is of the order of a few thousands square degrees such as DES or KiDS, 
or with CMB lensing maps with less noise such as ACT and SPT. High hopes are placed on upcoming surveys  to improve this measurement.

{\color{black} Note finally that the $\ell$-dependence of the IA contamination can cause some confusion in the  search for the scale-dependent features in the signal left by baryonic feedback and massive neutrinos. However, their signatures are likely to be  stronger than that of IA,  causing suppression of up to 50\% (see Fig.  \ref{fig:b_nu}), versus $\sim10$\% for IA,
 and therefore still measurable.}





\section{Conclusion}

Cross-correlation measurements between two independent cosmological datasets  are  potentially free from contamination by residual systematics. 
As such they can be treated  as distinct and possibly cleaner probes of the large scale structure. 
In this paper, we combine the public Planck 2015 lensing map with the CFHTLenS and RCSLenS data
and extract the cross-correlation signal with a significance of {\color{black}$\sigma=4.62$}.
The combined galaxy surveys cover a total unmasked area of 747.2 deg$^2$, a factor of four larger than previous measurements by H15, LH15 and K15. 
Whereas these three papers focused on the  Fourier-space estimator $C_{\ell}^{\kappa_{\rm CMB}\kappa_{\rm gal}}$, 
we present here the first configuration-space measurements.
We achieve this with two different estimators, $\xi^{\kappa_{\rm CMB}\kappa_{\rm gal}}$ and $\xi^{\kappa_{\rm CMB}\gamma_t}$,
and find an excellent agreement between the configuration-space and Fourier-space approaches, providing extra confidence in the accuracy of our results.
Since the estimators probe slightly different scales, they are in that sense complementary, and we achieve a small gain in SNR by combining them.
More importantly, the main advantage of these configuration-space estimators is that they are less affected by mask-induced effects.

The three previous $C_{\ell}^{\kappa_{\rm CMB}\kappa_{\rm gal}}$ measurements used different methods and datasets, 
and it is unclear which technique is  optimal. Moreover,  tracking potential systematic effects across them becomes highly challenging.
To address this important  issue, we repeated our analysis using five different pseudo-$C_{\ell}$ pipelines and investigated their impact on both the SNR and the cosmological results. 
We find important differences,  largely due to changes in the implementation of the masks, but these are not distinguishable in the data given the current statistical precision.
These pipelines are verified against a suite of high resolution $N$-body simulations, 
and favour  strategies in which the CMB and galaxy masks are {\it not} combined and applied to the two datasets,
as opposed to the approach used  in previous studies. 
Mask multiplication reduces the SNR and biases low the amplitude of the measured signal.



We finally compare our measurements against theoretical models that include massive neutrinos and baryonic feedback,
and show that these  can easily be confused with intrinsic alignment contamination or 
offsets in the source redshift distribution. For instance, neutrinos with mass $M_{\nu}-0.2$eV produce a suppression of about 10-15\% 
at all scales, just as the preferred IA models. We also estimate that our signal could be biased similarly by the photometric redshift outliers.
 Some baryonic feedback models like the OWL-AGN predict a significant suppression 
with a tilt that is more pronounced and harder to reproduce by other secondary signals. There is thus hope that future experiments will
offer significant constraining power on neutrinos and baryon physics, once we overcome the challenges related to the modelling  of IA contamination
and to the photometric redshift measurements.

\section*{Acknowledgements}
We would like to thank Duncan Hanson for his significant help during the early stages of this analysis,
and Yuuki Omori and  Gilbert Holder  for helpful discussions on cross-correlation measurements.
We acknowledge the help of Andy Taylor in the development of the PCL formalism and in the calculation of the extra statistical uncertainty. 
We extend our gratitude to the McGill Space Institute for its kind hospitality, where most of this work was completed,
and to  all the CFHTLenS team for having made public their high quality shear data.
We would also like to thank Matthias Bartelmann for being our external blinder, 
revealing which of the four RCSLenS catalogues analysed in this study was the true data.
We are grateful to the RCS2 team for planning the survey, applying for observing time, and conducting the observations. 
This work is based on observations obtained with MegaPrime/MegaCam, a joint project of CFHT and CEA/DAPNIA, at the Canada-France-Hawaii Telescope (CFHT) which is operated by the National Research Council (NRC) of Canada, the Institut National des Sciences de l'Univers of the Centre National de la Recherche Scientifique (CNRS) of France, and the University of Hawaii. This research used the facilities of the Canadian Astronomy Data Centre operated by the National Research Council of Canada with the support of the Canadian Space Agency. 
RCSLenS data processing was made possible thanks to significant computing support from the NSERC Research Tools and Instruments grant program.
We acknowledge use of the Canadian Astronomy Data Centre operated by the Dominion Astrophysical Observatory for the 
National Research Council of CanadaÕs Herzberg Institute of Astrophysics. 
Computations for the $N$-body simulations were performed in part on the Orcinus supercomputer at the WestGrid HPC consortium (www.westgrid.ca), 
in part  on the GPC supercomputer at the SciNet HPC Consortium. SciNet is funded by: the Canada Foundation for Innovation under the auspices of Compute Canada;
the Government of Ontario; Ontario Research Fund - Research Excellence; and the University of Toronto. 
JHD acknowledge support from the European Commission under a Marie-Sk{\l}odwoska-Curie European Fellowship (EU project 656869) and from the NSERC of Canada,
which also supports LvW and AH.
TT is supported by the Swiss National Science Foundation (SNSF).
HH is supported by an Emmy Noether grant (No. Hi 1495/2-1) of the Deutsche Forschungsgemeinschaft.
RN acknowledges support from the German Federal Ministry for Economic Affairs and Energy (BMWi) provided via DLR under project no.50QE1103.
AC, MV and CH acknowledge support from the European Research Council under FP7 grant number 279396 (MV), 240185 (AC) and 647112 (CH). 
MV is also supported by the Netherlands Organisation for Scientific Research (NWO) through grants 614.001.103.
TDK is supported by a Royal Society URF.

{\footnotesize
{\it Author Contributions:} All authors contributed to the development and writing of this paper. 
The authorship list reflects the lead authors of this paper (JHD, TT, AH,  LvW and MA), 
followed by two alphabetical groups. 
The first alphabetical group consists of the founding core team of the RCSLenS collaboration (lead by HH)
whose significant efforts produced the  final RCSLenS data products used in this analysis.
The second group includes external collaborators. 
JHD carried out the Fourier measurements, the simulation products and lead the analysis;
TT and AH developed the configuration-space estimator; LvW produced the $\kappa_{\rm gal}$ maps; MA developed the PCL infrastructure;
SA, JC and TM contributed to the CFHTLenS-VIPERS catalogue.}


\bibliographystyle{hapj}
\bibliography{cross_corr}


\appendix

\section{Forward modelling}
\label{sec:mode_mixing}

\begin{figure}
\begin{center}
\includegraphics[width=2.7in]{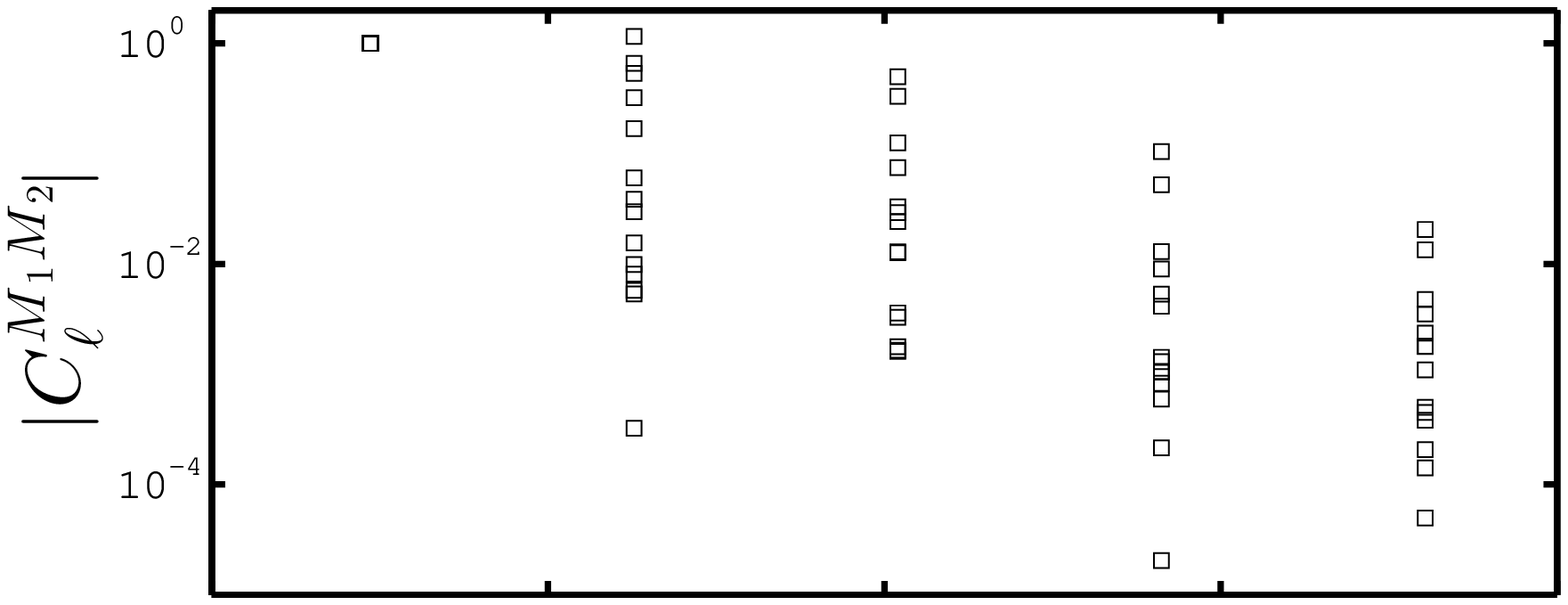}
\includegraphics[width=2.7in]{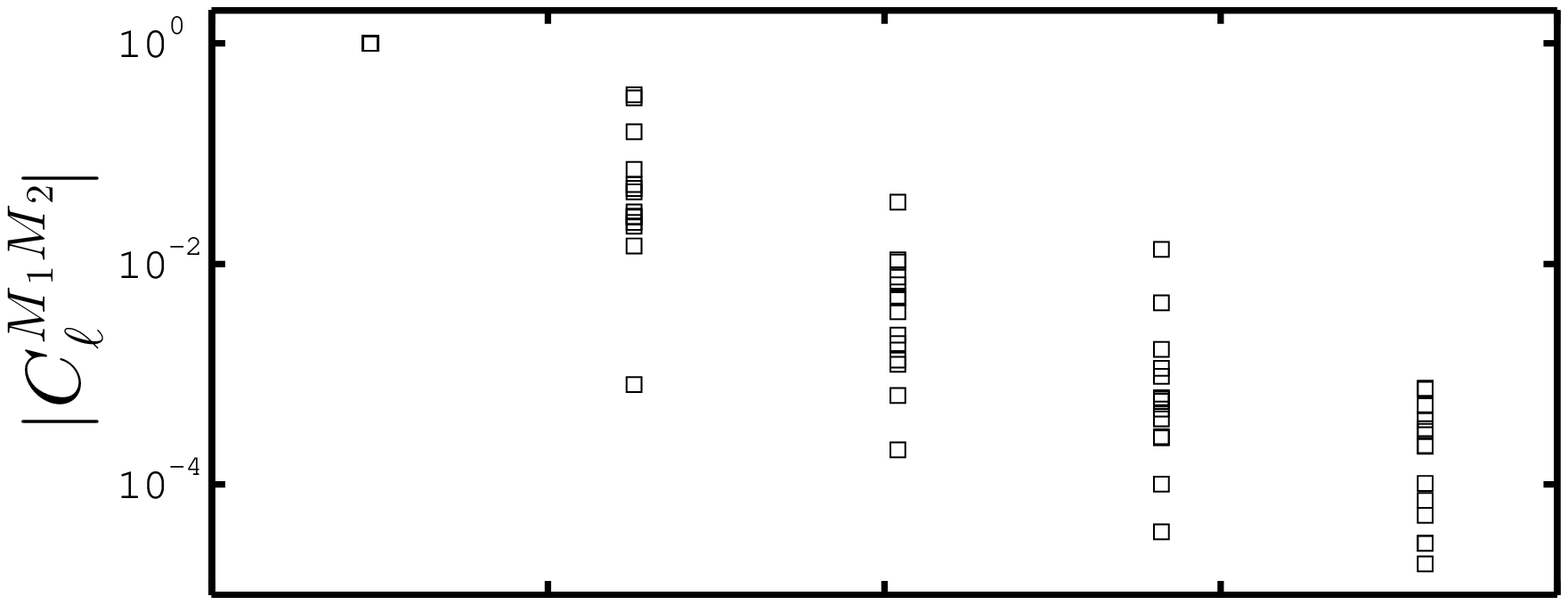}
\includegraphics[width=2.7in]{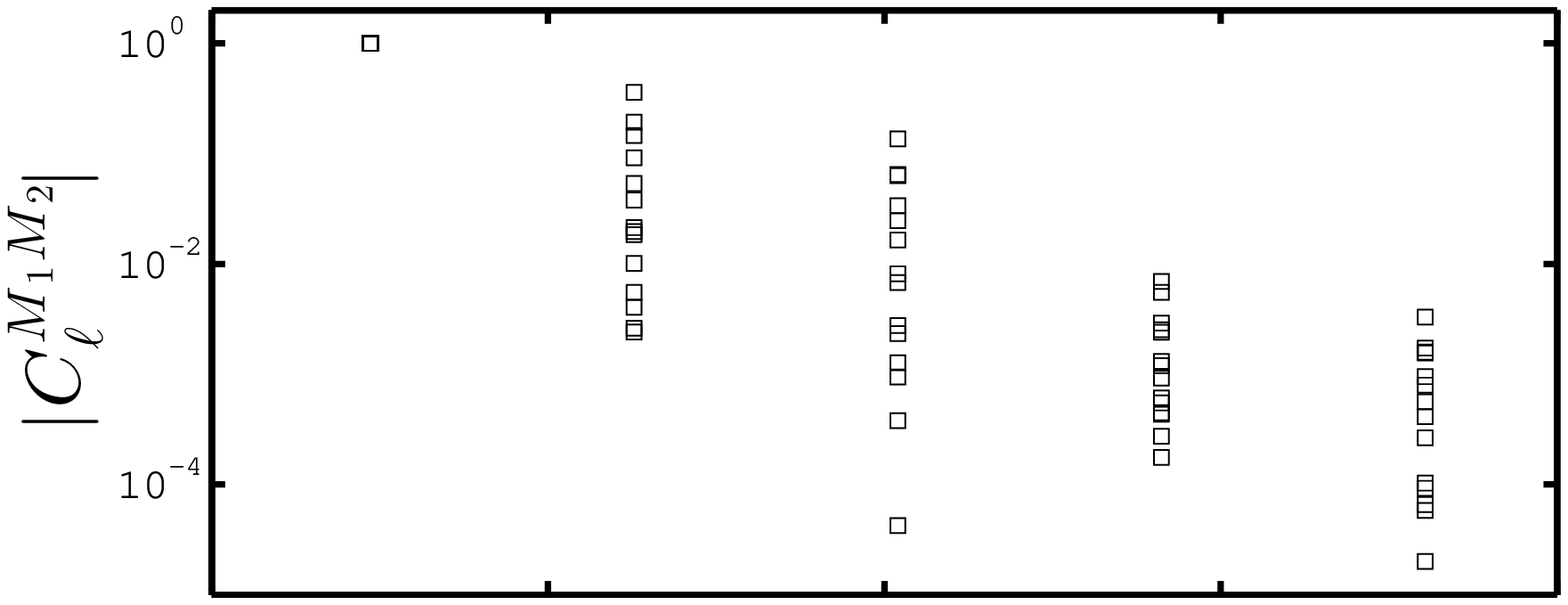}
\includegraphics[width=2.7in]{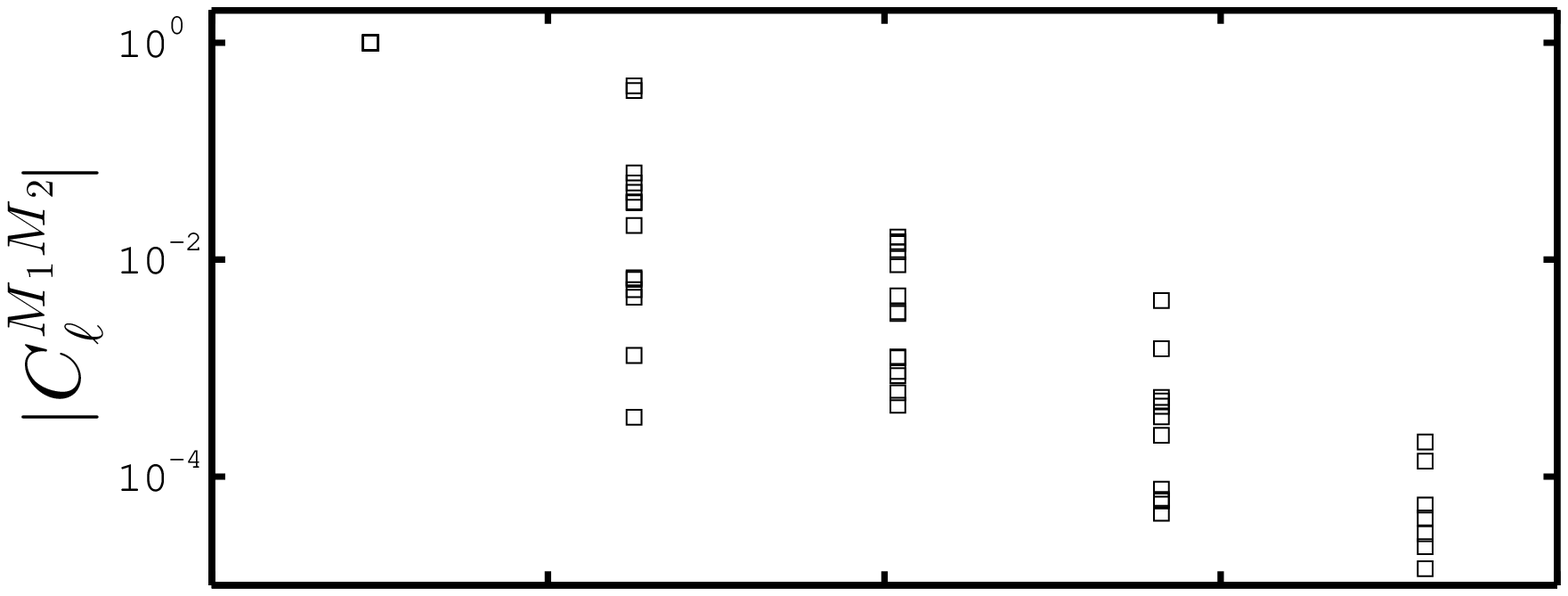}
\includegraphics[width=2.7in]{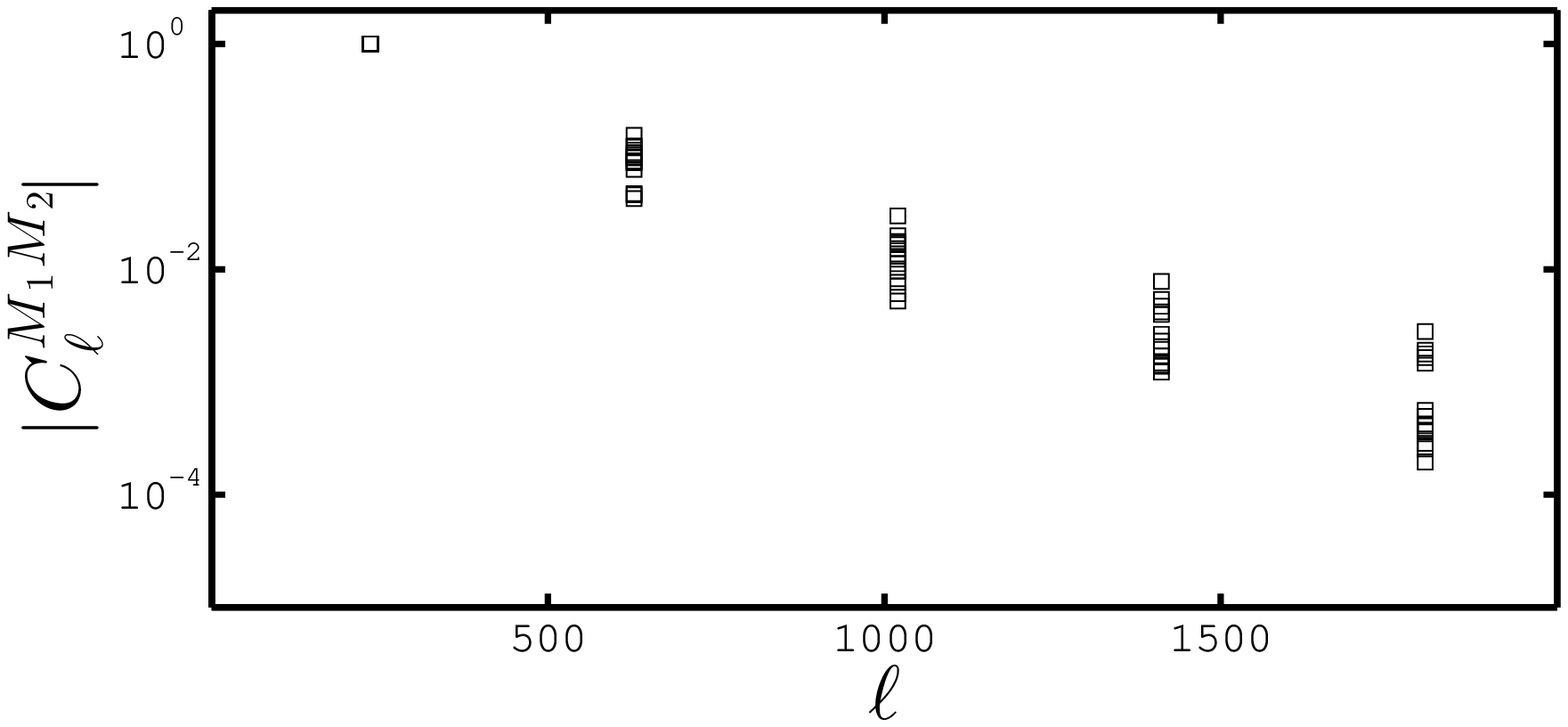}
\caption{ Cross-spectra between the RCSLenS masks and  the Planck 2015 masks, over-plotted for each of the 14 RCSLenS fields.
Top to bottom panels are for pipelines 1 to 5.
All measurements are normalized such that their lowest $\ell$-mode coincide, in order to show the level of mode-mixing that enters equation \ref{eq:mixmatrix}. 
The suppression of mode-mixing due to apodization is clearly visible here.
The two fields which show the largest amount of cross-correlation for $\ell>200$ (causing larger mode-mixing) are CDE0310 and CSP0320.}
\label{fig:cross_mask}
\end{center}
\end{figure}


{\color{black}This section describes our forward modelling algorithm, which turns a theoretical $C_{\ell}^{\kappa_{\rm CMB}\kappa_{\rm gal}}$ into 
a pseudo-$C_{\ell}$ that can then be compared against the Fourier-space measurements 
 detailed in  Section \ref{subsubsec:fourier}. At the core of this procedure is the calculation of the mode-mixing matrix, which is sensitive to the analysis masks $M_i$. 
Since these masks vary for different fields `$i$', the pseudo-$C_{\ell}$ are field-dependent quantities and are therefore  written as $\widetilde{C}_{\ell}^{\kappa_{\rm CMB}\kappa_{\rm gal},i}$.
The derivations presented here are adapted from the cosmic shear method described in \citet{Asgari2015} in the context of $\kappa_{\rm CMB}$-$\kappa_{\rm gal}$ cross-correlations. 

The very first step consists  
in interpolating the $C_{\ell}^{\kappa_{\rm CMB}\kappa_{\rm gal}}$ predictions (from equation \ref{eq:limber_cross})  in fine $\ell$-bins (these are  unique for each field, 
see Section \ref{subsubsec:fourier}).
This ensures that the binning process does not create biases between the data and theory.
 %
We next  combine these predictions with the field masks. 
For the sake of clarity,  let us assume first that the underlying $\kappa$ maps are unbiased, such that 
$\widetilde{C}_{\ell}^{\kappa_{\rm CMB}\kappa_{\rm gal},i}$  can be expressed as a convolution between the underlying $C_{\ell}^{\kappa_{\rm CMB}\kappa_{\rm gal}}$ and the Fourier transform of the mask, $W^{i}(\ell)$.
 When measuring the auto-spectrum, $C_{\ell}^{\kappa}$, this procedure can be  written as  \citep{2002ApJ...567....2H}:
 \begin{eqnarray}
 \widetilde{C}_{\ell}^{\kappa,i} &=& \left\langle \int \!\!  C_{\ell'}^{\kappa} \left|W^i({\boldsymbol \ell} - {\boldsymbol \ell}')\right|^2 \mbox{d}^2{\boldsymbol \ell'} \right\rangle_{\Omega},
 \end{eqnarray}
 where the angle brackets refer to the angular averaging of $\boldsymbol \ell$, i.e. $\langle ... \rangle_{\Omega} = (2\pi)^{-1} \int ... \mbox{d}\Omega$,
 and $\mbox{d}^2{\boldsymbol \ell'} = \ell \mbox{d}\ell \mbox{d}\Omega$. 
 We can generalize the above results to the cross-spectrum between any maps $\kappa_{1}$ and $\kappa_{2}$ as 
 \begin{eqnarray}
  \widetilde{C}_{\ell}^{\kappa_{1}\kappa_{2},i} &\equiv& \left\langle \int \!\!  C_{\ell'}^{\kappa_{1}\kappa_{2}} \left[W_1^i(\boldsymbol L)W_2^{i*}(\boldsymbol L)\right] \mbox{d}^2{\boldsymbol \ell'} \right\rangle_{\Omega},
 \end{eqnarray}
 where we defined ${\boldsymbol L} = {\boldsymbol \ell} - {\boldsymbol \ell}'$.
 Note that in the above two expressions, there are two angular integrations, running over $\Omega$ and $\Omega'$, the polar angles of the vectors $\boldsymbol \ell$ and  $\boldsymbol \ell'$ respectively.
 Assuming that  the true convergence cross-spectrum $C_{\ell}^{\kappa_{1}\kappa_{2}}$ is isotropic, it can be taken outside of the two angular integrations:
 \begin{eqnarray}
    \widetilde{C}_{\ell}^{\kappa_{1}\kappa_{2},i}  &=& \!\! \int \!\!  C_{\ell'}^{\kappa_{1}\kappa_{2}}\left[  \left\langle  W_1^i(\boldsymbol L)W^{i*}_2( \boldsymbol L)   \right\rangle_{\Omega, \Omega'} \right] \ell' \mbox{d}\ell'.
    \label{eq:ConvolveLimber}
     \end{eqnarray}
     Let us focus on the inner bracketed term, which is averaged over both $\Omega$ and $\Omega'$. 
     Under the change of variables $\Omega_L \equiv \Omega + \Omega'$ and $\eta \equiv \Omega - \Omega'$,  
     we can write d$\Omega$d$\Omega'$ = 2d$\Omega_L$d$\eta$, where $\Omega_L$ can be identified as  the polar angle of $\boldsymbol L$.
     In this $\{\Omega_L,\eta\}$ coordinate system, the double integration over the mask cross-spectra simplifies greatly: 
      \begin{eqnarray}
\left\langle   W_1^i(\boldsymbol L)W^{i*}_2(\boldsymbol L)  \right\rangle_{\Omega, \Omega'}& = &  \int \!\! \!\! \int \!\! W_1^i(\boldsymbol L)W^{i*}_2(\boldsymbol L) \mbox{d}\Omega \mbox{d}\Omega' \\
                                                                                        & = &2 \!\! \int_0^{\pi} \!\! \mbox{d}\eta \!\! \int \!\! W_1^i(\boldsymbol L)W^{i*}_2(\boldsymbol L) \mbox{d}\Omega_L \\
                                                                                        &=& 4\pi \!\! \int_0^{\pi} \!\! W_1^i(L)W^{i*}_2(L)\mbox{d}\eta .
       \label{eq:int_eta}                                                                                 
 \end{eqnarray}
     Written in this form, the angular dependence of $L$ vanishes;  
     the integrand is simply the angle-averaged cross-spectrum of the two masks, $C_{\ell}^{M_1 M_2}$, which is easy to compute as a separate step,
     for each of the 5 pipelines under investigation in this paper.  
     The mask cross-spectra are provided in Fig. \ref{fig:cross_mask}, for methods P1-5.
     These are measured separately for each of the 14 RCSLenS fields, rebinned in the final coarse binning scheme, and over-plotted in this figure. 
     Only three fields out of the 18 have cross-spectrum elements that are higher than 10\%, which indicates that the effects of mode-mixing, averaged over the 18 fields,
     should be minor. 
     The integral in equation \ref{eq:int_eta} runs over $\eta$, which can be identified -- from the change of variable described below equation \ref{eq:ConvolveLimber} -- 
     as the angle between  $\boldsymbol \ell$ and $\boldsymbol \ell'$.
     From the  law of cosines, we can write $L^2 = \ell^2 + \ell'^2 - 2\ell \ell' \mbox{cos}\eta$, which maps  $\eta$ to  $L$ given a pair  $\{\ell,\ell'\}$,  and finally perform the integral numerically. 
     The term on the left hand side of equation \ref{eq:int_eta} is the mode-mixing functional $\mathcal{M}^i(\ell,\ell') $ and applies in the continuous case.
     We can rewrite equation \ref{eq:ConvolveLimber} as:
 \begin{eqnarray}
    \widetilde{C}_{\ell}^{\kappa_{1}\kappa_{2},i}   &\equiv& \!\! \int \!\!  C_{\ell'}^{\kappa_{1}\kappa_{2}} \mathcal{M}^i(\ell,\ell') \ell' \mbox{d}\ell'.
 \end{eqnarray}
  Turning the integral into a discrete sum, and absorbing all normalization factors in this process, we finally obtain the mode-mixing matrix ${\rm M}_{\ell\ell'}^i$ as:
 \begin{eqnarray}                           
   \widetilde{C}_{\ell}^{\kappa_{1}\kappa_{2},i}   & =&   \sum_{\ell'} C_{\ell'}^{\kappa_1 \kappa_2} \mathcal{M}_{\ell\ell'}^i \ell' \Delta \ell' \\
                           &\equiv& {\rm M}_{{\ell}{\ell'}}^iC_{\ell'}^{\kappa_1 \kappa_2}.
   \label{eq:mixmatrix}
 \end{eqnarray}
 The smoothing of the maps can be included at this stage by multiplying the results by the appropriate kernels $\left|S_{1}^i(\ell)S_{2}^i(\ell)\right|$, which are typically Gaussian functions.
 To simplify the notation further, this term can be absorbed in the definition of $\rm M_{\ell \ell'}^i$. 
 We could then use equation \ref{eq:mixmatrix}  to propagate the effect of masking and smoothing on the theoretical models,
 rebin these theoretical pseudo-$C_{\ell}$ onto the coarse $\ell_c$ bins, and finally compare with the binned data.

     \begin{figure}
\begin{center}
\includegraphics[width=2.7in]{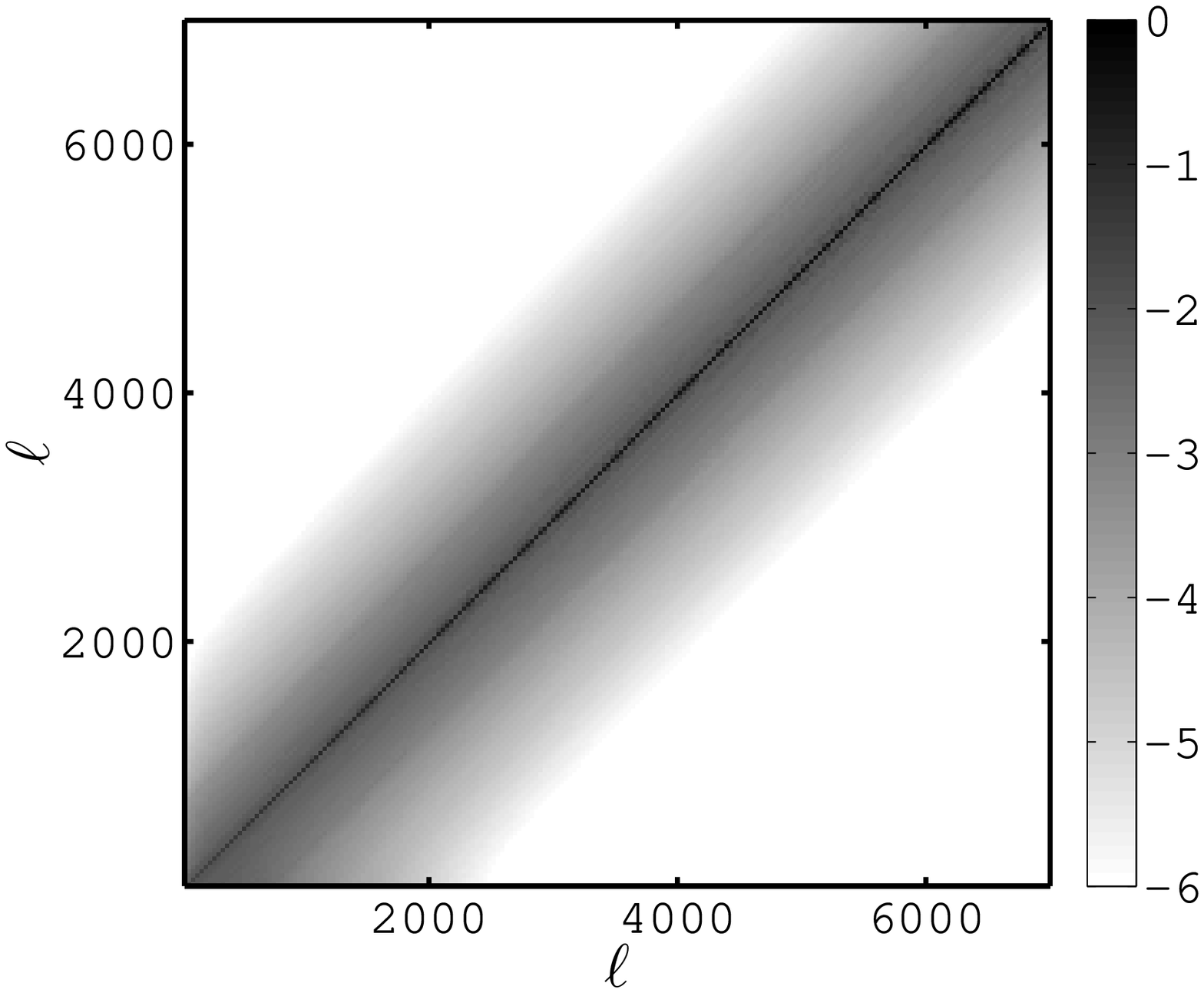}
\includegraphics[width=2.7in]{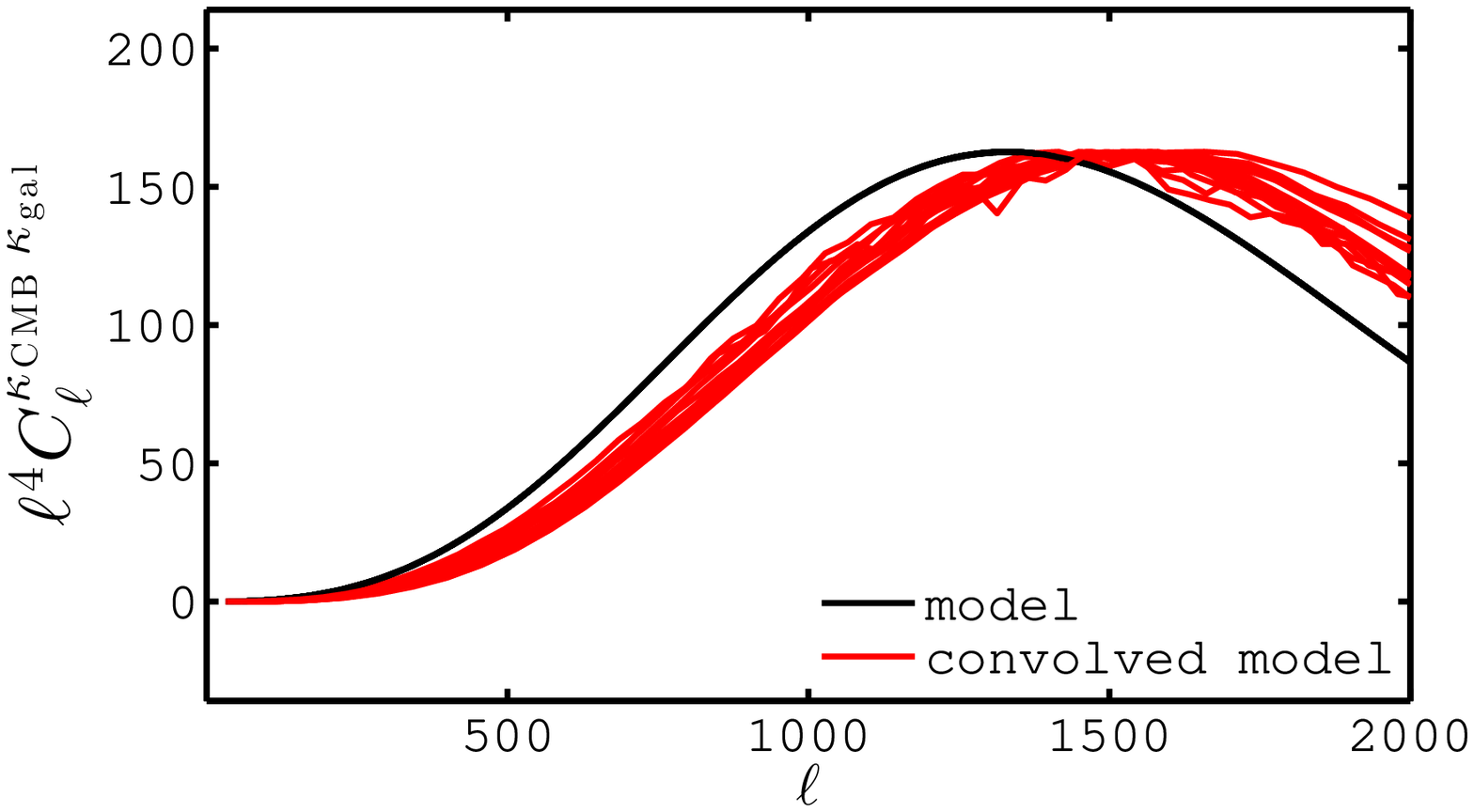}
\caption{(top:) Mode-mixing matrix for the largest RCSLenS field, shown in log$_{10}$ colour scale to increase the contrast, for pipeline P5.
The field smoothing term  $S(\ell)$ is not included here in order to highlight the mode-mixing part.
The off-diagonal elements are highly suppressed due to apodization.
(bottom:) Comparison of  the predictions with (red) and without  (black) applying the mode-mixing matrices, for the 14 RCSLenS fields.  }
\label{fig:mix_matrix}
\end{center}
\end{figure}

 As mentioned before, this procedure is exact only when the $\kappa$ maps are unbiased, such as for lensing simulations.
 Generally, the optimal mode-mixing matrix involves an additional level of complexity,  which arises from the fact that it is not the $\kappa_{\rm gal}$ maps that are masked,
 but the {\it shear} maps, ${\gamma_1, \gamma_2}$. The convergence maps are extracted by performing  KS over the masked shear catalogues. 
 The analyses mask are re-applied on the 
 reconstructed $\kappa_{\rm gal}$ maps, but it would be false to assume that there are no other effects.
 On the contrary,  the mask propagates {\it everywhere} on the convergence maps.
 The reconstructed map, $\widetilde{\kappa_{\rm gal}}$,  is therefore biased, but can be simply related  to the true map in Fourier space \citep{Asgari2015}: 
 \begin{eqnarray}
 \widetilde{\kappa_{\rm gal}}(\boldsymbol \ell) = \int \kappa_{\rm gal}(\boldsymbol \ell') W({\boldsymbol \ell} - {\boldsymbol \ell'}) \mbox{d}{\boldsymbol \ell'} \mbox{cos}2\phi_{\ell\ell'}
 \label{eq:kappa_bias}
 \end{eqnarray} 
 where $\phi_{\ell\ell'}$ is the angle between the vectors ${\boldsymbol \ell}$ and  ${\boldsymbol \ell'}$.
 In the $(\Omega_L,\eta)$ coordinates introduced above, this angle is exactly $\eta$.
 When $\kappa$ comes from the masked shear fields, we must replace  $\left[\kappa_{\rm gal}W({\boldsymbol \ell} - {\boldsymbol \ell'})\right]$ by $\left[\kappa_{\rm gal}W({\boldsymbol \ell} - {\boldsymbol \ell'})\mbox{cos}2\eta\right]$, which produces  an extra factor of $\mbox{cos}2{\eta}$  inside the inner term of equation \ref{eq:ConvolveLimber}.
 We are neglecting E/B-mode mixing in this calculation, which is justified since it was shown to be minimal for the CFHTLenS maps in \citet{VanWaerbeke2013}.
 
 Note that we apply this correction only to the $\kappa_{\rm gal}$ map; correcting the $\kappa_{\rm CMB}$ map requires 
 a modification to the quadratic estimator of \citet{2002ApJ...574..566H}, which is beyond the scope of this paper. 
In addition, this is probably not necessary given the current precision of the CMB lensing maps, which are noise-dominated. 
 In the end, we obtain our optimal mode-mixing matrix by replacing equation \ref{eq:int_eta} with 
  \begin{eqnarray}
\left\langle   W_1^i(\boldsymbol L)W^{i*}_2(\boldsymbol L)  \right\rangle_{\Omega, \Omega'}& = & 4\pi \!\! \int_0^{\pi} \!\! W_1^i(L)W^{i*}_2(L) \mbox{cos}2\eta \mbox{d}\eta 
       \label{eq:int_eta_corr}                                                                                 
 \end{eqnarray}
 }
 
 The mode-mixing matrix for pipeline P5 is presented in the upper panel of Fig. \ref{fig:mix_matrix}, in the case of the RCSLenS field CDE2143. 
 Once applied on the theoretical model, this matrix redistributes the power to different $\ell$-modes, as seen in the bottom panel.
 The forward-modelled predictions are visibly changing from field to field, which needs to be accounted for in a full pseudo-$C_{\ell}$ analysis.

\section{Pipeline Comparison}
\label{sec:pipelines}

{\color{black} We have shown in Section \ref{subsec:results} that the choice of analysis pipeline has repercussions on the SNR and on the measured signal.  
We compare in  Fig. \ref{fig:cmb_x_kappa_RCS_all_P} the $\hat{C}_{\ell}^{\kappa_{\rm CMB}\kappa_{\rm gal}}$ measurements from RCSLenS, for the 
5 pseudo-$C_{\ell}$ pipelines  listed in Table \ref{table:pipeline}.
Amongst them, P1 has the highest SNR hence is presented in the main body of this paper (Fig. \ref{fig:cmb_x_kappa_RCS}). 
The black squares represent the first data point, $\hat{C}_{\ell=227}^{\kappa_{\rm CMB}\kappa_{\rm gal}}$,
measured with methods P1 to P5. The red circles and blue triangles are for $\ell = 1015$ and $\ell = 1803$ respectively,
while the black crosses represent $\hat{C}_{\ell=1803}^{\kappa_{\rm CMB}\kappa_{\rm gal-B}}$, measured from the $\kappa_{\rm gal}$ B-mode maps.
Ideally, all points from a given symbol would align horizontally, however we see that it is not the case.
They are nevertheless  consistent within $1\sigma$, which gives us confidence that our conclusions are relatively  insensitive to the choice of pipeline.
As future surveys will gain in statistical precision, the error bars on this figure are expected to shrink enough that there will be room for tension 
across different  pipelines.  It will therefore be important to reproduce such a validation test to assert the robustness of the measurement.}




\begin{figure}
\begin{center}
\includegraphics[width=3.1in]{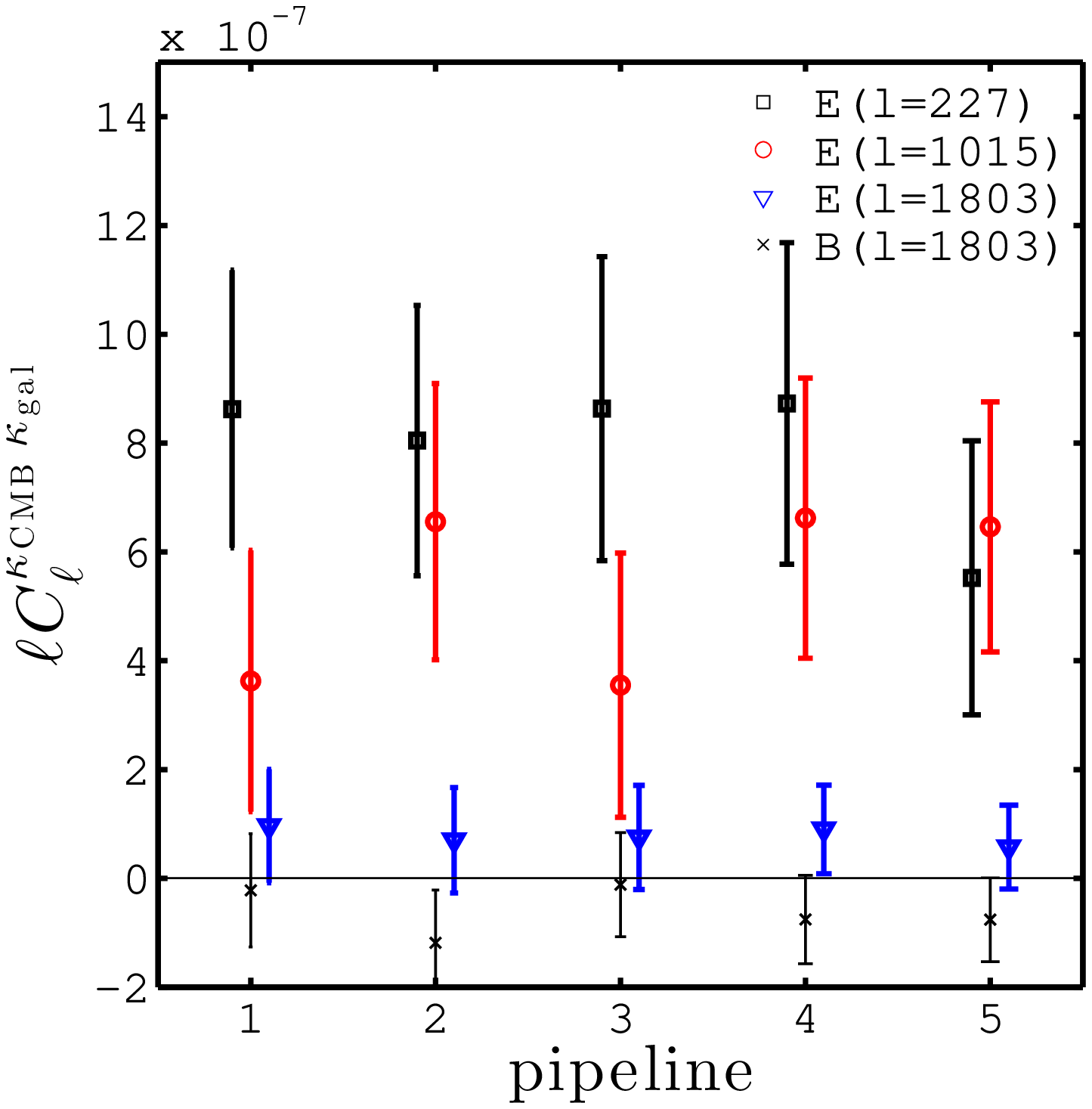}  
\caption{
 RCSLenS measurements of the $\hat{C}_{\ell}^{\kappa_{\rm CMB}\kappa_{\rm gal}}$ for all pipelines listed in Table \ref{table:pipeline}.
 E-modes are presented for multipoles $\ell = 227$, 1015  and 1803 in  squares, circles and triangles. B-modes are shown for the   $\ell = 1803$ with cross symbols. 
 All pipelines are consistent within $1\sigma$, however they result in important differences in significance as seen in Table \ref{table:stat}. }
\label{fig:cmb_x_kappa_RCS_all_P}
\end{center}
\end{figure}

\bsp

\label{lastpage}

\end{document}